\documentclass[twocolumn]{aastex701}

\usepackage{amsmath}

\begin{document}

\title{The Occurrence of Circumbinary Planets Around TESS M-dwarf Eclipsing Binaries}

\correspondingauthor{Dominic Oddo}
\email{doddo@unm.edu}

\author[0000-0002-2702-7700]{Dominic Oddo}
\affiliation{Department of Physics and Astronomy, The University of New Mexico, 210 Yale Blvd NE, Albuquerque, NM 87106, USA}
\email{doddo@unm.edu}

\author[0000-0003-2313-467X]{Diana Dragomir}
\affiliation{Department of Physics and Astronomy, The University of New Mexico, 210 Yale Blvd NE, Albuquerque, NM 87106, USA}
\email{dragomir@unm.edu}

\author[0000-0002-7595-6360]{David V. Martin}
\affiliation{Department of Physics and Astronomy, Tufts University, 574 Boston Ave, Medford, MA 02155, USA}
\email{david.martin@tufts.edu}

\author[0000-0002-7907-2634]{Wata Tubthong}
\affiliation{Department of Physics and Astronomy, Tufts University, 574 Boston Ave, Medford, MA 02155, USA}
\email{chanita.tubthong@tufts.edu}

\author[0000-0002-7127-7643]{Te Han}
\affiliation{Department of Physics and Kavli Institute for Astrophysics and Space Research, Massachusetts Institute of Technology, Cambridge, MA 02139, USA}
\email{tehan@mit.edu}

\begin{abstract}

We present the first determination of the occurrence of transiting circumbinary planets (CBPs) in TESS data. Since tight binaries represent a constrained environment relative to single stars, CBPs contribute to our understanding of planet formation and evolution. We applied our custom single-transit detection algorithm to the recently-published sample of TESS eclipsing binaries containing two M-dwarfs (M\&Ms), finding only one signal attributable to a new candidate planet-sized object. We further performed sensitivity analysis and applied corrections to account for the TESS window function to arrive at our survey completeness. We find a 95\% confidence level (CL) upper limit on the occurrence rate of coplanar CBPs with sizes in the range $2-16 R_\oplus$ and with orbital periods up to $600$ days of 33.0\%. If the mutual inclinations of CBPs are isotropically distributed with respect to the binaries, the occurrence rate upper limit increases to 40.2\%. Our occurrence rate estimates are limited in part by the misalignment of our short-period binaries relative to the line of sight.

\end{abstract}

\section{Introduction} \label{sec:intro}

The confirmation of more than 6,300 exoplanets at the time of writing\footnote{according to the NASA Exoplanet Archive: \url{https://exoplanetarchive.ipac.caltech.edu/}} has revealed a remarkable diversity of planetary architectures. Among the most compelling are circumbinary planets (CBPs), which are worlds that orbit two stars simultaneously. These systems serve as natural laboratories for studying the robustness and adaptability of planet formation and evolution. Their existence directly contradicts early theoretical concerns that planets could not form in such environments \citep{meschiari2012planet,paardekooper2012not}, especially in the regions close to the binary stability limit where they are found \citep{Welsh_2018,li2016uncovering}. Studying the properties and demographics of the transiting CBP population is therefore important to understanding planet migration \citep{thun2018migration,Penzlin_2021}. Finally, since CBPs are often in the habitable zones of their hosts, they represent an interesting case of planetary habitability \citep{Kane_2013,Eggl_2020}.

Space missions have transformed the study of CBPs \citep{Kostov_2023}. The NASA Kepler mission \citep{Borucki_2010} provided the first unambiguous detections of a dozen transiting circumbinary worlds, such as the seminal discovery of Kepler-16b \citep{Doyle_2011}, the Kepler-47 multi-planet system \citep{Orosz_2012_kepler47,Orosz_2019}, and one of the longest-period transiting planets known, Kepler-1647b \citep{Kostov_2016_kepler1647}. From the Kepler sample, it was estimated that the occurrence rate of co-planar CBPs with periods $<300$ days is approximately 10\% \citep{Armstrong_2014}. After Kepler, the NASA Transiting Exoplanet Survey Satellite (TESS; \citealt{ricker2015TESS}) mission has again reshaped the exoplanet landscape. TESS is conducting an all-sky survey of millions of the nearest and brightest stars across the sky \citep{Kunimoto_2022}, unlocking new possibilities for planet discovery, including two transiting CBPs so far \citep{Kostov_2020_toi1338,Kostov_2021}. Crucially for CBPs, TESS surveys possibly millions of new eclipsing binary star systems (EBs; \citealt{Kostov_2025}), which are the primary CBP hosts. 

The greatest bottleneck to the continued study of CBPs is that these planets are difficult to find, for numerous reasons. Conventional transit search methods (e.g., the Box-fitting Least Squares, or BLS, algorithm \citealt{Kovacs_2002}), which tend to rely on the stacking of periodic transit signals as a way to increase signal-to-noise ratio, do not work for transit signals which vary in periodicity, depth, and duration due to geometric and dynamical effects that are signature to binary systems. 3-body geometry, including the continuously-changing sky positions of the host stars causes CBP transits to appear aperiodically, and 3-body dynamics change the planet's orbit \citep{Martin_2015}. Further, unstable inner regions imply that transiting CBPs have longer orbital periods than other transiting planets. From a data perspective, photometrically variable tight binaries (due to, for example, increased activity) can both create false positive transit signals and mask true transits. Until this point, all known transiting CBPs have been detected by-eye\footnote{although detections of some individual transit events have been aided by photodynamical models predicting when transits ought to occur}. However, TESS's sheer volume of data makes by-eye searches of the entire catalog intractable, and it is predicted that many transiting CBPs exist in the full TESS sample \citep{Kostov_2020}.

Automated methods for detecting CBPs have been advanced, but such pursuits are not without their own challenges. There exists a body of quasi-periodic and/or dynamically-informed search methods across decades, demonstrating multiple attempts to overcome this problem \citep{Jenkins_CMDrac_1996,Doyle_2000,Ofir_2008,Carter_2013,Kostov_2013,Armstrong_2014,Klagyivik_2017,Windemuth_2019_CBPs,Martin_2021}. However, such methods may not be most suitable to a transit search in TESS data. First, dynamical searches are strongest with a long baseline, but a majority of TESS sources only have one sector of data (roughly 27 days) per two year mission, leaving dynamical models largely unconstrained. Second, these search methods tend to be computationally expensive, and searching TESS's incredibly large target base is nearly intractable. It is therefore necessary to utilize a single-transit search method \citep{Davies2026}, which offers much greater speed. In this work, we present a new single-transit search for CBPs, which can be applied readily to numerous light curves. It is notable that there are also methods to search for non-transiting circumbinary objects from their dynamical signatures via the analysis of eclipse timing variations \citep{Deeg_2000,Borkovits_2016,Esmer_2025}, but a majority of these detections (save \citet{Goldberg_2023}) have been controversial \citep{Zorotovic_2013,Kalomeni_2025}. More recent work has utilized eclipse times to constrain the apsidal precession of the binary \citep{Thornton2026}. We focus our efforts on transiting CBPs, which produce reliable signatures with proven followup methods.

In this work, we present the first search for circumbinary planets orbiting M\&M EBs, representing a new frontier in the continuing quest for small CBPs. We present our novel single-transit search methodology for the first time. We further perform injection and recovery testing of physical planets to place limits on the occurrence rate of small CBPs. In Section \ref{sec:sample_data}, we briefly outline our sample. In Section \ref{sec:search_method}, we describe our transit search methodology, while the results of the search and vetting are described in Section \ref{sec:results}. In Section \ref{sec:occurrence}, we describe our injection and recovery protocol and completeness calculations, yielding a calculation of the occurrence rates of CBPs orbiting M\&M EBs. In Section \ref{sec:disc}, we place these results in the context of other results and share implications. Concluding thoughts are in Section \ref{sec:conc}.

\section{Sample Selection and Data}\label{sec:sample_data}

EBs with known periods and eclipse properties are most suitable for CBP searches, since they are known short-period and edge-on systems. The most likely scenario is that CBPs are coplanar with the binary, which increases the chances of detection with transits \citep{Martin_2015}, but CBPs can exist stably in highly misaligned or polar orbits \citep{Baycroft_2025_Polar,Childs_2022,Smallwood_2025,RMartin_2023}. Such planets do not transit often and are thus difficult to find \citep{Chen_2022}.

The recent publication of a new catalog of 1,291\footnote{It was found that one row in the original catalog was duplicated} homogeneously-selected M\&M EBs \citep{Oddo_2026} with known orbital and physical properties enables a new search for CBPs (with sky positions shown in the top panel of Fig. \ref{fig:tess_coverage}). The fact that the masses and radii of these binaries are known means that planet properties can be calculated in physical, rather than relative, terms. Additionally, the fact that the catalog contains only low-mass main sequence M dwarfs enables sensitivity to smaller planets. However, this choice also introduces trade-offs that bear acknowledging. M-dwarf binaries are intrinsically less luminous than their higher-mass counterparts, meaning a magnitude-limited sample like ours preferentially captures the nearest or most luminous members of the class and is shallower in volume than an equivalent survey of K or G-type binaries. Furthermore, the binary fraction among M dwarfs is lower than for solar-type stars \citep{Offner_2023}, which limits the achievable sample size for a given survey footprint and constrains the statistical power of occurrence rate estimates. All relevant observational (i.e., sky positions and magnitudes) orbital properties (i.e., periods and eclipse properties) and physical properties (i.e., physical sizes, masses, and temperatures) are given in \citet{Oddo_2026} Tables 1-3. 

\subsection{Observations and data}

\begin{figure}
    \centering
    \includegraphics[width=0.98\linewidth]{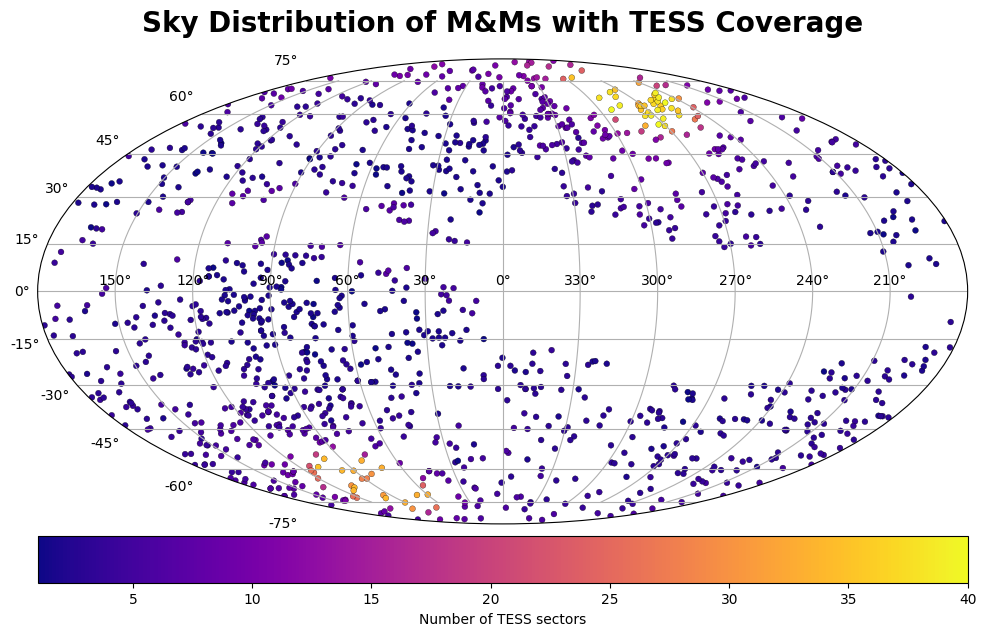}
    \includegraphics[width=0.98\linewidth]{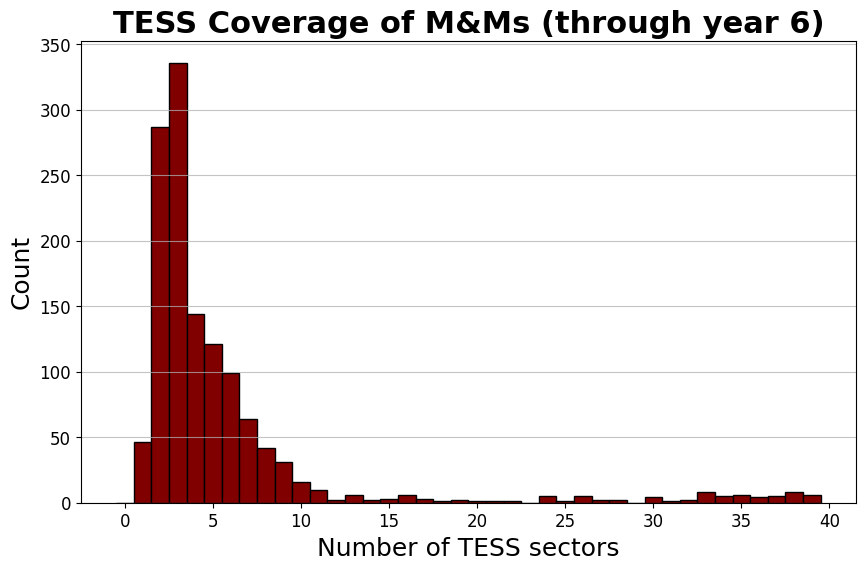}
    \caption{Top: Sky positions of the TESS M\&M sample (shown in a full-sky Mollweide projection), with number of single-sector TESS light curves through Year 6 as colorbar. Bottom: Histogram distribution of the number of single-sector TESS light curves through  Year 6 for our M\&M sample.}
    \label{fig:tess_coverage}
\end{figure}

In this work, we utilized light curves generated from TESS Full Frame Images (FFIs) from the start of the Primary Mission in 2018 through the end of Year 6, comprising data from Sector 1 up to Sector 83. 
The distribution of the number of TESS single-sector light curves through year 6 for our M\&M sample is shown in the bottom panel of Fig. \ref{fig:tess_coverage}. 
Note that \citet{Oddo_2026} only used data up to sector 55 (through the end of TESS Year 4). We utilize more TESS data such that the total baseline increases by two years, and that the median number of sectors per TIC rises from 2 to 3, and the average number of sectors for each source rises to 5.7. The total number of single-sector light curves we analyze across all TICs is 7422. We generated light curves for all TICs using the \texttt{tglc} package. TESS FFIs are stored at 30 minute (1800 second) cadence in Sectors 1-26, 10 minute (600 second) cadence in Sectors 27-55, and 200 second cadence in Sectors 56-83, which is relevant for calculation of transit signal-to-noise ratio since the number of points for a given duration will change. 

\section{Planet Candidate Identification Methodology}\label{sec:search_method}

In our efforts to detect new CBPs in TESS data, we developed a novel single transit search pipeline. The search pipeline proceeds in four major stages, with an additional fifth fitting step applied to planet candidate detections. The four major stages are as follows: 1. light curve loading and eclipse masking; 2. de-trending; 3. signal-to-noise ratio based transit event search; and 4. vetting of candidate signals. All targets are treated uniformly in this search scheme. In the following subsections, we will step through the pipeline stages. We verified this method as below by testing on the existing transiting TESS CBPs (TOI-1338 b and TIC 172900988 b). 

\subsection{Light curve loading and eclipse masking}

In the first stage, all sectors for the target are concatenated into one light curve. It is known that false transit-like events tend to occur near sector gaps (e.g. the starts/ends of sectors and mid-sector data downlink gaps), so data within 0.5 days of these sector gaps is masked. Additionally, spuriously deep single points, which can affect the calculation of transit depth downstream, are removed. A point is spuriously deep if it is outside a primary or secondary eclipse and is deeper than 75\% of the primary eclipse depth. This threshold is benchmarked to the primary eclipse depth because it is assumed that planet transits will be significantly shallower than the binary eclipses. 

Next, orbital and physical properties for the system from \citet{Oddo_2026} are queried and stored. Using these, points in the light curve which are identified as during primary or secondary eclipse are masked from the light curve. It is notable that the binary ephemerides in \citet{Oddo_2026} were calculated using only 4 years of TESS data, but 6 years of TESS data were used in this project. In most cases, a continuation of the linear ephemeris is sufficient to mask the binary eclipses. However, occasionally, the ephemeris becomes ``stale" or slightly incorrect in the final two years of TESS data. To account for this, we split the light curve into individual sectors, and in sectors beyond Sector 55, we manually recalculate the phase of both the primary and secondary eclipse, and mask according to the new phase center. This was applied uniformly to all targets. Finally, a dynamical simulation of the system using the real system properties at the start time of the light curve, including the binary period, eccentricity, and argument of periastron, component masses and radii, and mid-eclipse time, is generated using the \texttt{REBOUND} package \citep{Rein_2012}\footnote{\url{https://rebound.hanno-rein.de/}}. This simulation is used in the de-trending stage and in the injection and recovery stage.

\subsection{De-trending of variability}

Once the light curve is assembled and the eclipses are masked, the light curve must be de-trended of variability. The goal of this process is to flatten the light curve and remove systematic and physical variability which can obscure planet transits. EBs are naturally highly photometrically variable objects, and this variability can manifest due to many interesting physical or geometric phenomena (see e.g. \citet{Prsa_2018} for deeper discussion and modeling). Short-period EBs therefore require careful and rigorous de-trending. However, the ethos of de-trending means the process should be as ``gentle" as possible, so as not to remove real planet signals. Because of this, we de-trend our light curves in multiple steps, inspired by the de-trending scheme of \citet{Martin_2021}, which is designed specifically for CBP searches. 
At all stages, we use the \texttt{wotan} package \citep{Hippke_2019_Wotan}, which contains many de-trending schemes and is simple to implement.

The de-trending proceeds in the following stages. In the first de-trending stage, we perform a simple quadratic de-trending, wherein we split the light curve into continuous chunks, fit a quadratic to each chunk, and divide the flux by the fit. This removes significant trends and ramps in the data. In the second stage, we perform an ellipsoidal de-trending. Most (but not all) short-period eclipsing binaries exhibit ellipsoidal variations concurrent with the binary orbital period (or a harmonic of it), which manifests as sinusoidal variation. In this stage, we fit a sinusoid at half the binary period, and apply a Bayesian Information Criterion (BIC)-based decision for whether to divide by the sinusoidal trend. We divide by the sinusoidal trend if $\Delta \text{BIC} > 0$, where 
\begin{equation}
    \Delta \text{BIC} = \text{BIC}_{const} - \text{BIC}_{\sin}
\end{equation}
. In this case, $\text{BIC}_{const,\sin}$ are essentially the goodness of fit for the light curve to a constant vs the sinusoidal fit. 

In the next two steps, the \texttt{REBOUND} simulation generated during the light curve loading stage is used. Due to the continuous motion of the binary components, the transverse component of the velocity of the star is constantly changing. The star appears to be moving fastest relative to our line of sight closest to the time of eclipses. At each time step in the light curve, we estimate the duration of a representative planet transit based on the instantaneous velocity of the star at that time. For this stage, we assume a coplanar, prograde, circular planet with a constant velocity and a physical size of 6 R$_\oplus$, with an orbital period that is $6.27\times P_{\text{bin}}$, near where most of the Kepler CBPs were detected. With this information, we can estimate the durations of representative planet transits, which is used to inform the de-trending process more deeply than blind guessing. 

In the third de-trending stage, we apply an iterative cosine de-trending process. This follows the steps of Section 2.6 in \citet{Martin_2021}, and is itself similar to the cosine filtering with autocorrelation minimization (CoFiAM) algorithm of \citet{Kipping_2013} and implemented by \citet{Rodenbeck_2018}. The iterative cosine approach was developed for shallow transits with non-periodic transit timing signatures. Briefly, we iteratively de-trend the light curve with progressively shorter filter windows in steps. We use the maxima and minima of the transit durations from above to set the duration of the cosine de-trending window. At each window step, we check whether further cosine de-trending is necessary by computing a Lomb-Scargle periodogram \citep{Lomb_1976,Scargle_1982} and checking whether significant periodicity remains. The process stops when either no significant periodicity remains or when the shortest window is reached. 

The fourth stage is the variable duration de-trend step. This step is inspired by Section 2.7 of \citet{Martin_2021}. In this step, we apply a Tukey biweight sliding filter \citep{mosteller1977data} to copies of the light curve. We split the light curve into 48 copies, with each copy being de-trended with a different window length. The window length used for a copy decreases from three times the minimum transit duration from the array generated above down to 75\% of three times the maximum transit duration from above. The final version of the light curve is built from the 48 copies, where each point in the light curve is selected based on the duration of transit predicted at that time. For example, in places in the light curve where long transits are predicted to occur, a longer de-trending window is selected, and vice versa for locations where shorter transits are predicted to occur. 

After the variable duration de-trending, the light curve can still contain short-timescale instrumental artifacts, including gaps with ``hooks” and large jumps unassociated with a gap. We implement a kink-removal stage to address these artifacts. A gap with a hook occurs when a real data interruption coincides with a step in the baseline flux level, such that the light curve resumes at a different level than where it left off; we flag any gap longer than 2 hours where the mean flux in the 2-hour windows before and after differs by more than $1.5\sigma$. A large jump is defined similarly but without an associated gap: for each point, we compare the mean flux in the 2-hour windows immediately before and after it to the global mean of the light curve, and flag the point if both windows deviate from the global mean by more than $2\sigma$ in opposite directions (implying the total difference needs to be more than $4\sigma$); this sign-flip requirement isolates genuine bipolar discontinuities from one-sided variability already handled upstream. Once a kink is identified, we remove all data within 0.2 days on either side of the discontinuity. This step was the dominant source of point removal during the de-trending process, removing as many as about 10\% of the points in the light curve but usually not removing more than approximately 5\% of the points. At this point, the light curve has been eclipse-masked and de-trended of variability.

\begin{figure}
    \centering
    \includegraphics[width=0.98\linewidth]{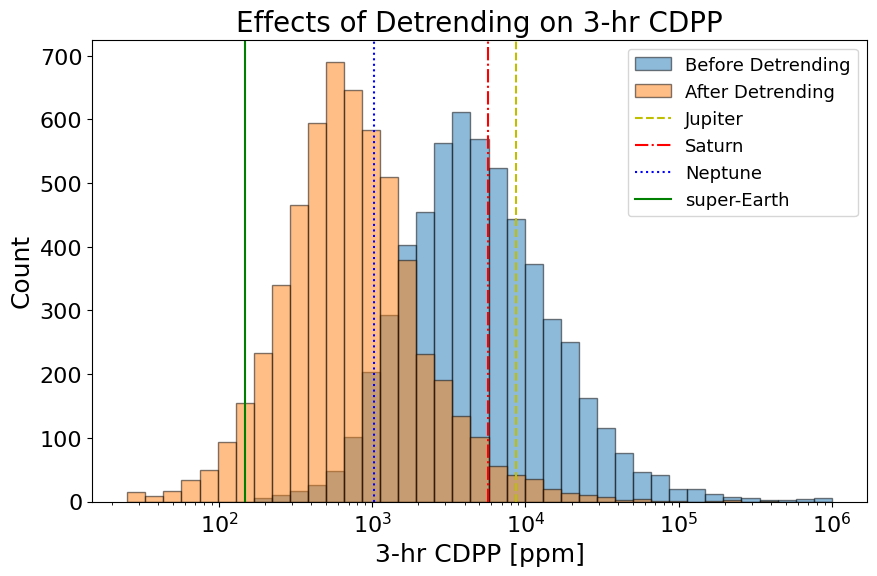}
    \caption{Comparison of CDPP of TESS light curves before the de-trending scheme (blue histogram) and after the de-trending scheme (orange histogram). CDPP required to detect a Jupiter, Saturn, Neptune, and super-Earth ($2 R_\oplus$ sized planet as yellow, red, blue, and green vertical lines, respectively, at our threshold of SNR~$=10$ orbiting a typical binary.}
    \label{fig:de-trending_CDPP}
\end{figure}

Figure. \ref{fig:de-trending_CDPP} shows the comparison of the combined differential photometric precision (CDPP; \citet{Christiansen_2012}), binned to 3 hours, of our light curves before, in blue, and after, in orange, de-trending. The overall distribution after de-trending has shifted to lower CDPP, demonstrating that noise in our light curves has decreased as a result of the de-trending process. Additionally, we plot the CDPP needed to detect a transit with a depth consistent with a Jupiter-, Saturn-, Neptune-, and super-Earth-sized planet around a typical binary (defined as a 0.5 M$_\odot$ primary with an equal mass secondary orbiting with a period of 3 days) as yellow dashed, red dot-dash, blue dotted, and solid green lines, respectively. After de-trending, approximately 8\% of our light curves are left of the super-Earth line.  

\subsection{SNR-based single transit search}

After the de-trending stage, we perform the transit search. We use a straightforward signal-to-noise ratio (SNR) based search, wherein the strength of the transit ``signal" divided by the ``noise" in the light curve is compared to a threshold. Since this is a non-periodic search, each individual transit event must surpass the signal-to-noise threshold. For our SNR equation, we adopt the general form of \citet{Villanueva_2019},
but for computational efficiency across our eight search durations, we
compute a single reference CDPP over a fixed 3-hr window rather than
re-measuring CDPP independently at each duration:
\begin{equation}
    \text{CDPP}_{\rm ref} = \sqrt{\frac{1}{N}\sum_j (M_j^{\rm ref} - \mu_M)^2}
\end{equation}
where $M_j^{\rm ref}$ is the running mean of the flux computed in a
fixed 3-hr window, $N$ is the number of valid windows, and $\mu_M$ is the mean
of the light curve. We rescale this reference value to each search
duration $\tau$ under the assumption of white (uncorrelated) noise,
\begin{equation}
    \sigma_{\rm eff}(\tau) = \text{CDPP}_{\rm ref}\sqrt{\frac{\tau_{\rm ref}}{\tau}}
\end{equation}
where $\tau_{\rm ref} = 3$ hr, and evaluate the signal-to-noise ratio
of a single transit as
\begin{equation}
    \text{SNR}(\tau) = \frac{\delta}{\sigma_{\rm eff}(\tau)}.
\end{equation}
We implicitly assume that noises scales as $1/\sqrt{\tau}$ between the 3-hr reference and the search duration. 

At this point, two meaningful aspects of the data are the level of noise in the light curve, and the sampling cadence, which changes depending on which sector is under examination. Presumably, de-trending has flattened the light curve such that the only noise remaining is point-to-point white noise. The sampling cadence of the light curve influences the number of points in a transit $N_{pts}$ and number of points in the window used to calculate the running mean. 

To treat the varying cadence of the light curve, we split the light curve into sectors at this point and search each sector individually. This means that the noise in the light curve is calculated on a per-sector basis. This is justified because intrinsic instrumental noise changes on a sector-by-sector basis. The advantage of this approach is that we are therefore more sensitive to smaller signals in sectors with lower intrinsic noise levels. Further, we can search the light curve for signals of varying durations. Functionally, we search using an array of 8 durations spaced evenly from 2 hours (4 points in the Primary Mission) to 10 hours (beyond which, de-trending tends to wipe out signals). For each duration, we calculate the number of points that are expected to be in a transit of that duration ($N_{pts,expected} = \text{int}(\frac{\text{duration}}{\text{sampling cadence}})$. We then calculate the depth of a box of the duration being searched at every point in the light curve. If any given time point has fewer than 60\% of the number of points expected for that duration, the time point is excluded (which accounts for gaps). This implicitly means that true partial transits with less than 60\% phase coverage will not be detected.

An important choice to be made at this stage is the SNR threshold selected. The threshold should be chosen such that False Positive events (such as spurious dips from an instrumental effect, or a random series of a few points lower than the mean) are not passed, but also so that real transit events are not excluded. During the Kepler mission, an SNR threshold of 7.3 for single transits was used \citep{Villanueva_2019}. However, after extensive testing, we found that this threshold was too low, and passed too many signals that were not real. We instead selected an SNR threshold of 10. Additionally, we check the number of signals detected in a given light curve. If a given sector has more than 3 unique signals ``detected", we throw out the detections as likely false positives. This effect will have a small completeness penalty, as a sector of data with a real transit signal in the presence of two or more threshold crossing events will still be removed. However, this is justified because the cases where this step was taken only occurred where variability was extremely high, and in these cases distinguishing between binary variability and a true planet transit would be difficult regardless.

\subsection{Vetting of candidate signals}

By this stage, candidate signals in the data have been identified, but it is important to vet them. Any given signal can have one of three origins: 1. a noise/systematic false alarm (FA); 2. an astrophysical false positive (FP); or 3. a planet candidate (PC). Examples of false alarms include instrumental noise, scattered light, or stellar pulsations, whereas false positives are due to genuine astrophysical eclipse or transit signals, but are not caused by an actual planet in the host system. These can include eclipsing binaries or other eclipsing objects which are off-target. Planet candidates are only those signals which are likely caused by a planet transiting in the host system.

Automated vetting is faster than by-eye for large catalogs and produces uniform, reproducible planet catalogs \citep{Kunimoto_2025}. We therefore use the automated vetting tool \texttt{LEO-vetter}\footnote{\url{https://github.com/mkunimoto/LEO-vetter}}. In short, \texttt{LEO} computes vetting metrics and then checks those metrics against a series of pass-fail thresholds. If a signal passes all checks, it is considered a planet candidate. If a signal does not pass, it is labeled as either a FA or an FP (or sometimes both), and no further action is pursued. A summary of all metrics and tests computed by \texttt{LEO} are discussed in \citet{Kunimoto_2025}. 

One of the strengths of \texttt{LEO} is that the thresholds can be customized to suit the needs of the search at hand. Because we are searching for single-transiting planets orbiting short-period binaries with a code that was designed primarily for periodic signals in single-star systems, we needed to change some of the thresholds and/or disable some of the tests. We disabled any tests which require multiple transits. This included the SNR consistency test and the ratio of the maximum of single event SNR measured at each time step across the light curve compared to the phase-folded SNR (characterized in \texttt{LEO} as \texttt{max\_SES\_to\_MES}), both of which were FA tests that require more than one transit. Further, we disable the odd-even test, which is a check of whether the odd numbered transit depths are significantly different from the even numbered transit depths, given that this FP test also requires multiple transits. We also disable the unphysical duration test because this is based on the assumption of a single star host, which is invalid in this case.  Finally, we disable the sinusoidal test. We expect that short-period binaries will exhibit sinusoidal variability, and we disable this test to mitigate risk of a true signal being excluded, especially in a short-period binary with an orbital period of $\lesssim 1$ d. Additionally, we modified the threshold for the asymmetry test. The transit asymmetry test is meant to account for ramp-like features, but given that the star's velocity is constantly changing, the transit may be slightly asymmetric. We therefore relax the threshold for the asymmetry test from 10 to 18. Finally, we relaxed the threshold for the candidate size test to 40 Earth radii (from 22 Earth radii). \texttt{LEO} assumes a single-star host, and although a threshold of 22 Earth radii is already generous, we still pass relatively deep signals at this stage. First, this search maintains a secondary science interest in detecting low-mass stellar tertiaries, which can be individually modeled and recategorized later. Second, flux dilution within short-period binary environments artificially suppresses observed transit depths, causing physically larger stellar companions to exhibit smaller apparent radii. A relaxed threshold of 40 Earth radii ensures that these diluted stellar-flux signals are not prematurely discarded by the automated vetting. We use the default values for all other tests and metrics, including the SNR test, the model fit test, the depth mean to median ratio (DMM) test, the Chases test, the uniqueness test, the transit shape test, the V-shape test, the significant secondary test, and the centroid offset test. One of the most powerful tests is the centroid offset test, where the average of in- and out-of-transit pixels surrounding the target is found from a candidate transit event's ephemerides and duration. The difference between the in- and out-of-transit images should appear starlike at the location of the transit source, assuming the transit is the explanation for any difference in flux. An offset could indicate that the candidate transit event occurs on a different source.

Functionally, \texttt{LEO} proceeds in two main stages, with the first stage being flux vetting and the second stage being pixel vetting. Flux vetting is based purely on the light curve itself, whereas during the pixel vetting stage, \texttt{LEO} downloads FFI cutouts around the target star and produces difference images to estimate the centroid offset. This same centroid offset was used to validate the eclipse features of the M\&M binaries in \citet{Oddo_2026}. Flux vetting takes of order a few seconds per sector, whereas pixel vetting takes approximately 10-30 seconds per sector. For computational efficiency, our code is set up to run all flux vetting tests first, then perform pixel vetting only if all checks are passed at the flux vetting stage. 

\subsection{Fitting of candidate signals}

In the final stage, the candidate signals identified by the search algorithm and vetted during the vetting stage above are fitted to a transit model. For computational efficiency, we select points in a window around the central transit time which is three times the predicted transit duration. We use the popular \texttt{batman} package \citep{Kreidberg2015batman}\footnote{available on Github: \url{https://lkreidberg.github.io/batman/docs/html/index.html}} to generate a physical model that is fitted to the light curve data. \texttt{batman} expects model parameters as in Table \ref{tab:batman_params}. We initialize the transit model with \texttt{t0} at the time marked by the search and fit within a 0.1 day range around the mid-transit time. From the duration of the transit and the instantaneous velocity of the star being transited, we can estimate the period of the planet. We do so by solving for planet period in \citet{Martin_2021} Equations 4 and 5. We also make the assumption that the planet is transiting across the primary star. The value found at this stage is used as an input for \texttt{per}. We use Kepler's Third Law to solve for the semi-major axis \texttt{a} at this period and divide by the stellar radius, which is also a fixed input. Period and semi-major axis are held constant because \texttt{batman} assumes the host is a single star, which is a poor assumption for our case and would bias the fit. Further, we calculate the radius of the planet (in units of stellar radii) from the transit depth and the binary mass ratio $q = \frac{M_2}{M_1}$ using \citet{Martin_2019} Equation 11, which is copied below:
\begin{equation}
    \frac{R_p}{R_s} = \sqrt{\delta(1+q^{3.5})}.
\end{equation}
This compensates for dilution from the secondary. We initialize with an edge-on configuration (\texttt{inc} = 90), zero eccentricity (\texttt{ecc} = 0), and longitude of periastron at $90^\circ$. Finally, we estimate limb-darkening parameters for the host star using the quadratic limb darkening law \citep{Mandel_2002}, which are included as model inputs (these are u$_{1,2o}$ in Table \ref{tab:batman_params}). 

\begin{table*}[]
    \centering
    \begin{tabular}{c|c|c}
        Parameter & Description & Prior \\
        \hline
        \texttt{t0} & time of mid-transit & $\mathcal{U}[t_{0}-0.05,t_{0}+0.05]$ \\
        \texttt{per} & orbital period & \textbf{fixed} \\
        \texttt{rp} & planet radius (in units of stellar radii) & $\mathcal{U}[0,1]$ \\
        \texttt{a} & semi-major axis (in units of stellar radii) & \textbf{fixed} \\
        \texttt{inc} & inclination (degrees) & $\mathcal{U}[80,95]$ \\
        \texttt{ecc} & orbital eccentricity & $\mathcal{U}[0,0.9]$ \\
        \texttt{w} & longitude of periastron (in degrees) & $\mathcal{U}[0,180]$ \\
        \texttt{limb\_dark} & limb-darkening model & ``quadratic" \\
        \texttt{u} & limb-darkening coefficients (u$_1$,u$_2$) & $\mathcal{N}(u_{1,2o},0.1)$
    \end{tabular}
    \caption{Table of \texttt{batman} model parameters to be fit, with descriptions and priors for the MCMC fitting scheme.}
    \label{tab:batman_params}
\end{table*}

The fitting process is in two main stages, with an initial least-squares minimization stage and a second Markov Chain Monte Carlo (MCMC) sampling stage, similar to Section 5.5.2 in \citet{Oddo_2023}. In both stages, the planet period, and semi-major axis are held fixed. First, we fit the light curve with an initial linear least-squares regression. This initial fitting stage yields an initial model, whose outputs are used as initial guesses in the MCMC sampling stage. In the MCMC sampling stage, we utilize the \texttt{emcee} python package \citep{Foreman_Mackey_2013}. The parameters being fitted for and their prior distributions are shown in Table \ref{tab:batman_params}. It is recommended to initialize walkers for the sampler in the amount of eight times the number of parameters being fit, so we initialize 56 walkers. We burned in for 200 steps, and then sampled for 1500 more steps. This approach yields posteriors, from which parameters and uncertainties for the fitted quantities can be obtained.

We can make some preliminary checks for both convergence and physical realism at this stage. To check for fit convergence, we check convergence time in the MCMC sampler. If the fit does not converge, the protocol is to retain the values from the least-squares regression stage and plot the transit model with a warning that the MCMC fit did not converge. In this case, the protocol is to modify the parameter ranges and run the MCMC for longer. In the alternative case where convergence is reached, there are two major ``physical reality" checks that we must perform at this stage. First, we must check that the period calculated above is stable for the binary. In the case that a candidate is found, we calculate the binary's critical stable semi-major axis from \citet{Holman_1999} and compare to the semi-major axis calculated from the duration of the transit. If the planet is in the unstable region (interior to the critical semi-major axis), the signal is likely a FP and is discounted. Second, we must check that the object is planet-sized (less than 1.8 R$_{Jup}$) based on the depth of the signal. We still retained objects which were likely stellar in nature. 

\section{Results from the Search}\label{sec:results}

We searched all 7422 single-sector light curves of TESS data in our M\&M EB sample. The search yielded 302 total unique events. All of these signals except for five individual events were identified as False Alarms or False Positives as determined by the vetting thresholds above, demonstrating the importance of vetting. More than two-thirds of the events (207 out of 302) were identified as FAs. Of the five non-FA/FP signals, four are from likely stellar companions, and one is from a new single-transiting CBP candidate. 

\subsection{Eclipsing tertiary stellar companions}

Four individual eclipse events which passed vetting were likely due to eclipsing tertiary stellar companions based on their depths. These events do not count towards our planet occurrence rate estimates. One event occurred on TIC 22457651 during sector 63 and three events occurred in three different sectors on TIC 137809567, including in sectors 21, 41, and 75. 

\subsubsection{TIC 22457651}

\begin{figure*}
    \centering
    \includegraphics[width=0.85\linewidth]{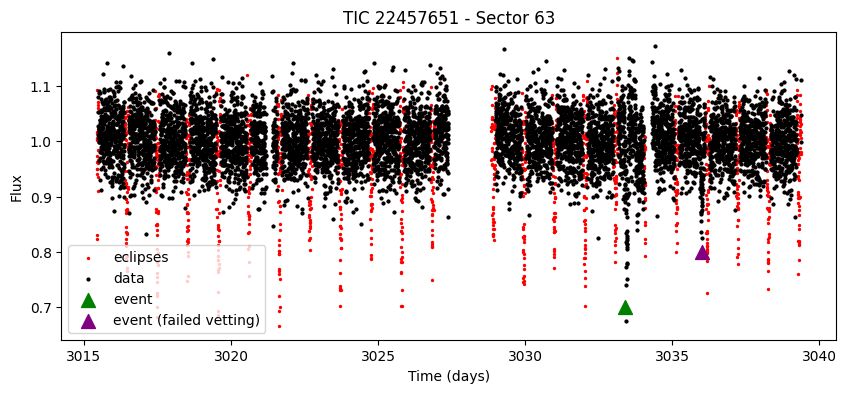}
    \caption{The additional eclipse event identified in the sector 63 light curve of TIC 22457651, demarcated by a green triangle. Another possible eclipse event was identified in the light curve, which is demarcated by the purple triangle, but this event was shown to be off-target by pixel vetting.}
    \label{fig:TIC22457651_S63}
\end{figure*}

TIC 22457651 is a near-equal mass ($m_{A,B} \approx 0.43$ M$_\odot$) 2.08-day binary in the Southern hemisphere (RA = $186.784534^\circ$, Dec = $-19.371479^\circ$) with an apparent magnitude in the TESS band of 14.9. The system has been observed by TESS once in each of its Primary Mission, Extended Mission, and Second Extended Mission, including in sectors 10, 37, and 63. Two events were flagged by the transit search in sector 63. The first event was at $t0 = 3033.41$ and the second was at $t0 = 3036.02$ (both events are shown in Fig. \ref{fig:TIC22457651_S63}). While the first event passed vetting, the second did not, so no further action was pursued. No similar events appear in earlier TESS sectors. The event is approximately as deep as the primary eclipses and is V-shaped, but is clearly not in the same phase as the binary. Based on the depth, $R_p/R_s$ was found to be 0.632, corresponding to a 0.30 $R_\odot$ tertiary. With just one additional eclipse feature in the light curve and a very dim source, prospects for followup for this system are not good. 

\subsubsection{TIC 137809567}

\begin{figure}
    \centering
    \includegraphics[width=.98\linewidth]{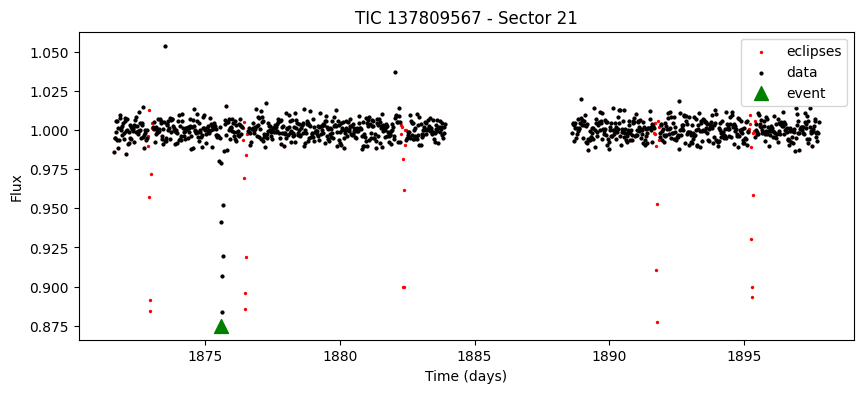}
    \includegraphics[width=.98\linewidth]{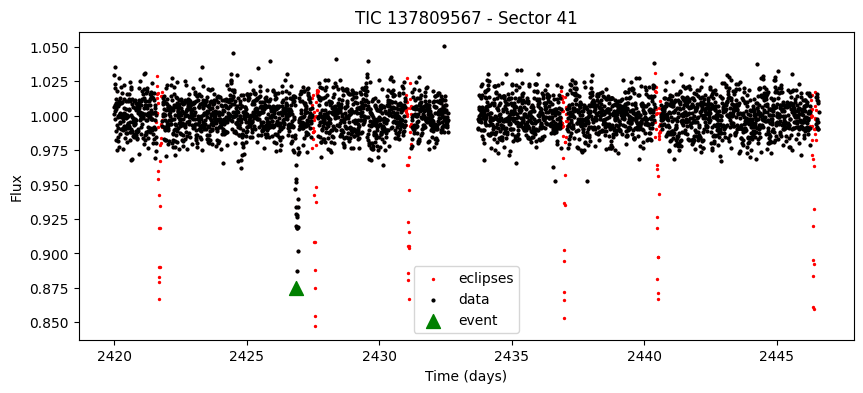}
    \includegraphics[width=.98\linewidth]{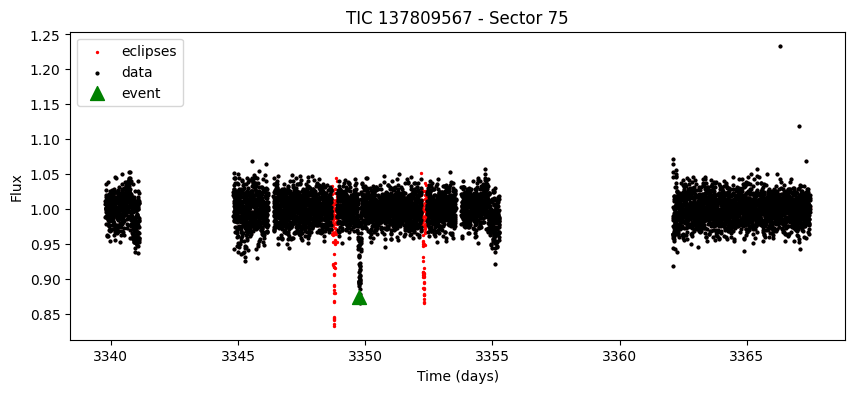}
    \caption{Eclipse events identified in the light curves of TIC 137809567. Sectors 21 (top), 41 (middle), and 75 (bottom) are shown. Eclipse events are demarcated by the green triangles.}
    \label{fig:TIC137809567}
\end{figure}

TIC 137809567 is a near-equal mass ($m_A = 0.512$ M$_\odot$, $m_B = 0.479$ M$_\odot$, $q = 0.94$) 9.4-day binary with a moderate eccentricity ($e=0.196$) in the Northern hemisphere (RA = $177.501579^\circ$, Dec = $56.231941^\circ$) with an apparent magnitude in the TESS band of 14.3. In the first six years of the TESS mission, the system was observed in six sectors spanning from the Primary Mission to the Second Extended Mission. Our search identified three additional eclipse events in three different sectors, at $t_1 = 1875.63$ in sector 21, $t_2 = 2426.91$ in sector 41, and $t_3 = 3349.78$ in sector 75, which are shown in Fig. \ref{fig:TIC137809567}. The three events are similar to one another in depth (approximately 10\% deep) and duration (around 2 hours) and are separated by approximately 550 days from the first event to the second and approximately 922.9 days from the second event to the third. The events are too deep to be due to a planet-sized object, hinting at a tertiary stellar companion.

The timings of these three events could match with a periodicity of approximately 184.27 days. However, with this candidate period, an eclipse event would be expected around $t0 = 2612.67$, which is at the beginning of sector 48. Although our \texttt{tglc} light curve starts after this time, a version of the sector 48 light curve exists from the MIT Quick-Look Pipeline (QLP). No eclipse event appears in the QLP light curve at this time, despite all other events showing up in their respective QLP light curves.

\begin{figure}
    \centering
    \includegraphics[width=0.98\linewidth]{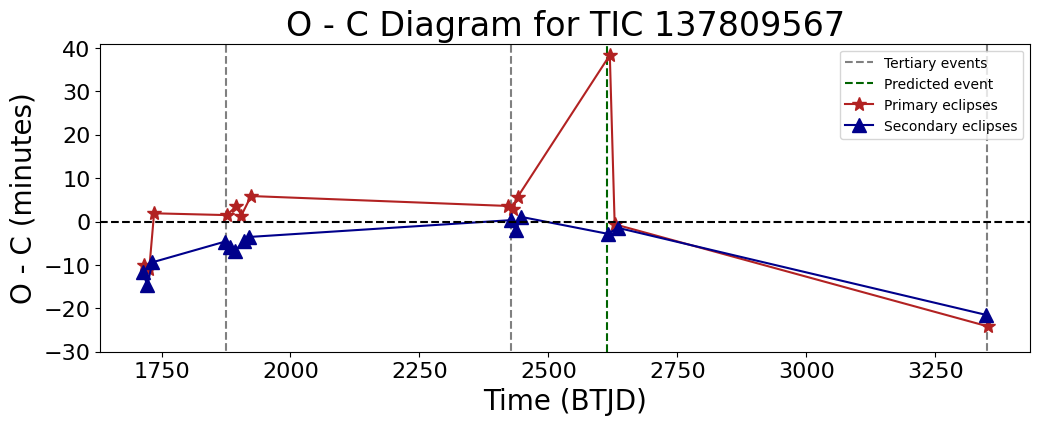}
    \caption{The O - C diagram for TIC 137809567 primary (dark red stars) and secondary (dark blue triangles) eclipses. The times of the three eclipses are marked with green lines, and the "missing`` eclipse time is marked in red with a vertical line, which nearly coincides with a spike in the primary eclipse timing difference.}
    \label{fig:TIC137809567_OC}
\end{figure}

Other intriguing clues about this system come from analysis of the eclipses and its available astrometric data. Variations in the times of the binary eclipses, known as eclipse timing variation (ETV analysis), can inform our understanding of these systems. There are many quantifiable effects that cause ETVs, but there are two which are the most important. First is the Light Travel Time effect, which is a geometric effect due to the binary center of mass being perturbed by a third bound object, causing delays or advances in the signal. Second is the direct dynamical perturbation of the times of eclipses from the presence of a third body. Beyond these two major indicators of third body detection, ETVs can reveal internal stellar physics, including mass transfer, tidal interactions, or magnetic activity signals. Therefore, precise measurement of the binary eclipse times is relevant. To do so, we fitted a simple gaussian model to each binary primary and secondary eclipse and recorded the Observed time of eclipse. Then we compared these to the expected time of the eclipses Calculated from a linear ephemeris. The result is the O - C (Observed minus Calculated) diagram, which is shown in Fig. \ref{fig:TIC137809567_OC}. Interestingly, we see a spike in the primary eclipse timing difference approximately 6 days after the time of the missing third-body eclipse. This could be evidence for a dynamical interaction between the primary star and a massive third body in the system. 

The final line of evidence comes from the ESA Gaia astrometric data \citep{Gaia_2016}. Gaia records a Renormalized Unit Weight Error (RUWE) for each source, which is a metric for how well the source fits to a single star astrometric solution based on monitoring the photocenter of the system. The typical expected RUWE value is around 1 for a single star, and most sources in the sky have a RUWE near this value. A high RUWE is indicative of a poor fit to a single-star model, with anything more than 1.25 in the Third Data Release usually indicating that the source is part of a separated binary or higher order system. The Gaia RUWE for this source is greater than 9, indicating a very poor fit. We know the central source is already a tight binary, but such a tight binary would not induce any detectable change to the system photo-center \citep{Belokurov_2020}, so any poor fit is likely due to a third body in the system which perturbs the photo-center of the system. 

In all, TIC 137809567 presents some interesting hints as a candidate hierarchical triple system. The three deep eclipses, the missing eclipse, the spike in the eclipse timings, and the abnormally high RUWE all point to an additional body in the system. More work is required to reveal the true nature of this system, but such work is beyond the current scope of a CBP occurrence rate. 

\subsection{Candidate TIC 5012780.01}

\begin{figure}
    \centering
    \includegraphics[width=0.95\linewidth]{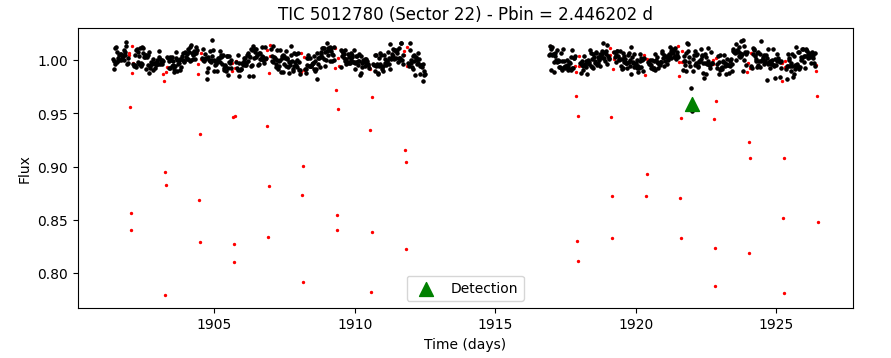}
    \includegraphics[width=0.95\linewidth]{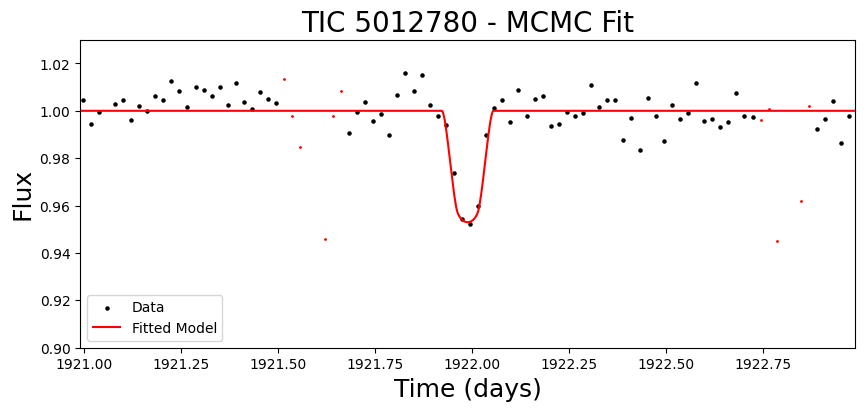}
    \caption{Sector 22 light curve for TIC 5012780, showing the detected single-transit event (green triangle) in the top panel. In the bottom panel, we zoom in on the detected event and show the best-fit MCMC transit model.}
    \label{fig:placeholder}
\end{figure}

One transit event with a depth consistent with a planetary-sized object passed both flux and pixel vetting. The event occurred at $t_o \sim 1922$ in sector 22 on TIC 5012780 with a depth of $\delta = 0.049$ (nearly 5\%) and a duration of 2.69 hours. The signal passed all FA and FP checks, though notably with an offset of 14.5 arcseconds from the host star which is just below the \texttt{LEO} threshold of 15 arcseconds. With this offset, we proceed with caution, noting that additional followup to confirm this detection will be necessary.

TIC 5012780 consists of a $M_A = 0.521 M_\odot$ and a $M_B = 0.423 M_\odot$ ($q = 0.812$) stellar pair in a $P_{bin} = 2.44$ day orbit. From the given transit depth and duration at the given time in the light curve, we calculate a period of 33.27 days, corresponding to a planet-binary period ratio of $\frac{P_p}{P_{bin}} = 13.64$. At the given depth, the signal corresponds to a planet with a size of 15.316 Earth radii, or 1.36 Jupiter radii. With a roughly $\sim30$ day period and a super-Jupiter size, this candidate has properties similar to TOI-2295 b \citep{Heidari_2025} or somewhat similar to Kepler-39 b \citep{Bonomo_2017}. If validated or confirmed, this system would be the shortest-period binary host to host a CBP, which itself would be the CBP with the shortest period known to date.

TIC 5012780.01 presents a challenge to the traditional pileup evident in the Kepler transiting CBP literature. The binary's semi-major axis is $a_{bin} = 0.03485$ AU. With a measured binary eccentricity of $e= 0.0133$ and a mass ratio $q=0.812$, the stability limit is $a_{crit} = 0.0852$ AU ($\frac{a_{crit}}{a_{bin}} = 2.44$), which is calculated from \citet{Holman_1999}. With the measured orbital period, the candidate planet's semi-major axis is $a_p = 0.1986$ AU, which is 5.70 times the binary semi-major axis. Even though TIC 5012780.01 is the shortest-period candidate CBP known to date, its distance from its host's stability limit is unlike most of the Kepler population \citep{Welsh_2018}, save for Kepler-1647 AB b\citep{Kostov_2016_kepler1647}. Indeed, this system may be more like the emerging BEBOP population beyond the stability limit \citep{Baycroft_2025}. This could suggest that in ultra-compact binary systems (i.e. those with orbital periods less than 7 days), the mechanism that halts inward migration may decouple from the dynamical stability limit, or that the system has sufficient room for interior, yet-undetected companions \citep{Quarles_2018}. As a test of the additional yet-undetected companions hypothesis, we generated a dynamical model of the system with \texttt{REBOUND}. We initialized the binary along with two planets, including a Jupiter-mass planet at the proposed semi-major axis of $a_p = 0.1986$ AU and a 10 Earth mass at $3\times$ the binary semi-major axis, near the stability limit. We integrated for 200,000 years, finding that the orbital elements of the planets did not vary significantly over this period and that both planets remain bound, demonstrating that such a system would be long-term stable.

With only a single transit and no additional data backing this detection up, it cannot be ruled a discovery. Although the signal is vetted, additional followup is necessary to further validate this signal. We will explore occurrence rate statistics both with and without this candidate.


\section{Calculation of Occurrence Rates}\label{sec:occurrence}

The fundamental goal of this analysis is to reach an estimate of the occurrence rate, which is the average number of planets per star ($f$) within a specific parameter space of radius ($R_p$) and orbital period ($P$). Because no transit survey is perfectly sensitive, we must account for the observational completeness of our search. We define the total detection completeness, $\epsilon(R_p,P)$, as the probability that a planet with given physical and orbital properties would both produce transits during TESS observations and be successfully recovered by our detection pipeline.

To quantify $\epsilon$, we factor it into two primary components: detection sensitivity and the probability of transit observation. The sensitivity $S(R_p,P)$, derived from our injection and recovery simulations, measures the probability that a transit signal of a certain depth and duration can be distinguished from stellar and instrumental noise. The probability of transit observation, otherwise known as the window function $W(P)$, is calculated via N-body simulations to determine the fraction of runs that result in detectable transits within the specific TESS duty cycle for each host binary. Therefore, $\epsilon(R_p,P) = S(R_p,P) \times W(P)$ in our formulation.

By applying the Inverse Detection Efficiency Method (IDEM) \citep{Howard_2012,Christiansen_2015,Sayeed_2025}, we can weight detected candidates by the reciprocal of the completeness at specific $(R_p,P)$ coordinates:
\begin{equation}
    f(R_p,P) = \frac{1}{N_*} \sum_{i=1}^{k} \frac{1}{\epsilon(R_{p,i},P_i)}
\end{equation}
where $N_*$ is the total number of systems searched and $k$ is the number of detections.

\subsection{Injection and recovery testing}\label{subsec:inj_rec}

We determined the sensitivity of our search empirically through injection and recovery testing. This process involves the introduction of synthetic planetary signals into the observed TESS light curves, which are then passed through the same de-trending and searching stages used for the primary candidate search as in Section \ref{sec:search_method}. For computational efficiency, a synthetic signal is defined as successfully ``recovered" if it 1. passes the initial search algorithm's SNR threshold of 10.0, 2. a downstream trapezoidal fit within 0.05 days of the injected mid-transit time verifies that the recovered depth and duration match the injected parameters to within 40\%, and 3. passes vetting with \texttt{LEO}, excluding pixel vetting. Injected signals are not passed through the secondary pixel vetting or the MCMC fitting.

For each host binary, we generated synthetic planets with radii selected from a uniform distribution ranging from 2 to 16 Earth radii and orbital periods drawn from a log-uniform distribution from 10 to 600 days. We calculated the critical stability period for each binary from \citealt{Holman_1999} and checked that the period draw was stable; if the period draw was not stable, we resampled until a stable period was drawn. This implies that for binaries with stability limits beyond the innermost period, the whole range could not be sampled, but rather only the range of dynamically stable periods. While numerical methods from \citet{Georgakarakos_2024} suggest that planets with periods shorter than 10 days can be stable, we adopted a conservative lower limit of 10 days. To justify this cut, we ran dynamical simulations testing the stability of short period planets. We found that a planet on a 10 day orbit around a 0.5 day nearly-equal mass binary is reliably stable across multiple trials. Planets on shorter orbits around the shortest-period binaries may not be stable and might not exist in the first place \citep{Martin_2015_tight,Munoz_2015,Hamers_2016}. A test of limits of the shortest-period CBPs orbiting the shortest-period binaries with all relevant physics included should be a line of future work. 

Since CBPs do not follow a strictly linear ephemeris, we utilized the \texttt{REBOUND} N-body integrator to simulate the dynamical evolution of each system. We measured the times at which the planet in the simulation passed in front of each star and the relative velocities of the bodies at this time, providing a schedule of transit mid-times and durations. The light curve profiles were modeled using the \texttt{batman} package, adopting quadratic limb-darkening coefficients which were calculated based on the primary star's effective temperature and surface gravity. The observed transit depth is suppressed by the flux of the non-transited stellar companion. We scaled the synthetic transit depths so that they were shallower compared to transits in a single-star system by the binary flux ratio ($F_{A,B} /(F_A+F_B)$) to accurately account for the untransited companion's flux.

To prevent the synthetic signals from being obscured by the binary eclipses, which are masked during the search phase, we implemented an eclipse-avoidance heuristic. If a generated planet's transits overlapped significantly with the primary or secondary eclipses of the binary, the orbital parameters were redrawn and the simulation was re-run until there was no overlap. This ensured that our sensitivity measurements accurately reflect the pipeline's ability to find planetary transits in the searchable ``out-of-eclipse" portions of the light curve.

\begin{figure*}
    \centering
    \includegraphics[width=\linewidth]{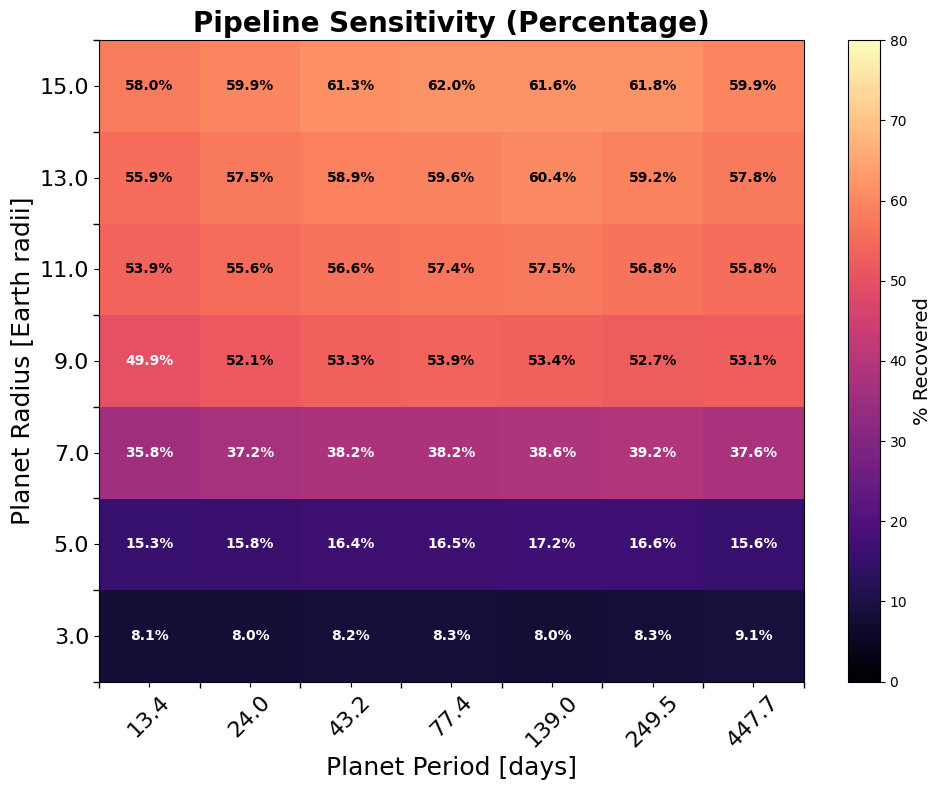}
    \caption{Our injection and recovery grid for all injections. Grid cells are colored according to  pipeline sensitivity, $S(R_p,P)$, expressed as a percentage. Each transit injection is counted individually, so sensitivity here is the overall percentage of transits detected in a given period and radius bin.}
    \label{fig:inj_rec_all}
\end{figure*}

We performed a total of more than 1 million transit injections across all systems in the catalog. The overall sensitivity map as a function of planet period and radius is shown in Fig. \ref{fig:inj_rec_all} as a grid of 7 uniformly-spaced bins in planet radius and 7 log-uniformly-spaced bins in planet orbital period. Since each transit counts individually, sensitivity here is the overall percentage of transits detected for planets within the given bin range. Our results show that sensitivity generally scales with planet radius, where larger planets (which produce deeper, higher-SNR signals) are more likely to be detected. Our detection sensitivity is bounded around $\sim60$\% because of a variety of factors, including inherent noise in the light curves, signal attenuation from our multi-step de-trending that can suppress transit features (particularly long-duration signals), the removal of sectors where more than 3 unique signals passed the detection threshold, and the requirement that the recovered depths and durations match injected values to within 40\%. 

\subsection{Detection probability}\label{subsec:window}

\begin{figure}
    \centering
    \includegraphics[width=0.96\linewidth]{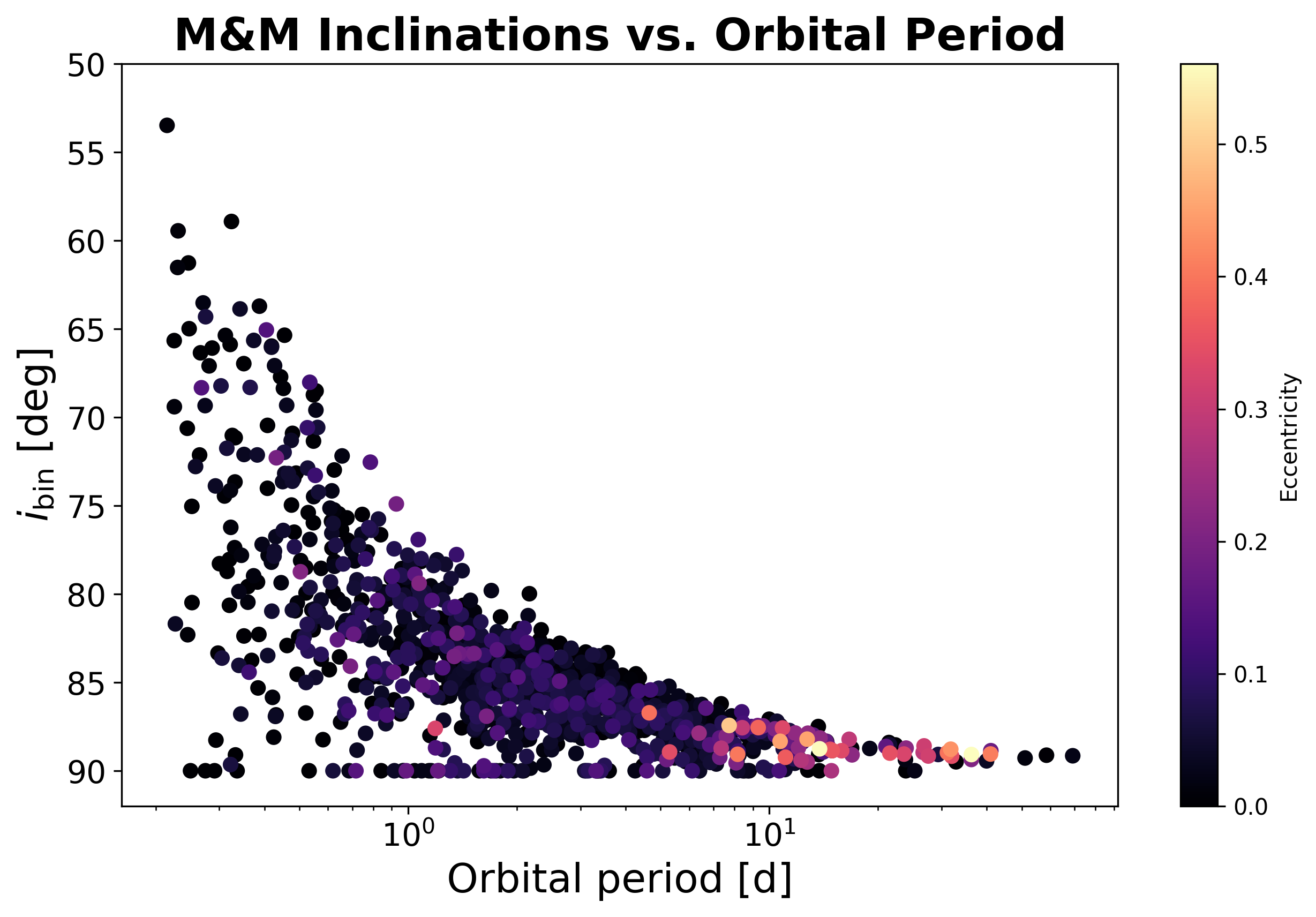}
    \caption{Calculated mutual inclinations $i_\text{bin}$ (in degrees) of M\&M EBs plotted against orbital period (in days, on a log scale) from eclipse durations and depths.}
    \label{fig:inclination_v_period}
\end{figure}

We must quantify the probability that a planet on a given orbit would produce transits during the specific times TESS was observing its host, which is known as the window function $W(P)$. TESS's sky coverage through time is complicated, and the observing strategy has continued to evolve. Further, in circumbinary systems, the window function is significantly more complex than in single-star systems due to the aforementioned aperiodicity of transits and transit timing and duration variations. Consequently, we cannot use a simple formulaic approach; instead, we used a dynamical approach to determine the observational coverage.

To account for system mutual inclination relative to our line of sight, we derived a per-system inclination from the primary and secondary eclipse widths and depths. Eclipse width and depth provide two independent geometric constraints on the system impact parameter $b$ and hence $i$, but the two are most sensitive in complementary regimes. Eclipse width is more constraining for grazing eclipses, whereas depth is more constraining for low-$b$ eclipses. Inclination can be calculated from eclipse widths through a system of equations. Impact parameter $b$ during a primary eclipse ($b_p$) or secondary eclipse ($b_s$) is 
\begin{equation}
    b_{p,s} = \frac{a\cos i}{R_\text{sum}}\frac{1-e^2}{1\pm e\sin\omega}
\end{equation}
and the eclipse width $T$ during a primary eclipse ($T_p$) or secondary eclipse ($T_s$) is 
\begin{equation}
    T_{p,s} =\frac{P}{\pi}\left[\frac{R_\text{sum}}{a}\frac{\sqrt{1-b_{p,s}^2}}{\sin i} \right] \cdot \frac{\sqrt{1-e^2}}{1\pm e \sin \omega}.
\end{equation}
In this form, solving for $i$ requires a numerical method, for which we used Brent's method \citep{brent1973algorithms}, implemented in \texttt{scipy.optimize}.
Alternatively, we can solve for binary inclination using the depths by calculating the amount of overlap between two uniform disks \citep{Mandel_2002}. 
We solve both methods for each system and adopt the depth-based solution where the width-based equation admits no physical solution. Because individual measurement uncertainties are not available for the eclipse widths, we instead propagate an assumed uniform fractional uncertainty of 20\%. Across our sample, this yields most systems as having inclinations in the range $80-90^\circ$, with a median of $i_{\text{bin}}=85.7^\circ$. Our derived binary inclinations are shown in Fig. \ref{fig:inclination_v_period}. For 72 systems (5.6\% of the sample), neither channel returned a solution, so in this case we assume an edge-on system ($i_{\text{bin}}=90^\circ$).

For every EB in our sample, we performed a Monte Carlo simulation to calculate the window function. We initialized a three-body system using the \texttt{REBOUND} N-body integrator, adopting the measured masses, radii, and orbital elements of the host binary. For each simulation, we injected a synthetic planet with a period drawn from the same log-uniform distribution as in the injection and recovery step and a randomized initial orbital phase. To account for dispersion in the mutual inclinations of CBPs with respect to the binaries, we initialized planet trials with mutual inclinations according to two models. In our fiducial, coplanar case, we initialized planets with mutual inclinations sampled from a Rayleigh distribution centered at $1.5^\circ$, similar to \citet{li2016uncovering,Windemuth_2020,Mogan_2025}. We chose a Rayleigh distribution primarily because it is naturally nonnegatively-valued and is suitable to describe low-inclination systems \citep{He_2020,Masuda_2020}. In the second case, we initialized planets with isotropically-distributed mutual inclinations. We integrated the system over the total time baseline of the TESS observations for that specific TIC. 

To identify when transits occur, we monitored the relative positions of the planet and the stars within the simulation. A transit was recorded whenever the planet passed in front of a stellar disk from the observer's perspective. Due to our eclipse-masking protocol and the fact that a small proportion of points were masked during our de-trending procedure, we compared the predicted transit times against the data points that survived the eclipse-removal and de-trending processes. A transit was counted if more than 60\% of the transit overlapped with the data, according to the observing cadence. We ran 1,000 realizations of this process for each binary for both mutual inclination cases, varying the planet phase each time, recording the planet period, the number of total simulated transits, and the number of transits coinciding with TESS data. In doing so, we calculated the window function $W(P)$. This formulation of the window function captures both the geometric probability (whether the planet transits at all) and the temporal duty cycle (whether TESS was looking when it happened). 

\begin{figure}
    \centering
    \includegraphics[width=0.98\linewidth]{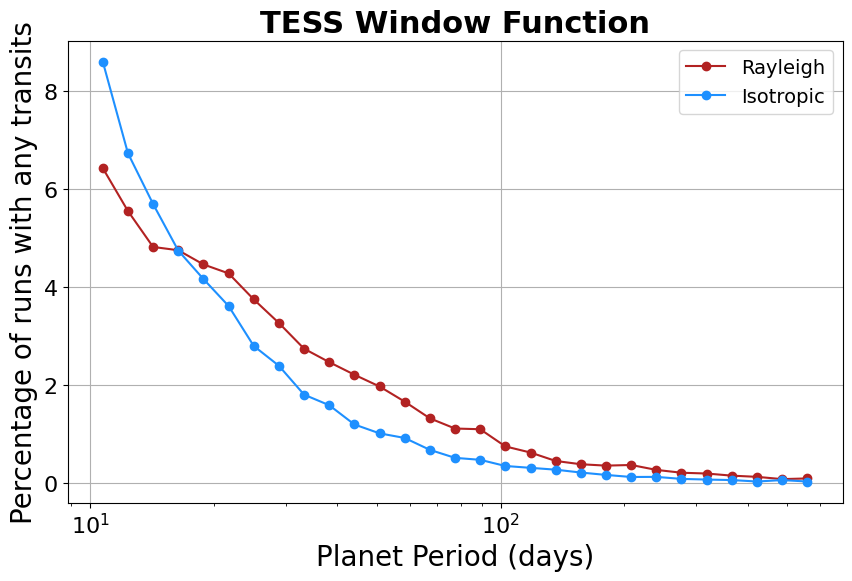}
    \caption{Any-transit percentage, which is our window function $W(P)$, in both our fiducial (Rayleigh) inclination case (red) and our isotropic mutual inclination case (blue).}
    \label{fig:window_function}
\end{figure}

We show our window function determination in Fig. \ref{fig:window_function}. Rather than calculating the overall percentage of transits per period bin that were found in TESS timestamps, we defined the ``any-transit" fraction as our window function, which is the fraction of runs per period bin with at least one found transit, which is consistent with our search methodology that only requires one transit for detection. Due to the widely varying inclinations of our binaries, the window function in both cases is low overall, topping out only around 7\% at the shortest orbital periods in the coplanar case. Interestingly, at the shortest orbital periods in our range, planets with isotropically-distributed mutual inclinations are more likely to transit. This is likely due to forced misalignment in the coplanar case; if the planet is aligned with the orbital plane of the binary, but the binary is misaligned with the observer's view (Fig. \ref{fig:inclination_v_period}), the planet is unlikely to transit. The window function then decreases to near zero for increasing planet period in both cases, with the isotropic case decreasing more quickly at longer periods, which matches expectations. We note there are two effects working together to reduce the any-transit fractions with increasing period: the limited time coverage of each target (and to first order, the 27-day duration of each sector), and the decreasing geometric transit probability. 

The final detection completeness for each bin in our occurrence rate grid is the product of this window function and the recovery fraction from our injection and recovery testing.

\subsection{Occurrence rates}

\begin{figure*}
    \centering
    \includegraphics[width=0.95\linewidth]{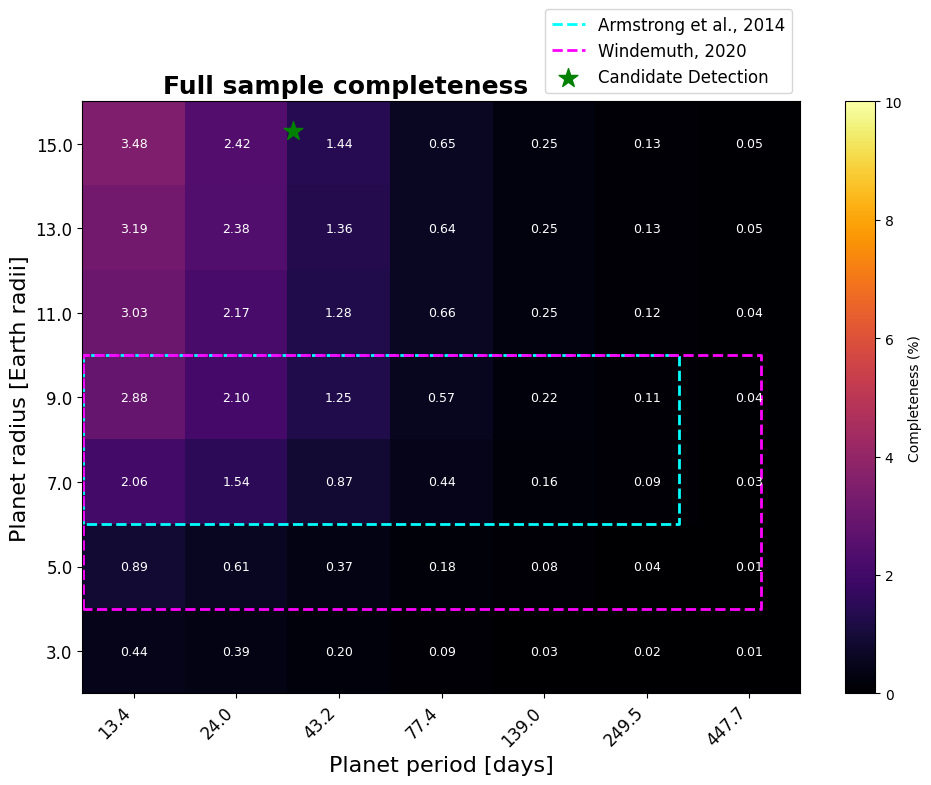}
    \caption{Completeness map for our search, in the same bins as the sensitivity grid. Completeness is expressed as a percentage. We also overplot our single detection (green star), as well as the \citet{Armstrong_2014} (cyan) and \citet{Windemuth_2020} (magenta) regions for direct comparison.}
    \label{fig:completeness_map}
\end{figure*}

To infer the intrinsic occurrence rate of circumbinary planets across the M\&M sample, we synthesized our pipeline sensitivity and dynamical window function (assuming coplanar as our fiducial case) for each system into a per-system completeness map. This included rescaling the window function bins from Fig. \ref{fig:window_function} such that the period bins were the same as in Fig. \ref{fig:inj_rec_all}. The per-system completeness $\epsilon_{\text{system}}$ is defined as
\begin{equation}
    \epsilon_{\text{system}}(R_p,P) = S_{\text{system}}(R_p,P)\times W_{\text{system}}(P)
\end{equation}
such that the system-averaged completeness $\bar{\epsilon}$ is
\begin{equation}
    \bar{\epsilon} = \frac{1}{N_{*}}\sum_{\text{system}}\epsilon_{\text{system}}(R_p,P)
\end{equation}
The averaged completeness map for all systems is shown in Fig. \ref{fig:completeness_map}. 

In the limit of low detections (e.g. $k=0$ or $k=1$), it is robust to employ a Poisson likelihood framework to constrain the occurrence rate $f$. For a survey of $N_*$ binaries with an aggregate detection efficiency $\eta$ marginalized over the parameter space of interest, the expected number of detections $\lambda$ is
\begin{equation}
    \lambda = f \cdot N_* \cdot \eta
\end{equation}
where $\eta$ is the integrated average of $\epsilon(R_p,P)$ over the specified ranges of $R_p$ and $P$ as 
\begin{equation}
    \eta = \frac{1}{\Delta (\log P) \Delta R_p} \iint \epsilon(R_p,P) dR_p d(\log P).
\end{equation}
Here, the integration is over $\log P$ rather $P$ alone given that our bins are uniform in $\log P$.

To determine the occurrence rate upper limit $f_{up}$ in the absence of detections ($k=0$), we utilize a frequentist confidence interval for a Poisson process \citep{Gehrels_1986}. The upper limit on the number of expected events $\lambda_{up}$ for a given confidence level $CL$ (where $CL = 1-\alpha$) is found by solving
\begin{equation}
    P(k=0 | \lambda_{up}) = e^{-\lambda_{up}} = \alpha.
\end{equation}
We can rearrange for the occurrence rate upper limit to obtain
\begin{equation}
    f_{up} = \frac{-\ln(\alpha)}{N_* \eta}.
\label{eqn:occurrence_upper}
\end{equation}
This formulation allows for flexible calculations of the upper limit depending on the desired degree of certainty (e.g., $1\sigma$ or 68.3\% CL, $2\sigma$ or 95\% CL, or $3\sigma$ or 99.7\% CL\footnote{therefore, for the $1\sigma$, $2\sigma$, and $3\sigma$ cases, the corresponding $\alpha$ values are 0.317, 0.050, and 0.003, respectively}). Although there is variation in reporting of these statistics, we choose to report our results as 95\% CL upper bounds, which is consistent with similar works in the field \citep{Armstrong_2014,Windemuth_2020}.

Given the candidate status of our detection, we present our results under two distinct scenarios to bound the possible population density. In the conservative scenario, we assume no true planet detections ($k=0$) within our searched parameter space. Using the 95\% confidence interval for a Poisson process with zero events, we find a global ($10\text{d}\leq P \leq 600\text{d}, 2R_\oplus\leq R_p \leq 16 R_\oplus$) occurrence upper limit of 33.0\%. We present a suite of area-averaged occurrence constraints in Table \ref{tab:occurrences}. When we include our primary candidate, TIC 5012780.01 (with P=33.27 d, $\text{R}_p = 15.3 \text{R}_\oplus$) as a confirmed detection ($k=1$), we can move from an upper limit to a measurement of the occurrence rate. Under this scenario, we calculate a point-estimate occurrence rate of 11.0\%, with a 95\% CL upper bound of 52.2\%. Given the current uncertainty in the status of the candidate TIC 5012780.01, we conservatively take the zero detection case as our fiducial scenario.

\begin{table}
\centering
\begin{tabular}{ccc}
R$_p$ range & P$_p$ range & Occurrence Constraint  \\
($R_\oplus$) & (d)    & (percent)                  \\ \hline \hline
2-16     & 10-600     & 33.0\footnote{global occurrence rate constraint}                       \\
         & 10-300     & 27.6                       \\
         & 10-200     & 24.6                       \\
         & 10-100     & 19.7                       \\ \hline
2-14     & 10-600     & 35.9                       \\
         & 10-300     & 30.1                       \\
         & 10-200     & 26.8                       \\
         & 10-100     & 21.4                       \\ \hline
2-12     & 10-600     & 40.2                       \\
         & 10-300     & 33.7                       \\
         & 10-200     & 30.0                       \\
         & 10-100     & 24.0                       \\ \hline
2-10     & 10-600     & 47.6                       \\
         & 10-300     & 39.9                       \\
         & 10-200     & 35.5                       \\
         & 10-100     & 28.4                       \\ \hline
2-8      & 10-600     & 65.6                       \\
         & 10-300     & 55.0                       \\
         & 10-200     & 49.0                       \\
         & 10-100     & 39.3                       \\ \hline
2-6      & 10-600     & 110.9                      \\ 
         & 10-300     & 92.9                       \\
         & 10-200     & 82.8                       \\
         & 10-100     & 66.6                       \\ \hline 
2-4      & 10-600     & 158.3                      \\
         & 10-300     & 132.7                      \\
         & 10-200     & 118.4                      \\
         & 10-100     & 94.5                      
\end{tabular}
\caption{Area-averaged occurrence constraints for various parameter regions, organized by radius first, then orbital period.}
\label{tab:occurrences}
\end{table}

\begin{figure}
    \centering
    \includegraphics[width=\linewidth]{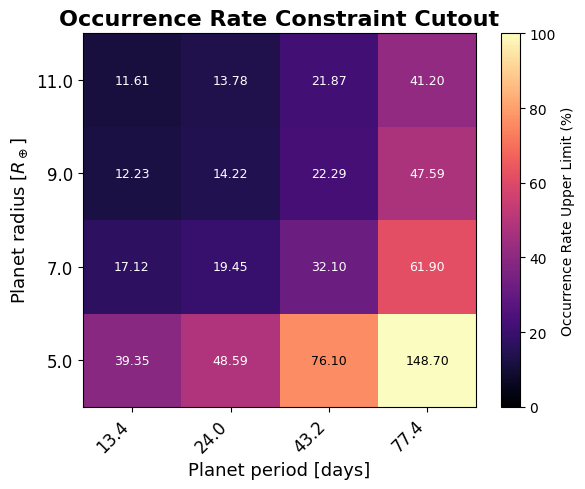}
    \caption{A cutout of our bin-by-bin occurrence rate constraints. }
    \label{fig:occurrence_cutout}
\end{figure}

From equation \ref{eqn:occurrence_upper}, we can calculate the bin-by-bin upper limits on our occurrence rate, which is useful for estimating the occurrence rate constraint in a localized region of parameter space. To do so, we recalculate the number of eligible binaries in a given period bin\footnote{therefore, we are dividing by the number of eligible binaries in the bin to calculate the occurrence constraint} and use the completeness evaluated in that period-radius bin as in Fig. \ref{fig:completeness_map}. Given that our completeness varies by two orders of magnitude across the global parameter space, the localized bin-by-bin occurrence rate constraint varies greatly as well. We show a cutout of our occurrence rate constraints in bins where occurrence is most well-constrained in Fig. \ref{fig:occurrence_cutout}. This is not the same as the area-averaged occurrence rate constraints in Table \ref{tab:occurrences}, nor can the area-averaged occurrence rate for a region be determined directly from the localized bin-by-bin occurrence rate constraints.

We also calculate our occurrence rate in the isotropic mutual inclinations case. In this case, we calculate our completeness in the same way as above, but use the window function runs where planets were initialized with isotropically-distributed mutual inclinations (the red curve in Fig. \ref{fig:window_function}). We find a 95\% CL occurrence rate upper limit of 40.2\% if planet inclinations are isotropically-distributed with respect to their hosts.

\section{Discussion}\label{sec:disc}

Our search yielded two eclipsing tertiary stellar companion candidates (TIC 22457651 and TIC 137809567) and a new singly-transiting CBP candidate, TIC 5012780.01. By constraining the sensitivity of our search pipeline and quantifying the probability of transit detection, we calculated search completeness and, for the first time, an occurrence rate of transiting CBPs in TESS data. 

\subsection{Occurrence rate comparison}

\subsubsection{Comparison to existing CBP occurrence rates}
Early speculations of CBP occurrence rates by \citet{Welsh_2012} upon the detection of the transiting Kepler-34 and Kepler-35 planets led to a lower bound of 1\% (although the authors did not perform a rigorous completeness analysis in that work), while subsequent analyses from various detection methods have suggested a range of values. The BEBOP radial velocity survey has placed constraints between 3\% and 23\% for giant planets at long periods \citep{Martin_2019_BEBOP,Baycroft_2024}.

Our results are most directly comparable to \citet{Armstrong_2014} and \citet{Windemuth_2020} because both works performed automated searches of the Kepler EB catalog. \citet{Armstrong_2014} utilized a sample of 1,735 EBs and employed a two-stage search algorithm that combined a box-fitting search for individual events with a periodicity test weighted by maximum geometric transit timing variations. They derived an occurrence rate of $10_{-6.5}^{+18}$\% for planets with periods less than 300 days and sizes in the range $6-10 R_\oplus$. \citet{Windemuth_2020} refined this search on a smaller, better-characterized sample of 728 binaries using the Quasiperiodic Automated Transit Search (QATS) algorithm \citep{Carter_2013}. QATS offers an advantage by using a dynamic window to account for the timing and duration variations without a prescribed TTV limit. Using this approach, they estimated a higher occurrence rate of $18_{-5}^{+7}$\% in their sample of detached binaries for planets in the $4-10 R_\oplus$ range out to periods of approximately 480 days \footnote{The author actually specifies the occurrence rate as a function of the critical stability period of the binary $P_c$. Their maximum period is $35\times P_c$, which roughly corresponds to 480 days.}.

To facilitate a direct comparison between the TESS M\&M population and the higher-mass Kepler population, we recalculate our occurrence rate within the specific parameter bounds defined by these previous works. Within the \citet{Armstrong_2014} bounds (planet orbital periods less than 300 days and planet radii in the range $6-10 R_\oplus$), we calculate a 95\% CL upper limit of 25.4\% for the coplanar case. Within the \citet{Windemuth_2020} bounds (planet orbital periods below 480 days and planet radii in the range $4-10 R_\oplus$), we calculate a 95\% CL upper limit of 36.6\% for the coplanar case. While these limits are consistent with the rates measured in the Kepler EB population, they are as of yet insufficiently constraining to confirm or rule out a similar occurrence rate for M\&Ms. 

\subsubsection{Comparison to occurrence rates of planets around single M dwarfs}

Planets around single M dwarfs represent a natural reference point for this population. Giant exoplanets around single M dwarfs (GEMS) are notoriously rare, with occurrence rates ranging from $\lesssim1\%$ at short periods \citep{Endl_2006,Mignon_2025,Glusman_2026} to roughly 6\% out to 1,000 days \citep{Sabotta_2021}. Meanwhile, it is known that single M dwarfs are prolific hosts of small planets at both short periods \citep{Ment2023} and at periods up to 256 days \citep{Dressing_2015,Hsu2020}. In the region examined here ($10\text{d} < P_p < 256\text{d}$, $2.0R_\oplus < R_p < 4.0 R_\oplus$), Kepler M dwarfs host $0.46_{-0.11}^{+1.19}$ small planets per star, and potentially as many as 2.8 planets per star depending on the assumed prior and treatment of unconstrained bins \citep{Hsu2020}. In the same region, we find a 95\% CL upper limit of 127\%. We interpret this as a relatively uninformative bound, and this result does not necessarily imply that each M\&M hosts a a small CBP. There are currently no observed transiting CBPs with radii smaller than $3 R_\oplus$, making this uncharted observational territory. Whether the small CBP occurrence rate around M\&Ms tracks, exceeds, or falls below even the already-rich single M-dwarf rate remains an open question that our data cannot yet address. 

The push towards smaller CBPs and characterizing how common they are carries important implications for how they form. In single star systems, the location of rocky planet formation is debated. Rocky planets may form \textit{in situ} \citep{Chachan_2023,Clement_2025}, or they may form in a more distant part of the disk and migrate inwards \citep{Sanchez_2024}, where they are detected in abundance. Since tight binary environments are more constrained than single stars, these systems represent \textit{natural laboratories} to study planet formation. Specifically, because close-in orbits are prohibited in circumbinary architectures, CBPs are thought to form in more distant regions and migrate inwards \citep{thun2018migration}. Therefore, CBPs provide a new lens to study planet formation locations, and a new constraint on the occurrence of small CBPs could therefore be informative as to which planet formation channel is dominant in this regime.

\subsection{Planet formation in the M\&M context}

The current estimate of M\&M CBP occurrence leaves open the question of whether planet formation proceeds differently in tight binary environments. Protoplanetary disks around tight binaries have been shown to be aligned with the binary (including in the M\&M context \citep{Czekala2019,Czekala2021}), but have lower surface densities than those around single stars \citep{Cox2017, Flaherty2025}. Further, tight binaries with high mass ratios host circumbinary disks less frequently \citep{Kounkel_2019}, reducing the raw material available for planet formation from the outset.  If planets form via core accretion, this deficit of solid material is particularly consequential for giant planet formation, which requires rapid assembly of a massive core before dispersal of the gaseous disk \citep{Laughlin_2004}. Pebble accretion, which relies on the assembly of mm- to cm-sized particles rather than planetesimal collisions, offers a potentially more efficient formation pathway and has been shown to produce CBPs in simulations around solar-type binaries \citep{Coleman2023, Coleman2024}. Whether this mechanism operates efficiently in the low-mass binary regime remains untested. Theoretical work modeling pebble accretion in M\&M circumbinary disks would provide a valuable prediction for the expected small planet occurrence rate.

\subsubsection{The impact of triples and very short-period binaries}

The TESS M\&M sample is biased toward very short-period binaries, with a median orbital period of only 3.4 days \citep{Oddo_2026}, compared to the Kepler EB sample which contains a higher proportion of longer-period EBs. Distinguishing between a true suppression of the intrinsic CBP occurrence rate or a merely reduced sensitivity to planets around short-period EBs is difficult with the current data, but several mechanisms suggest the former may be at least partially true.

High-order stellar multiplicity is common among tight binaries, with as many as 96\% of spectroscopic binaries with periods less than 3 days having a tertiary companion \citep{Tokovinin_2006}. Further, about 25\% of all Kepler EBs are in triple systems \citep{Rappaport2013, Borkovits_2016, Borkovits_2025}. Stellar tertiaries are thought to harden tight binaries via the von Zeipel-Kozai-Lidov mechanism \citep{Fabrycky_2007}, which has deleterious implications for CBPs caught in between \citep{Munoz_2015, Hamers_2016}. Even without a tertiary, short-period binaries may harden through disk-driven angular momentum exchange \citep{Bate_2000, Bate_2002, Zhao2013} or post-disk tidal processes including coupled stellar-tidal evolution \citep{Fleming2018}, tidal shrinkage \citep{Mogan_2025}, and capture into apsidal resonance \citep{Farhat_2025, Liu_2026}, all of which can destabilize primordial planets. However, the caveat is that none of these processes have yet been modeled specifically for M\&M binaries.

\citet{Oddo_2026} found a compressed circularization period and a notable deficiency of binaries with periods below 1d in the M\&M sample relative to other Kepler and TESS EB populations. This pattern could be consistent with more efficient tidal dissipation in low-mass binaries. Rather than stalling in a long-lived contact phase, ultra-short-period M\&Ms may undergo rapid, unstable orbital decay leading to coalescence. For any primordial CBP, such swift tidal shrinkage of the central binary could be catastrophic, and represents a destruction mechanism that may be uniquely efficient in the M\&M regime. More modeling is required to test this hypothesis.

\subsubsection{The role of system metallicity}

The role of metallicity in shaping the M\&M planet population warrants further investigation. Giant planets are known to preferentially orbit metal-rich hosts \citep{Gan2025}, but if tight binaries form via turbulent fragmentation, they may be depleted of planet-forming material since this process is favored at low metallicities \citep{Offner_2010}. Therefore, the metallicity distribution of the M\&M sample can be a key diagnostic for future study. Currently, the median metallicity of the M\&M sample is near-solar based on preliminary broadband SED fitting \citep{Oddo_2026}. A dedicated characterization of the M\&M sample with high-resolution near-infrared spectroscopy would sharpen the metallicity distribution, providing a necessary observational anchor for planet formation models.

\subsection{The impact of system geometry}

A central methodological finding of this work is that the geometric effect of binary inclination is a dominant source of incompleteness in our survey, outweighing the pipeline sensitivity losses quantified by injection and recovery testing. Although EBs are by definition observed at near-edge-on inclinations, the distribution of inclinations within the M\&M sample (Fig. \ref{fig:inclination_v_period}) is sufficiently broad that the fraction of orbital configurations yielding a detectable planet transit is small. This geometric dilution drives our global upper limit well above what the pipeline sensitivity alone would suggest, and explains why our constraints within the \citet{Armstrong_2014} and \citet{Windemuth_2020} parameter bounds are less restrictive than those derived from the Kepler EB samples. Future surveys seeking to improve on these constraints should prioritize systems with the most precisely characterized binary inclinations, where the window function can be calculated with greater confidence, or target larger samples to compensate for the geometric losses statistically.

\subsection{Caveats and limitations}

\subsubsection{Mutual inclination assumptions}

Our nominal occurrence rate calculation assumes a Rayleigh distribution for mutual inclinations centered at $1.5^\circ$. While a truncated Gaussian has on occasion been employed in the literature \citep{Armstrong_2014}, the Rayleigh distribution is physically motivated because it remains non-negative while effectively modeling the low-but-extant probability of higher mutual inclinations. Theoretical work suggests that CBPs orbiting highly eccentric binaries may evolve toward a stable polar alignment \citep{RMartin_2017,RMartin_2018,Lubow2024}, but coplanar disks are more likely for tighter binaries \citep{Czekala2019}. Because our search pipeline is optimized for transiting planets, a potential population of polar planets would remain largely invisible to our current methodology. Future studies may further reveal that a Rayleigh distribution centered at $1.5^\circ$ as a prior is not perfectly correct.

\subsubsection{Modeling transits in the short-period regime}

Unlike \citet{Armstrong_2014} and \citet{Windemuth_2020}, which excluded semi-detached and contact binaries, our sample includes systems with $P_{bin} < 1$ day. In this extreme regime, the stellar shapes can be distorted, which may mean that the radially symmetric approximation used by the \texttt{batman} transit model breaks down. Furthermore, a substantial portion of the light curve is often masked during binary eclipses, reducing our duty cycle for detecting short-duration transits. We expect that the complex transit profiles expected in semi-detached or contact systems (i.e. high-morphology systems) require more sophisticated modeling. Future work should leverage more complete tools like \texttt{PHOEBE} \citep{Prsa_2005,Prsa_2016}, the Eclipsing Light Curve (ELC) code \citep{Orosz_2000,Orosz_2019}, or a brightness-mapping tool such as \texttt{starry}\footnote{\url{https://starry.readthedocs.io/en/latest/}} to account for non-spherical stars and the rapid orbital motion of the hosts during transit.  

\subsubsection{Sample completeness}

The M\&M EBs utilized in this work are a high-confidence sample of bona fide low-mass systems \citep{Oddo_2026}. However, the multi-stage filtering process used to reach this sample introduced complexities in quantifying the underlying population completeness. The original catalog was derived from a neural-network-based search of millions of TESS FFI light curves \citep{Powell_2021,Kostov_2025}. While CNNs are highly efficient at identifying eclipse-like signals in large datasets, they often possess non-linear selection functions that are difficult to characterize. As noted in the catalog's construction, the initial classifier exhibited a high false-positive rate (up to 86\%), and it remains unknown how many true M\&M binaries were discarded as false negatives. Consequently, the results reported here represent the occurrence of planets within a vetted subsample rather than a complete sample. 

Further, the sensitivity of this work is predominantly constrained by the $T < 15$ magnitude limit of the parent sample. For M-dwarf systems, which are intrinsically faint, a $T < 15$ limit preferentially samples the most proximal or most luminous members of the class. The subsequent application of $T_{eff} < 4200$ K and color-color cuts ($BP - RP > 1.97$; $G - K > 3.013$) was essential to ensure a pure sample of M-dwarf binaries. While these thresholds were chosen to be conservative, incompleteness at the faint end of the magnitude limit remains unquantified. Because the total number of M\&M binaries in the TESS field that could have been in the sample is unknown, our reported occurrence rate is technically the occurrence rate per detected, vetted EB. Detailed testing of the binary detection pipeline would be required to convert this constraint into more absolute terms for the Galactic (or even nearby) M-dwarf population. While a formal completeness correction for the parent EB catalog is beyond the scope of this work, several factors suggest the sample is largely representative of the TESS M\&M EB population. The high degree of overlap between the initial neural network detections and existing EB catalogs \citep{Kostov_2025} suggests that the neural network-based search was broadly sensitive to EBs across the orbital periods considered here. 

\section{Conclusions}\label{sec:conc}

This work presents the first systematic determination of the occurrence rate of transiting CBPs using TESS data. The starting population of hosts we chose was M\&M EBs. In this work, we applied our custom single-transit detection pipeline to the TESS M\&M EBs and calculated the survey completeness to arrive at our occurrence rate estimate.

We summarize the results and conclusions of this work as follows:
\begin{itemize}
    \item Our transit search yielded five transit signals which passed our vetting steps, four of which are due to likely tertiary eclipsing stellar companions, and only one of which is due to a planet-sized object;
    \item We present the detection of the singly-transiting super-Jupiter CBP candidate TIC 5012780.01. If confirmed, this planet would have the shortest orbital period of any CBP to date and would be the only known transiting CBP with a host whose orbital period is less than 7 days. More follow-up is necessary to confirm the existence of this candidate;
    \item We performed both sensitivity analysis and geometric corrections to account for the TESS window function to find our survey completeness and calculate M\&M CBP occurrence rates. We find a 95\% CL upper limit on the occurrence rate of coplanar CBPs ($2-16R_\oplus$) of 33.0\%, which is our fiducial case. When including our primary candidate, TIC 5012780.01, the measured occurrence rate is 11.0\% (with a 95\% CL upper limit of 52.2\%);
    \item Our occurrence limits are sensitive to the assumed underlying mutual inclination distribution. Although the likeliest scenario is that CBPs are preferentially coplanar, if they are instead isotropically distributed, our upper limit rises to 40.2\%;
    \item Our completeness was predominantly limited by system geometry. The wide distribution of binary inclinations substantially reduced the window function. Definitively testing whether the M\&M CBP occurrence rate is suppressed relative to the Kepler EB population will require a substantially larger M\&M sample, a sample with more edge-on inclinations, or both.
\end{itemize}

The results presented here define several pathways for future investigation towards a more general census of TESS CBPs. A primary area for development lies in the continuing refinement of automated detection techniques. While our single-transit search methodology proved effective for identifying giant candidates, future iterations should examine additional de-trending algorithms, such as Gaussian Process regression (GPs), as well as more robust false positive identification, such as \citet{Davies2026}. 

This work sets the stage for the continued investigation of CBPs with future space telescopes. Due to their dynamically-active environments, CBPs may comprise a substantial portion of the free-floating planet population \citep{Coleman2024_freefloating}, to which the recently-launched Nancy Grace Roman Space Telescope (aka Roman) will be sensitive. With its ability to reach deeper than ever before, Roman is expected to discover more than 1000 small transiting planets around M dwarfs \citep{Tamburo_2023AJ}, and Roman may have the potential to detect M\&M CBPs.

Finally, the transiting companions detected by our search, including our primary candidate TOI 5012780.01, demand dedicated follow-up. Additional false positive verification through high resolution imaging and reconnaissance spectroscopy is necessary. This should be followed by high-resolution near-infrared spectroscopy, which can more deeply characterize these systems to recover precise masses.  

\section{Acknowledgments}

D.O. acknowledges support from and is primarily supported by NASA FINESST award number 80NSSC23K1448.

D. D. acknowledges support from the TESS Guest Investigator Program grant 80NSSC23K0769.

This research was supported by the UNM Office of the Vice President of Research.

We would like to thank the UNM Center for Advanced Research Computing, supported in part by the National Science Foundation, for providing the research computing resources used in this work.

This paper includes data collected by the TESS mission. Funding for the TESS mission is provided by NASA's Science Mission Directorate.

We acknowledge the use of public TESS data from pipelines at the TESS Science Office and at the TESS Science Processing Operations Center. Resources supporting this work were provided by the NASA High-End Computing (HEC) Program through the NASA Advanced Supercomputing (NAS) Division at Ames Research Center for the production of the SPOC data products.

\bibliography{bibliography}{}

@article{meschiari2012planet,
  title={Planet formation in circumbinary configurations: turbulence inhibits planetesimal accretion},
  author={Meschiari, Stefano},
  journal={The Astrophysical Journal Letters},
  volume={761},
  number={1},
  pages={L7},
  year={2012},
  publisher={IOP Publishing}
}

@article{paardekooper2012not,
  title={How not to build Tatooine: the difficulty of in situ formation of circumbinary planets Kepler 16b, Kepler 34b, and Kepler 35b},
  author={Paardekooper, Sijme-Jan and Leinhardt, Zo{\"e} M and Th{\'e}bault, Philippe and Baruteau, Cl{\'e}ment},
  journal={The Astrophysical journal letters},
  volume={754},
  number={1},
  pages={L16},
  year={2012},
  publisher={IOP Publishing}
}

@Inbook{Welsh_2018,
author="Welsh, William F.
and Orosz, Jerome A.",
editor="Deeg, Hans J.
and Belmonte, Juan Antonio",
title="Two Suns in the Sky: The Kepler Circumbinary Planets",
bookTitle="Handbook of Exoplanets ",
year="2018",
publisher="Springer International Publishing",
address="Cham",
pages="2749--2768",
isbn="978-3-319-55333-7",
doi="10.1007/978-3-319-55333-7_34",
url="https://doi.org/10.1007/978-3-319-55333-7_34"
}

@article{li2016uncovering,
  title={Uncovering circumbinary planetary architectural properties from selection biases},
  author={Li, Gongjie and Holman, Matthew J and Tao, Molei},
  journal={The Astrophysical Journal},
  volume={831},
  number={1},
  pages={96},
  year={2016},
  publisher={IOP Publishing}
}

@article{thun2018migration,
  title={Migration of planets in circumbinary discs},
  author={Thun, Daniel and Kley, Wilhelm},
  journal={Astronomy \& Astrophysics},
  volume={616},
  pages={A47},
  year={2018},
  publisher={EDP Sciences}
}

@ARTICLE{Penzlin_2021,
       author = {{Penzlin}, Anna B.~T. and {Kley}, Wilhelm and {Nelson}, Richard P.},
        title = "{Parking planets in circumbinary discs}",
      journal = {\aap},
         year = 2021,
        month = jan,
       volume = {645},
          eid = {A68},
        pages = {A68},
          doi = {10.1051/0004-6361/202039319},
archivePrefix = {arXiv},
       eprint = {2012.03651},
 primaryClass = {astro-ph.EP},
       adsurl = {https://ui.adsabs.harvard.edu/abs/2021A&A...645A..68P}
}

@ARTICLE{Kane_2013,
       author = {{Kane}, Stephen R. and {Hinkel}, Natalie R.},
        title = "{On the Habitable Zones of Circumbinary Planetary Systems}",
      journal = {\apj},
         year = 2013,
        month = jan,
       volume = {762},
       number = {1},
          eid = {7},
        pages = {7},
          doi = {10.1088/0004-637X/762/1/7},
archivePrefix = {arXiv},
       eprint = {1211.2812},
 primaryClass = {astro-ph.EP},
       adsurl = {https://ui.adsabs.harvard.edu/abs/2013ApJ...762....7K}
}

@ARTICLE{Eggl_2020,
       author = {{Eggl}, Siegfried and {Georgakarakos}, Nikolaos and {Pilat-Lohinger}, Elke},
        title = "{Habitable Zones in Binary Star Systems: A Zoology}",
      journal = {Galaxies},
         year = 2020,
        month = sep,
       volume = {8},
       number = {3},
          eid = {65},
        pages = {65},
          doi = {10.3390/galaxies8030065},
archivePrefix = {arXiv},
       eprint = {2009.00144},
 primaryClass = {astro-ph.EP},
       adsurl = {https://ui.adsabs.harvard.edu/abs/2020Galax...8...65E}
}

@ARTICLE{Kostov_2023,
       author = {{Kostov}, Veselin B.},
        title = "{Transiting Circumbinary Planets in the Era of Space-Based Photometric Surveys}",
      journal = {Universe},
         year = 2023,
        month = oct,
       volume = {9},
       number = {10},
          eid = {455},
        pages = {455},
          doi = {10.3390/universe9100455},
       adsurl = {https://ui.adsabs.harvard.edu/abs/2023Univ....9..455K}
}

@ARTICLE{Borucki_2010,
       author = {{Borucki}, William J. and {Koch}, David and {Basri}, Gibor and {Batalha}, Natalie and {Brown}, Timothy and {Caldwell}, Douglas and {Caldwell}, John and {Christensen-Dalsgaard}, J{\o}rgen and {Cochran}, William D. and {DeVore}, Edna and {Dunham}, Edward W. and {Dupree}, Andrea K. and {Gautier}, Thomas N. and {Geary}, John C. and {Gilliland}, Ronald and {Gould}, Alan and {Howell}, Steve B. and {Jenkins}, Jon M. and {Kondo}, Yoji and {Latham}, David W. and {Marcy}, Geoffrey W. and {Meibom}, S{\o}ren and {Kjeldsen}, Hans and {Lissauer}, Jack J. and {Monet}, David G. and {Morrison}, David and {Sasselov}, Dimitar and {Tarter}, Jill and {Boss}, Alan and {Brownlee}, Don and {Owen}, Toby and {Buzasi}, Derek and {Charbonneau}, David and {Doyle}, Laurance and {Fortney}, Jonathan and {Ford}, Eric B. and {Holman}, Matthew J. and {Seager}, Sara and {Steffen}, Jason H. and {Welsh}, William F. and {Rowe}, Jason and {Anderson}, Howard and {Buchhave}, Lars and {Ciardi}, David and {Walkowicz}, Lucianne and {Sherry}, William and {Horch}, Elliott and {Isaacson}, Howard and {Everett}, Mark E. and {Fischer}, Debra and {Torres}, Guillermo and {Johnson}, John Asher and {Endl}, Michael and {MacQueen}, Phillip and {Bryson}, Stephen T. and {Dotson}, Jessie and {Haas}, Michael and {Kolodziejczak}, Jeffrey and {Van Cleve}, Jeffrey and {Chandrasekaran}, Hema and {Twicken}, Joseph D. and {Quintana}, Elisa V. and {Clarke}, Bruce D. and {Allen}, Christopher and {Li}, Jie and {Wu}, Haley and {Tenenbaum}, Peter and {Verner}, Ekaterina and {Bruhweiler}, Frederick and {Barnes}, Jason and {Prsa}, Andrej},
        title = "{Kepler Planet-Detection Mission: Introduction and First Results}",
      journal = {Science},
         year = 2010,
        month = feb,
       volume = {327},
       number = {5968},
        pages = {977},
          doi = {10.1126/science.1185402},
       adsurl = {https://ui.adsabs.harvard.edu/abs/2010Sci...327..977B}
}

@ARTICLE{Doyle_2011,
       author = {{Doyle}, Laurance R. and {Carter}, Joshua A. and {Fabrycky}, Daniel C. and {Slawson}, Robert W. and {Howell}, Steve B. and {Winn}, Joshua N. and {Orosz}, Jerome A. and {P{\v{r}}sa}, Andrej and {Welsh}, William F. and {Quinn}, Samuel N. and {Latham}, David and {Torres}, Guillermo and {Buchhave}, Lars A. and {Marcy}, Geoffrey W. and {Fortney}, Jonathan J. and {Shporer}, Avi and {Ford}, Eric B. and {Lissauer}, Jack J. and {Ragozzine}, Darin and {Rucker}, Michael and {Batalha}, Natalie and {Jenkins}, Jon M. and {Borucki}, William J. and {Koch}, David and {Middour}, Christopher K. and {Hall}, Jennifer R. and {McCauliff}, Sean and {Fanelli}, Michael N. and {Quintana}, Elisa V. and {Holman}, Matthew J. and {Caldwell}, Douglas A. and {Still}, Martin and {Stefanik}, Robert P. and {Brown}, Warren R. and {Esquerdo}, Gilbert A. and {Tang}, Sumin and {Furesz}, Gabor and {Geary}, John C. and {Berlind}, Perry and {Calkins}, Michael L. and {Short}, Donald R. and {Steffen}, Jason H. and {Sasselov}, Dimitar and {Dunham}, Edward W. and {Cochran}, William D. and {Boss}, Alan and {Haas}, Michael R. and {Buzasi}, Derek and {Fischer}, Debra},
        title = "{Kepler-16: A Transiting Circumbinary Planet}",
      journal = {Science},
         year = 2011,
        month = sep,
       volume = {333},
       number = {6049},
        pages = {1602},
          doi = {10.1126/science.1210923},
archivePrefix = {arXiv},
       eprint = {1109.3432},
 primaryClass = {astro-ph.EP},
       adsurl = {https://ui.adsabs.harvard.edu/abs/2011Sci...333.1602D}
}

@ARTICLE{Doyle_2000,
       author = {{Doyle}, Laurance R. and {Deeg}, Hans J. and {Kozhevnikov}, Valerij P. and {Oetiker}, Brian and {Mart{\'\i}n}, Eduardo L. and {Blue}, J. Ellen and {Rottler}, Lee and {Stone}, Remington P.~S. and {Ninkov}, Zoran and {Jenkins}, Jon M. and {Schneider}, Jean and {Dunham}, Edward W. and {Doyle}, Moira F. and {Paleologou}, Efthimious},
        title = "{Observational Limits on Terrestrial-sized Inner Planets around the CM Draconis System Using the Photometric Transit Method with a Matched-Filter Algorithm}",
      journal = {\apj},
         year = 2000,
        month = may,
       volume = {535},
       number = {1},
        pages = {338-349},
          doi = {10.1086/308830},
archivePrefix = {arXiv},
       eprint = {astro-ph/0001177},
 primaryClass = {astro-ph},
       adsurl = {https://ui.adsabs.harvard.edu/abs/2000ApJ...535..338D}
}

@ARTICLE{Kostov_2016_kepler1647,
       author = {{Kostov}, Veselin B. and {Orosz}, Jerome A. and {Welsh}, William F. and {Doyle}, Laurance R. and {Fabrycky}, Daniel C. and {Haghighipour}, Nader and {Quarles}, Billy and {Short}, Donald R. and {Cochran}, William D. and {Endl}, Michael and {Ford}, Eric B. and {Gregorio}, Joao and {Hinse}, Tobias C. and {Isaacson}, Howard and {Jenkins}, Jon M. and {Jensen}, Eric L.~N. and {Kane}, Stephen and {Kull}, Ilya and {Latham}, David W. and {Lissauer}, Jack J. and {Marcy}, Geoffrey W. and {Mazeh}, Tsevi and {M{\"u}ller}, Tobias W.~A. and {Pepper}, Joshua and {Quinn}, Samuel N. and {Ragozzine}, Darin and {Shporer}, Avi and {Steffen}, Jason H. and {Torres}, Guillermo and {Windmiller}, Gur and {Borucki}, William J.},
        title = "{Kepler-1647b: The Largest and Longest-period Kepler Transiting Circumbinary Planet}",
      journal = {\apj},
         year = 2016,
        month = aug,
       volume = {827},
       number = {1},
          eid = {86},
        pages = {86},
          doi = {10.3847/0004-637X/827/1/86},
archivePrefix = {arXiv},
       eprint = {1512.00189},
 primaryClass = {astro-ph.EP},
       adsurl = {https://ui.adsabs.harvard.edu/abs/2016ApJ...827...86K}
}

@ARTICLE{Orosz_2019,
       author = {{Orosz}, Jerome A. and {Welsh}, William F. and {Haghighipour}, Nader and {Quarles}, Billy and {Short}, Donald R. and {Mills}, Sean M. and {Satyal}, Suman and {Torres}, Guillermo and {Agol}, Eric and {Fabrycky}, Daniel C. and {Jontof-Hutter}, Daniel and {Windmiller}, Gur and {M{\"u}ller}, Tobias W.~A. and {Hinse}, Tobias C. and {Cochran}, William D. and {Endl}, Michael and {Ford}, Eric B. and {Mazeh}, Tsevi and {Lissauer}, Jack J.},
        title = "{Discovery of a Third Transiting Planet in the Kepler-47 Circumbinary System}",
      journal = {\aj},
         year = 2019,
        month = may,
       volume = {157},
       number = {5},
          eid = {174},
        pages = {174},
          doi = {10.3847/1538-3881/ab0ca0},
archivePrefix = {arXiv},
       eprint = {1904.07255},
 primaryClass = {astro-ph.EP},
       adsurl = {https://ui.adsabs.harvard.edu/abs/2019AJ....157..174O}
}

@ARTICLE{Orosz_2012_kepler47,
       author = {{Orosz}, Jerome A. and {Welsh}, William F. and {Carter}, Joshua A. and {Fabrycky}, Daniel C. and {Cochran}, William D. and {Endl}, Michael and {Ford}, Eric B. and {Haghighipour}, Nader and {MacQueen}, Phillip J. and {Mazeh}, Tsevi and {Sanchis-Ojeda}, Roberto and {Short}, Donald R. and {Torres}, Guillermo and {Agol}, Eric and {Buchhave}, Lars A. and {Doyle}, Laurance R. and {Isaacson}, Howard and {Lissauer}, Jack J. and {Marcy}, Geoffrey W. and {Shporer}, Avi and {Windmiller}, Gur and {Barclay}, Thomas and {Boss}, Alan P. and {Clarke}, Bruce D. and {Fortney}, Jonathan and {Geary}, John C. and {Holman}, Matthew J. and {Huber}, Daniel and {Jenkins}, Jon M. and {Kinemuchi}, Karen and {Kruse}, Ethan and {Ragozzine}, Darin and {Sasselov}, Dimitar and {Still}, Martin and {Tenenbaum}, Peter and {Uddin}, Kamal and {Winn}, Joshua N. and {Koch}, David G. and {Borucki}, William J.},
        title = "{Kepler-47: A Transiting Circumbinary Multiplanet System}",
      journal = {Science},
         year = 2012,
        month = sep,
       volume = {337},
       number = {6101},
        pages = {1511},
          doi = {10.1126/science.1228380},
archivePrefix = {arXiv},
       eprint = {1208.5489},
 primaryClass = {astro-ph.SR},
       adsurl = {https://ui.adsabs.harvard.edu/abs/2012Sci...337.1511O}
}

@ARTICLE{Armstrong_2014,
       author = {{Armstrong}, D.~J. and {Osborn}, H.~P. and {Brown}, D.~J.~A. and {Faedi}, F. and {G{\'o}mez Maqueo Chew}, Y. and {Martin}, D.~V. and {Pollacco}, D. and {Udry}, S.},
        title = "{On the abundance of circumbinary planets}",
      journal = {\mnras},
         year = 2014,
        month = oct,
       volume = {444},
       number = {2},
        pages = {1873-1883},
          doi = {10.1093/mnras/stu1570},
archivePrefix = {arXiv},
       eprint = {1404.5617},
 primaryClass = {astro-ph.EP},
       adsurl = {https://ui.adsabs.harvard.edu/abs/2014MNRAS.444.1873A}
}

@article{ricker2015TESS,
author = {George R. Ricker and Joshua N. Winn and Roland Vanderspek and David W. Latham and Gáspár  Á. Bakos and Jacob L. Bean and Zachory K. Berta-Thompson and Timothy M. Brown and Lars Buchhave and Nathaniel R. Butler and R. Paul Butler and William J. Chaplin and David B. Charbonneau and Jørgen Christensen-Dalsgaard and Mark Clampin and Drake Deming and John P. Doty and Nathan De Lee and Courtney Dressing and Edward W. Dunham and Michael Endl and François Fressin and Jian Ge and Thomas Henning and Matthew J. Holman and Andrew W. Howard and Shigeru Ida and Jon M. Jenkins and Garrett Jernigan and John Asher Johnson and Lisa Kaltenegger and Nobuyuki Kawai and Hans Kjeldsen and Gregory Laughlin and Alan M. Levine and Douglas Lin and Jack J. Lissauer and Phillip MacQueen and Geoffrey Marcy and Peter R. McCullough and Timothy D. Morton and Norio Narita and Martin Paegert and Enric Palle and Francesco Pepe and Joshua Pepper and Andreas Quirrenbach and Stephen A. Rinehart and Dimitar Sasselov and Bun’ei Sato and Sara Seager and Alessandro Sozzetti and Keivan G. Stassun and Peter Sullivan and Andrew Szentgyorgyi and Guillermo Torres and Stephane Udry and Joel Villasenor},
title = {{Transiting Exoplanet Survey Satellite}},
volume = {1},
journal = {Journal of Astronomical Telescopes, Instruments, and Systems},
number = {1},
publisher = {SPIE},
pages = {1 -- 10},
year = {2014},
doi = {10.1117/1.JATIS.1.1.014003},
URL = {https://doi.org/10.1117/1.JATIS.1.1.014003}
}

@ARTICLE{Kunimoto_2022,
       author = {{Kunimoto}, Michelle and {Winn}, Joshua and {Ricker}, George R. and {Vanderspek}, Roland K.},
        title = "{Predicting the Exoplanet Yield of the TESS Prime and Extended Missions through Years 1-7}",
      journal = {\aj},
         year = 2022,
        month = jun,
       volume = {163},
       number = {6},
          eid = {290},
        pages = {290},
          doi = {10.3847/1538-3881/ac68e3},
archivePrefix = {arXiv},
       eprint = {2202.03656},
 primaryClass = {astro-ph.EP},
       adsurl = {https://ui.adsabs.harvard.edu/abs/2022AJ....163..290K}
}

@ARTICLE{Kostov_2025,
       author = {{Kostov}, Veselin B. and {Powell}, Brian P. and {Fornear}, Aline U. and {Di Fraia}, Marco Z. and {Gagliano}, Robert and {Jacobs}, Thomas L. and {de Lambilly}, Julien S. and {Durantini Luca}, Hugo A. and {Majewski}, Steven R. and {Omohundro}, Mark and {Orosz}, Jerome and {Rappaport}, Saul A. and {Salik}, Ryan and {Short}, Donald and {Welsh}, William and {Alexandrov}, Svetoslav and {da Silva}, Cledison Marcos and {Dunning}, Erika and {G{\"u}hne}, Gerd and {Huten}, Marc and {Hyogo}, Michiharu and {Iannone}, Davide and {Lee}, Sam and {Magliano}, Christian and {Sharma}, Manya and {Tarr}, Allan and {Yablonsky}, John and {Acharya}, Sovan and {Adams}, Fred and {Barclay}, Thomas and {Montet}, Benjamin T. and {Mullally}, Susan and {Olmschenk}, Greg and {Pr{\v{s}}a}, Andrej and {Quintana}, Elisa and {Wilson}, Robert and {Balcioglu}, Hasret and {Kruse}, Ethan and {The Eclipsing Binary Patrol Collaboration}},
        title = "{The TESS Ten Thousand Catalog: 10,001 Uniformly Vetted and Validated Eclipsing Binary Stars Detected in Full-frame Image Data by Machine Learning and Analyzed by Citizen Scientists}",
      journal = {\apjs},
         year = 2025,
        month = aug,
       volume = {279},
       number = {2},
          eid = {50},
        pages = {50},
          doi = {10.3847/1538-4365/ade2d8},
archivePrefix = {arXiv},
       eprint = {2506.05631},
 primaryClass = {astro-ph.SR},
       adsurl = {https://ui.adsabs.harvard.edu/abs/2025ApJS..279...50K}
}

@ARTICLE{Sanchez_2024,
       author = {{Sanchez}, M. and {van der Marel}, N. and {Lambrechts}, M. and {Mulders}, G.~D. and {Guerra-Alvarado}, O.~M.},
        title = "{Rocky planet formation in compact disks around M dwarfs}",
      journal = {\aap},
         year = 2024,
        month = sep,
       volume = {689},
          eid = {A236},
        pages = {A236},
          doi = {10.1051/0004-6361/202450350},
archivePrefix = {arXiv},
       eprint = {2406.07760},
 primaryClass = {astro-ph.EP},
       adsurl = {https://ui.adsabs.harvard.edu/abs/2024A&A...689A.236S}
}

@ARTICLE{Chachan_2023,
       author = {{Chachan}, Yayaati and {Lee}, Eve J.},
        title = "{Small Planets around Cool Dwarfs: Enhanced Formation Efficiency of Super-Earths around M Dwarfs}",
      journal = {\apjl},
         year = 2023,
        month = jul,
       volume = {952},
       number = {1},
          eid = {L20},
        pages = {L20},
          doi = {10.3847/2041-8213/ace257},
archivePrefix = {arXiv},
       eprint = {2305.00803},
 primaryClass = {astro-ph.EP},
       adsurl = {https://ui.adsabs.harvard.edu/abs/2023ApJ...952L..20C}
}

@ARTICLE{Clement_2025,
       author = {{Clement}, Matthew S. and {Quintana}, Elisa V. and {Stevenson}, Kevin B.},
        title = "{On the Local Formation of the TRAPPIST-1 Exoplanets}",
      journal = {\aj},
         year = 2025,
        month = jan,
       volume = {169},
       number = {1},
          eid = {16},
        pages = {16},
          doi = {10.3847/1538-3881/ad8e65},
archivePrefix = {arXiv},
       eprint = {2411.02529},
 primaryClass = {astro-ph.EP},
       adsurl = {https://ui.adsabs.harvard.edu/abs/2025AJ....169...16C}
}

@ARTICLE{Martin_2015,
       author = {{Martin}, David V. and {Triaud}, Amaury H.~M.~J.},
        title = "{Circumbinary planets - why they are so likely to transit}",
      journal = {\mnras},
         year = 2015,
        month = may,
       volume = {449},
       number = {1},
        pages = {781-793},
          doi = {10.1093/mnras/stv121},
archivePrefix = {arXiv},
       eprint = {1501.03631},
 primaryClass = {astro-ph.EP},
       adsurl = {https://ui.adsabs.harvard.edu/abs/2015MNRAS.449..781M}
}

@ARTICLE{Martin_2015_tight,
       author = {{Martin}, David V. and {Mazeh}, Tsevi and {Fabrycky}, Daniel C.},
        title = "{No circumbinary planets transiting the tightest Kepler binaries - a possible fingerprint of a third star}",
      journal = {\mnras},
         year = 2015,
        month = nov,
       volume = {453},
       number = {4},
        pages = {3554-3567},
          doi = {10.1093/mnras/stv1870},
archivePrefix = {arXiv},
       eprint = {1505.05749},
 primaryClass = {astro-ph.EP},
       adsurl = {https://ui.adsabs.harvard.edu/abs/2015MNRAS.453.3554M}
}

@ARTICLE{Kostov_2020,
       author = {{Kostov}, Veselin B. and {Welsh}, William F. and {Haghighipour}, Nader and {Quarles}, Billy and {Agol}, Eric and {Doyle}, Laurance and {Fabrycky}, Daniel C. and {Li}, Gongjie and {Martin}, David V. and {Mills}, Sean and {Mazeh}, Tsevi and {Orosz}, Jerome A. and {Powell}, Brian P.},
        title = "{Multiple Transits during a Single Conjunction: Identifying Transiting Circumbinary Planetary Candidates from TESS}",
      journal = {\aj},
         year = 2020,
        month = oct,
       volume = {160},
       number = {4},
          eid = {174},
        pages = {174},
          doi = {10.3847/1538-3881/abad40},
archivePrefix = {arXiv},
       eprint = {2008.02756},
 primaryClass = {astro-ph.EP},
       adsurl = {https://ui.adsabs.harvard.edu/abs/2020AJ....160..174K}
}

@ARTICLE{Jenkins_CMDrac_1996,
       author = {{Jenkins}, Jon M. and {Doyle}, Laurance R. and {Cullers}, D.~K.},
        title = "{A Matched Filter Method for Ground-Based Sub-Noise Detection of Terrestrial Extrasolar Planets in Eclipsing Binaries: Application to CM Draconis}",
      journal = {\icarus},
         year = 1996,
        month = feb,
       volume = {119},
       number = {2},
        pages = {244-260},
          doi = {10.1006/icar.1996.0018},
       adsurl = {https://ui.adsabs.harvard.edu/abs/1996Icar..119..244J}
}

@ARTICLE{Ofir_2008,
       author = {{Ofir}, A.},
        title = "{An algorithm for photometric identification of transiting circumbinary planets}",
      journal = {\mnras},
         year = 2008,
        month = jul,
       volume = {387},
       number = {4},
        pages = {1597-1604},
          doi = {10.1111/j.1365-2966.2008.13336.x},
archivePrefix = {arXiv},
       eprint = {0805.1116},
 primaryClass = {astro-ph},
       adsurl = {https://ui.adsabs.harvard.edu/abs/2008MNRAS.387.1597O}
}

@ARTICLE{Carter_2013,
       author = {{Carter}, Joshua A. and {Agol}, Eric},
        title = "{The Quasiperiodic Automated Transit Search Algorithm}",
      journal = {\apj},
         year = 2013,
        month = mar,
       volume = {765},
       number = {2},
          eid = {132},
        pages = {132},
          doi = {10.1088/0004-637X/765/2/132},
archivePrefix = {arXiv},
       eprint = {1210.5136},
 primaryClass = {astro-ph.EP},
       adsurl = {https://ui.adsabs.harvard.edu/abs/2013ApJ...765..132C}
}

@ARTICLE{Kostov_2013,
       author = {{Kostov}, V.~B. and {McCullough}, P.~R. and {Hinse}, T.~C. and {Tsvetanov}, Z.~I. and {H{\'e}brard}, G. and {D{\'\i}az}, R.~F. and {Deleuil}, M. and {Valenti}, J.~A.},
        title = "{A Gas Giant Circumbinary Planet Transiting the F Star Primary of the Eclipsing Binary Star KIC 4862625 and the Independent Discovery and Characterization of the Two Transiting Planets in the Kepler-47 System}",
      journal = {\apj},
         year = 2013,
        month = jun,
       volume = {770},
       number = {1},
          eid = {52},
        pages = {52},
          doi = {10.1088/0004-637X/770/1/52},
archivePrefix = {arXiv},
       eprint = {1210.3850},
 primaryClass = {astro-ph.EP},
       adsurl = {https://ui.adsabs.harvard.edu/abs/2013ApJ...770...52K}
}

@ARTICLE{Klagyivik_2017,
       author = {{Klagyivik}, P. and {Deeg}, H.~J. and {Cabrera}, J. and {Csizmadia}, Sz. and {Almenara}, J.~M.},
        title = "{Limits to the presence of transiting circumbinary planets in CoRoT Data}",
      journal = {\aap},
         year = 2017,
        month = jun,
       volume = {602},
          eid = {A117},
        pages = {A117},
          doi = {10.1051/0004-6361/201628244},
archivePrefix = {arXiv},
       eprint = {1610.08471},
 primaryClass = {astro-ph.EP},
       adsurl = {https://ui.adsabs.harvard.edu/abs/2017A&A...602A.117K}
}

@ARTICLE{Windemuth_2019_CBPs,
       author = {{Windemuth}, Diana and {Agol}, Eric and {Carter}, Josh and {Ford}, Eric B. and {Haghighipour}, Nader and {Orosz}, Jerome A. and {Welsh}, William F.},
        title = "{An automated method to detect transiting circumbinary planets}",
      journal = {\mnras},
         year = 2019,
        month = nov,
       volume = {490},
       number = {1},
        pages = {1313-1324},
          doi = {10.1093/mnras/stz2637},
archivePrefix = {arXiv},
       eprint = {1909.07443},
 primaryClass = {astro-ph.EP},
       adsurl = {https://ui.adsabs.harvard.edu/abs/2019MNRAS.490.1313W}
}

@ARTICLE{Martin_2021,
       author = {{Martin}, David V. and {Fabrycky}, Daniel C.},
        title = "{Searching for Small Circumbinary Planets. I. The STANLEY Automated Algorithm and No New Planets in Existing Systems}",
      journal = {\aj},
         year = 2021,
        month = sep,
       volume = {162},
       number = {3},
          eid = {84},
        pages = {84},
          doi = {10.3847/1538-3881/abeab1},
archivePrefix = {arXiv},
       eprint = {2101.03186},
 primaryClass = {astro-ph.EP},
       adsurl = {https://ui.adsabs.harvard.edu/abs/2021AJ....162...84M}
}

@ARTICLE{Davies2026,
       author = {{Davies}, Benjamin D.~R. and {Brown}, David J.~A. and {Gill}, Samuel and {French}, Jenni R.},
        title = "{Finding Circumbinary Planets: A Semi-Automated Transit Search of TESS Eclipsing Binaries}",
      journal = {arXiv e-prints},
         year = 2026,
        month = apr,
          eid = {arXiv:2604.09435},
        pages = {arXiv:2604.09435},
          doi = {10.48550/arXiv.2604.09435},
archivePrefix = {arXiv},
       eprint = {2604.09435},
 primaryClass = {astro-ph.EP},
       adsurl = {https://ui.adsabs.harvard.edu/abs/2026arXiv260409435D}
}

@ARTICLE{Deeg_2000,
       author = {{Deeg}, H.~J. and {Doyle}, L.~R. and {Kozhevnikov}, V.~P. and {Blue}, J.~E. and {Mart{\'\i}n}, E.~L. and {Schneider}, J.},
        title = "{A search for Jovian-mass planets around CM Draconis using eclipse minima timing.}",
      journal = {\aap},
         year = 2000,
        month = jun,
       volume = {358},
        pages = {L5-L8},
          doi = {10.48550/arXiv.astro-ph/0003391},
archivePrefix = {arXiv},
       eprint = {astro-ph/0003391},
 primaryClass = {astro-ph},
       adsurl = {https://ui.adsabs.harvard.edu/abs/2000A&A...358L...5D}
}

@ARTICLE{Esmer_2025,
       author = {{Murat Esmer}, Ekrem and {Daylan}, Tansu},
        title = "{Eclipse Timing Variations of Circumbinary Substellar Objects in TESS Data}",
      journal = {arXiv e-prints},
         year = 2025,
        month = mar,
          eid = {arXiv:2503.20900},
        pages = {arXiv:2503.20900},
          doi = {10.48550/arXiv.2503.20900},
archivePrefix = {arXiv},
       eprint = {2503.20900},
 primaryClass = {astro-ph.EP},
       adsurl = {https://ui.adsabs.harvard.edu/abs/2025arXiv250320900M}
}

@ARTICLE{Borkovits_2016,
       author = {{Borkovits}, T. and {Hajdu}, T. and {Sztakovics}, J. and {Rappaport}, S. and {Levine}, A. and {B{\'\i}r{\'o}}, I.~B. and {Klagyivik}, P.},
        title = "{A comprehensive study of the Kepler triples via eclipse timing}",
      journal = {\mnras},
         year = 2016,
        month = feb,
       volume = {455},
       number = {4},
        pages = {4136-4165},
          doi = {10.1093/mnras/stv2530},
archivePrefix = {arXiv},
       eprint = {1510.08272},
 primaryClass = {astro-ph.SR},
       adsurl = {https://ui.adsabs.harvard.edu/abs/2016MNRAS.455.4136B}
}

@ARTICLE{Kalomeni_2025,
       author = {{Kalomeni}, Belinda and {Yakut}, Kadri},
        title = "{Comprehensive analysis of CM Draconis: eclipse timing variations driven by either a third body or stellar magnetic activity}",
      journal = {\mnras},
         year = 2025,
        month = aug,
       volume = {541},
       number = {3},
        pages = {2553-2565},
          doi = {10.1093/mnras/staf1166},
archivePrefix = {arXiv},
       eprint = {2507.17035},
 primaryClass = {astro-ph.SR},
       adsurl = {https://ui.adsabs.harvard.edu/abs/2025MNRAS.541.2553K}
}

@ARTICLE{Goldberg_2023,
       author = {{Goldberg}, Max and {Fabrycky}, Daniel and {Martin}, David V. and {Albrecht}, Simon and {Deeg}, Hans J. and {Nowak}, Grzegorz},
        title = "{A 5M$_{Jup}$ non-transiting coplanar circumbinary planet around Kepler-1660AB}",
      journal = {\mnras},
         year = 2023,
        month = nov,
       volume = {525},
       number = {3},
        pages = {4628-4641},
          doi = {10.1093/mnras/stad2568},
archivePrefix = {arXiv},
       eprint = {2308.09255},
 primaryClass = {astro-ph.EP},
       adsurl = {https://ui.adsabs.harvard.edu/abs/2023MNRAS.525.4628G}
}

@ARTICLE{Zorotovic_2013,
       author = {{Zorotovic}, M. and {Schreiber}, M.~R.},
        title = "{Origin of apparent period variations in eclipsing post-common-envelope binaries}",
      journal = {\aap},
         year = 2013,
        month = jan,
       volume = {549},
          eid = {A95},
        pages = {A95},
          doi = {10.1051/0004-6361/201220321},
archivePrefix = {arXiv},
       eprint = {1211.5356},
 primaryClass = {astro-ph.SR},
       adsurl = {https://ui.adsabs.harvard.edu/abs/2013A&A...549A..95Z}
}

@ARTICLE{Thornton2026,
       author = {{Thornton}, Margo and {Montet}, Benjamin T. and {White}, Riley and {Shao}, Arden and {Kumar}, Diya T.},
        title = "{Detection of 27 candidate circumbinary planets through apsidal precession of eclipsing binaries observed by TESS}",
      journal = {\mnras},
         year = 2026,
        month = may,
       volume = {548},
       number = {3},
          eid = {stag515},
        pages = {stag515},
          doi = {10.1093/mnras/stag515},
archivePrefix = {arXiv},
       eprint = {2512.07934},
 primaryClass = {astro-ph.EP},
       adsurl = {https://ui.adsabs.harvard.edu/abs/2026MNRAS.548ag515T}
}

@ARTICLE{Baycroft_2025_Polar,
       author = {{Baycroft}, Thomas A. and {Sairam}, Lalitha and {Triaud}, Amaury H.~M.~J. and {Correia}, Alexandre C.~M.},
        title = "{Evidence for a polar circumbinary exoplanet orbiting a pair of eclipsing brown dwarfs}",
      journal = {Science Advances},
         year = 2025,
        month = apr,
       volume = {11},
       number = {16},
          eid = {eadu0627},
        pages = {eadu0627},
          doi = {10.1126/sciadv.adu0627},
archivePrefix = {arXiv},
       eprint = {2504.12209},
 primaryClass = {astro-ph.EP},
       adsurl = {https://ui.adsabs.harvard.edu/abs/2025SciA...11..627B}
}

@ARTICLE{Smallwood_2025,
       author = {{Smallwood}, Jeremy L. and {Li}, Ya-Ping and {Deng}, Hongping and {Franchini}, Alessia},
        title = "{Circumbinary accretion as a diagnostic for binary-disc misalignment}",
      journal = {\mnras},
         year = 2025,
        month = feb,
       volume = {536},
       number = {4},
        pages = {3431-3444},
          doi = {10.1093/mnras/stae2768},
archivePrefix = {arXiv},
       eprint = {2412.10653},
 primaryClass = {astro-ph.EP},
       adsurl = {https://ui.adsabs.harvard.edu/abs/2025MNRAS.536.3431S}
}

@ARTICLE{Childs_2022,
       author = {{Childs}, Anna C. and {Martin}, Rebecca G.},
        title = "{Misalignment of Terrestrial Circumbinary Planets as an Indicator of Their Formation Mechanism}",
      journal = {\apjl},
         year = 2022,
        month = mar,
       volume = {927},
       number = {1},
          eid = {L7},
        pages = {L7},
          doi = {10.3847/2041-8213/ac574f},
archivePrefix = {arXiv},
       eprint = {2202.10495},
 primaryClass = {astro-ph.EP},
       adsurl = {https://ui.adsabs.harvard.edu/abs/2022ApJ...927L...7C}
}

@ARTICLE{RMartin_2023,
       author = {{Martin}, Rebecca G. and {Lubow}, Stephen H. and {Vallet}, David and {Anugu}, Narsireddy and {Gies}, Douglas R.},
        title = "{AC Her: Evidence of the First Polar Circumbinary Planet}",
      journal = {\apjl},
         year = 2023,
        month = nov,
       volume = {957},
       number = {2},
          eid = {L28},
        pages = {L28},
          doi = {10.3847/2041-8213/ad0730},
archivePrefix = {arXiv},
       eprint = {2310.19600},
 primaryClass = {astro-ph.EP},
       adsurl = {https://ui.adsabs.harvard.edu/abs/2023ApJ...957L..28M}
}

@ARTICLE{Chen_2022,
       author = {{Chen}, Zirui and {Kipping}, David},
        title = "{The number of transits per epoch for transiting misaligned circumbinary planets}",
      journal = {\mnras},
         year = 2022,
        month = jul,
       volume = {513},
       number = {4},
        pages = {5162-5173},
          doi = {10.1093/mnras/stac1246},
archivePrefix = {arXiv},
       eprint = {2112.00966},
 primaryClass = {astro-ph.EP},
       adsurl = {https://ui.adsabs.harvard.edu/abs/2022MNRAS.513.5162C}
}

@ARTICLE{Oddo_2026,
       author = {{Oddo}, Dominic and {Dragomir}, Diana and {Powell}, Brian P. and {Kostov}, Veselin B. and {Han}, Te},
        title = "{A Catalog of M\&M Eclipsing Binaries with TESS}",
      journal = {\apj},
         year = 2026,
        month = jan,
       volume = {996},
       number = {1},
          eid = {82},
        pages = {82},
          doi = {10.3847/1538-4357/ae0c0f},
archivePrefix = {arXiv},
       eprint = {2508.13941},
 primaryClass = {astro-ph.SR},
       adsurl = {https://ui.adsabs.harvard.edu/abs/2026ApJ...996...82O}
}

@ARTICLE{Rein_2012,
       author = {{Rein}, H. and {Liu}, S.-F.},
        title = "{REBOUND: an open-source multi-purpose N-body code for collisional dynamics}",
      journal = {\aap},
         year = 2012,
        month = jan,
       volume = {537},
          eid = {A128},
        pages = {A128},
          doi = {10.1051/0004-6361/201118085},
archivePrefix = {arXiv},
       eprint = {1110.4876},
 primaryClass = {astro-ph.EP},
       adsurl = {https://ui.adsabs.harvard.edu/abs/2012A&A...537A.128R}
}

@BOOK{Prsa_2018,
       author = {{Pr{\v{s}}a}, Andrej},
        title = "{Modeling and Analysis of Eclipsing Binary Stars; The theory and design principles of PHOEBE}",
        publisher="{IOP Publishing}",
         year = 2018,
          doi = {10.1088/978-0-7503-1287-5},
       adsurl = {https://ui.adsabs.harvard.edu/abs/2018maeb.book.....P}
}

@ARTICLE{Hippke_2019_Wotan,
       author = {{Hippke}, Michael and {David}, Trevor J. and {Mulders}, Gijs D. and
         {Heller}, Ren{\'e}},
        title = "{W{\={o}}tan: Comprehensive Time-series Detrending in Python}",
      journal = {\aj},
         year = "2019",
        month = "Oct",
       volume = {158},
       number = {4},
          eid = {143},
        pages = {143},
          doi = {10.3847/1538-3881/ab3984},
archivePrefix = {arXiv},
       eprint = {1906.00966},
 primaryClass = {astro-ph.EP},
       adsurl = {https://ui.adsabs.harvard.edu/abs/2019AJ....158..143H}
}

@ARTICLE{Kipping_2013,
       author = {{Kipping}, D.~M. and {Hartman}, J. and {Buchhave}, L.~A. and {Schmitt}, A.~R. and {Bakos}, G. {\'A}. and {Nesvorn{\'y}}, D.},
        title = "{The Hunt for Exomoons with Kepler (HEK). II. Analysis of Seven Viable Satellite-hosting Planet Candidates}",
      journal = {\apj},
         year = 2013,
        month = jun,
       volume = {770},
       number = {2},
          eid = {101},
        pages = {101},
          doi = {10.1088/0004-637X/770/2/101},
archivePrefix = {arXiv},
       eprint = {1301.1853},
 primaryClass = {astro-ph.EP},
       adsurl = {https://ui.adsabs.harvard.edu/abs/2013ApJ...770..101K}
}

@ARTICLE{Rodenbeck_2018,
       author = {{Rodenbeck}, Kai and {Heller}, Ren{\'e} and {Hippke}, Michael and {Gizon}, Laurent},
        title = "{Revisiting the exomoon candidate signal around Kepler-1625 b}",
      journal = {\aap},
         year = 2018,
        month = sep,
       volume = {617},
          eid = {A49},
        pages = {A49},
          doi = {10.1051/0004-6361/201833085},
archivePrefix = {arXiv},
       eprint = {1806.04672},
 primaryClass = {astro-ph.EP},
       adsurl = {https://ui.adsabs.harvard.edu/abs/2018A&A...617A..49R}
}

@ARTICLE{Lomb_1976,
       author = {{Lomb}, N.~R.},
        title = "{Least-Squares Frequency Analysis of Unequally Spaced Data}",
      journal = {\apss},
         year = 1976,
        month = feb,
       volume = {39},
       number = {2},
        pages = {447-462},
          doi = {10.1007/BF00648343},
       adsurl = {https://ui.adsabs.harvard.edu/abs/1976Ap&SS..39..447L}
}

@ARTICLE{Scargle_1982,
       author = {{Scargle}, J.~D.},
        title = "{Studies in astronomical time series analysis. II. Statistical aspects of spectral analysis of unevenly spaced data.}",
      journal = {\apj},
         year = 1982,
        month = dec,
       volume = {263},
        pages = {835-853},
          doi = {10.1086/160554},
       adsurl = {https://ui.adsabs.harvard.edu/abs/1982ApJ...263..835S}
}

@book{mosteller1977data,
  title={Data Analysis and Regression: A Second Course in Statistics},
  author={Mosteller, F. and Tukey, J.W.},
  isbn={9780201048544},
  lccn={lc76015465},
  series={Addison-Wesley series in behavioral science},
  url={https://books.google.com/books?id=pGlHAAAAMAAJ},
  year={1977},
  publisher={Addison-Wesley Publishing Company}
}

@article{Christiansen_2012,
	doi = {10.1086/668847},
	url = {https://doi.org/10.1086/668847},
	year = 2012,
	month = {dec},
	publisher = {{IOP} Publishing},
	volume = {124},
	number = {922},
	pages = {1279--1287},
	author = {Jessie L. Christiansen and Jon M. Jenkins and Douglas A. Caldwell and Christopher J. Burke and Peter Tenenbaum and Shawn Seader and Susan E. Thompson and Thomas S. Barclay and Bruce D. Clarke and Jie Li and Jeffrey C. Smith and Martin C. Stumpe and Joseph D. Twicken and Jeffrey Van Cleve},
	title = {The Derivation, Properties, and Value of Kepler's Combined Differential Photometric Precision},
	journal = {Publications of the Astronomical Society of the Pacific}
}

@ARTICLE{Villanueva_2019,
       author = {{Villanueva}, Jr., Steven and {Dragomir}, Diana and {Gaudi}, B. Scott},
        title = "{An Estimate of the Yield of Single-transit Planetary Events from the Transiting Exoplanet Survey Satellite}",
      journal = {\aj},
         year = 2019,
        month = feb,
       volume = {157},
       number = {2},
          eid = {84},
        pages = {84},
          doi = {10.3847/1538-3881/aaf85e},
archivePrefix = {arXiv},
       eprint = {1805.00956},
 primaryClass = {astro-ph.EP},
       adsurl = {https://ui.adsabs.harvard.edu/abs/2019AJ....157...84V}
}

@ARTICLE{Dressing_2015,
       author = {{Dressing}, Courtney D. and {Charbonneau}, David},
        title = "{The Occurrence of Potentially Habitable Planets Orbiting M Dwarfs Estimated from the Full Kepler Dataset and an Empirical Measurement of the Detection Sensitivity}",
      journal = {\apj},
         year = 2015,
        month = jul,
       volume = {807},
       number = {1},
          eid = {45},
        pages = {45},
          doi = {10.1088/0004-637X/807/1/45},
archivePrefix = {arXiv},
       eprint = {1501.01623},
 primaryClass = {astro-ph.EP},
       adsurl = {https://ui.adsabs.harvard.edu/abs/2015ApJ...807...45D}
}

@ARTICLE{Kunimoto_2025,
       author = {{Kunimoto}, Michelle and {Bryson}, Steve and {Jaffee}, Drayson and {Rowe}, Jason F. and {Daylan}, Tansu and {Giacalone}, Steven and {Lissauer}, Jack J. and {Matesic}, Michael R.~B. and {Mullally}, Susan E. and {Eschen}, Yoshi Nike Emilia},
        title = "{LEO-Vetter: Fully Automated Flux- and Pixel-level Vetting of TESS Planet Candidates to Support Occurrence Rates}",
      journal = {\aj},
         year = 2025,
        month = nov,
       volume = {170},
       number = {5},
          eid = {280},
        pages = {280},
          doi = {10.3847/1538-3881/ae070a},
archivePrefix = {arXiv},
       eprint = {2509.10619},
 primaryClass = {astro-ph.EP},
       adsurl = {https://ui.adsabs.harvard.edu/abs/2025AJ....170..280K}
}

@ARTICLE{Kreidberg2015batman,
       author = {{Kreidberg}, Laura},
        title = "{batman: BAsic Transit Model cAlculatioN in Python}",
      journal = {\pasp},
         year = 2015,
        month = nov,
       volume = {127},
       number = {957},
        pages = {1161},
          doi = {10.1086/683602},
archivePrefix = {arXiv},
       eprint = {1507.08285},
 primaryClass = {astro-ph.EP},
       adsurl = {https://ui.adsabs.harvard.edu/abs/2015PASP..127.1161K}
}

@ARTICLE{Mandel_2002,
       author = {{Mandel}, Kaisey and {Agol}, Eric},
        title = "{Analytic Light Curves for Planetary Transit Searches}",
      journal = {\apjl},
         year = 2002,
        month = dec,
       volume = {580},
       number = {2},
        pages = {L171-L175},
          doi = {10.1086/345520},
archivePrefix = {arXiv},
       eprint = {astro-ph/0210099},
 primaryClass = {astro-ph},
       adsurl = {https://ui.adsabs.harvard.edu/abs/2002ApJ...580L.171M}
}

@ARTICLE{Martin_2019,
       author = {{Martin}, David V.},
        title = "{The binary mass ratios of circumbinary planet hosts}",
      journal = {\mnras},
         year = 2019,
        month = sep,
       volume = {488},
       number = {3},
        pages = {3482-3491},
          doi = {10.1093/mnras/stz959},
archivePrefix = {arXiv},
       eprint = {1904.04832},
 primaryClass = {astro-ph.EP},
       adsurl = {https://ui.adsabs.harvard.edu/abs/2019MNRAS.488.3482M}
}

@ARTICLE{Oddo_2023,
       author = {{Oddo}, Dominic and {Dragomir}, Diana and {Brandeker}, Alexis and {Osborn}, Hugh P. and {Collins}, Karen and {Stassun}, Keivan G. and {Astudillo-Defru}, Nicola and {Bieryla}, Allyson and {Howell}, Steve B. and {Ciardi}, David R. and {Quinn}, Samuel and {Almenara}, Jose M. and {Brice{\~n}o}, C{\'e}sar and {Collins}, Kevin I. and {Col{\'o}n}, Knicole D. and {Conti}, Dennis M. and {Crouzet}, Nicolas and {Furlan}, Elise and {Gan}, Tianjun and {Gnilka}, Crystal L. and {Goeke}, Robert F. and {Gonzales}, Erica and {Harris}, Mallory and {Jenkins}, Jon M. and {Jensen}, Eric L.~N. and {Latham}, David and {Law}, Nicholas and {Lund}, Michael B. and {Mann}, Andrew W. and {Massey}, Bob and {Murgas}, Felipe and {Ricker}, George and {Relles}, Howard M. and {Rowden}, Pamela and {Schwarz}, Richard P. and {Schlieder}, Joshua and {Shporer}, Avi and {Seager}, Sara and {Srdoc}, Gregor and {Torres}, Guillermo and {Twicken}, Joseph D. and {Vanderspek}, Roland and {Winn}, Joshua N. and {Ziegler}, Carl},
        title = "{Characterization of a Set of Small Planets with TESS and CHEOPS and an Analysis of Photometric Performance}",
      journal = {\aj},
         year = 2023,
        month = mar,
       volume = {165},
       number = {3},
          eid = {134},
        pages = {134},
          doi = {10.3847/1538-3881/acb4e3},
archivePrefix = {arXiv},
       eprint = {2301.08162},
 primaryClass = {astro-ph.EP},
       adsurl = {https://ui.adsabs.harvard.edu/abs/2023AJ....165..134O}
}

@article{Foreman_Mackey_2013,
	doi = {10.1086/670067},
	url = {https://doi.org/10.1086/670067},
	year = 2013,
	month = {mar},
	publisher = {{IOP} Publishing},
	volume = {125},
	number = {925},
	pages = {306--312},
	author = {Daniel Foreman-Mackey and David W. Hogg and Dustin Lang and Jonathan Goodman},
	title = {$\less$tt$\greater$emcee$\less$/tt$\greater$: The {MCMC} Hammer},
	journal = {Publications of the Astronomical Society of the Pacific}
}

@ARTICLE{Holman_1999,
       author = {{Holman}, Matthew J. and {Wiegert}, Paul A.},
        title = "{Long-Term Stability of Planets in Binary Systems}",
      journal = {\aj},
         year = 1999,
        month = jan,
       volume = {117},
       number = {1},
        pages = {621-628},
          doi = {10.1086/300695},
archivePrefix = {arXiv},
       eprint = {astro-ph/9809315},
 primaryClass = {astro-ph},
       adsurl = {https://ui.adsabs.harvard.edu/abs/1999AJ....117..621H}
}

@ARTICLE{Belokurov_2020,
       author = {{Belokurov}, Vasily and {Penoyre}, Zephyr and {Oh}, Semyeong and {Iorio}, Giuliano and {Hodgkin}, Simon and {Evans}, N. Wyn and {Everall}, Andrew and {Koposov}, Sergey E. and {Tout}, Christopher A. and {Izzard}, Robert and {Clarke}, Cathie J. and {Brown}, Anthony G.~A.},
        title = "{Unresolved stellar companions with Gaia DR2 astrometry}",
      journal = {\mnras},
         year = 2020,
        month = aug,
       volume = {496},
       number = {2},
        pages = {1922-1940},
          doi = {10.1093/mnras/staa1522},
archivePrefix = {arXiv},
       eprint = {2003.05467},
 primaryClass = {astro-ph.SR},
       adsurl = {https://ui.adsabs.harvard.edu/abs/2020MNRAS.496.1922B}
}

@ARTICLE{Heidari_2025,
       author = {{Heidari}, N. and {H{\'e}brard}, G. and {Martioli}, E. and {Eastman}, J.~D. and {Jackson}, J.~M. and {Delfosse}, X. and {Jord{\'a}n}, A. and {Correia}, A.~C.~M. and {Sousa}, S. and {Dragomir}, D. and {Forveille}, T. and {Boisse}, I. and {Giacalone}, S.~A. and {D{\'\i}az}, R.~F. and {Brahm}, R. and {Almasian}, D. and {Almenara}, J.~M. and {Bieryla}, A. and {Barkaoui}, K. and {Baker}, D. and {Barros}, S.~C.~C. and {Bonfils}, X. and {Carmona}, A. and {Collins}, K.~A. and {Cort{\'e}s-Zuleta}, P. and {Deleuil}, M. and {Demangeon}, O.~D.~S. and {Edwards}, B. and {Eberhardt}, J. and {Espinoza}, N. and {Eisner}, N. and {Feliz}, D.~L. and {Frommer}, A.~C. and {Fukui}, A. and {Grau}, F. and {Gupta}, A.~F. and {Hara}, N. and {Hobson}, M.~J. and {Henning}, T. and {Howell}, S.~B. and {Jenkins}, J.~M. and {Kiefer}, F. and {LaCourse}, D.~M. and {Laskar}, J. and {Law}, N. and {Mann}, A.~W. and {Murgas}, F. and {Moutou}, C. and {Narita}, N. and {Palle}, E. and {Relles}, H.~M. and {Stassun}, K.~G. and {Serrano Bell}, J. and {Schwarz}, R.~P. and {Srdoc}, G. and {Str{\o}m}, P.~A. and {Safonov}, B. and {Sarkis}, P. and {Schlecker}, M. and {Tala Pinto}, M. and {Pepper}, J. and {Rojas}, F.~I. and {Twicken}, J.~D. and {Trifonov}, T. and {Villanueva}, S. and {Watkins}, C.~N. and {Winn}, J.~N. and {Ziegler}, C.},
        title = "{Characterization of seven transiting systems, including four warm Jupiters from SOPHIE and TESS}",
      journal = {\aap},
         year = 2025,
        month = feb,
       volume = {694},
          eid = {A36},
        pages = {A36},
          doi = {10.1051/0004-6361/202451519},
archivePrefix = {arXiv},
       eprint = {2412.08527},
 primaryClass = {astro-ph.EP},
       adsurl = {https://ui.adsabs.harvard.edu/abs/2025A&A...694A..36H}
}

@ARTICLE{Bonomo_2017,
       author = {{Bonomo}, A.~S. and {Desidera}, S. and {Benatti}, S. and {Borsa}, F. and {Crespi}, S. and {Damasso}, M. and {Lanza}, A.~F. and {Sozzetti}, A. and {Lodato}, G. and {Marzari}, F. and {Boccato}, C. and {Claudi}, R.~U. and {Cosentino}, R. and {Covino}, E. and {Gratton}, R. and {Maggio}, A. and {Micela}, G. and {Molinari}, E. and {Pagano}, I. and {Piotto}, G. and {Poretti}, E. and {Smareglia}, R. and {Affer}, L. and {Biazzo}, K. and {Bignamini}, A. and {Esposito}, M. and {Giacobbe}, P. and {H{\'e}brard}, G. and {Malavolta}, L. and {Maldonado}, J. and {Mancini}, L. and {Martinez Fiorenzano}, A. and {Masiero}, S. and {Nascimbeni}, V. and {Pedani}, M. and {Rainer}, M. and {Scandariato}, G.},
        title = "{The GAPS Programme with HARPS-N at TNG . XIV. Investigating giant planet migration history via improved eccentricity and mass determination for 231 transiting planets}",
      journal = {\aap},
         year = 2017,
        month = jun,
       volume = {602},
          eid = {A107},
        pages = {A107},
          doi = {10.1051/0004-6361/201629882},
archivePrefix = {arXiv},
       eprint = {1704.00373},
 primaryClass = {astro-ph.EP},
       adsurl = {https://ui.adsabs.harvard.edu/abs/2017A&A...602A.107B}
}

@PHDTHESIS{Baycroft_2025,
       author = {{Baycroft}, Thomas},
        title = "{Tools to Detect and Characterise Circumbinary Exoplanets}",
       school = {University of Birmingham, School of Physics and Astronomy},
         year = 2025,
        month = jul,
       adsurl = {https://ui.adsabs.harvard.edu/abs/2025PhDT.........9B}
}

@ARTICLE{Quarles_2018,
       author = {{Quarles}, B. and {Satyal}, S. and {Kostov}, V. and {Kaib}, N. and {Haghighipour}, N.},
        title = "{Stability Limits of Circumbinary Planets: Is There a Pile-up in the Kepler CBPs?}",
      journal = {\apj},
         year = 2018,
        month = apr,
       volume = {856},
       number = {2},
          eid = {150},
        pages = {150},
          doi = {10.3847/1538-4357/aab264},
archivePrefix = {arXiv},
       eprint = {1802.08868},
 primaryClass = {astro-ph.EP},
       adsurl = {https://ui.adsabs.harvard.edu/abs/2018ApJ...856..150Q}
}

@ARTICLE{Howard_2012,
       author = {{Howard}, Andrew W. and {Marcy}, Geoffrey W. and {Bryson}, Stephen T. and {Jenkins}, Jon M. and {Rowe}, Jason F. and {Batalha}, Natalie M. and {Borucki}, William J. and {Koch}, David G. and {Dunham}, Edward W. and {Gautier}, III, Thomas N. and {Van Cleve}, Jeffrey and {Cochran}, William D. and {Latham}, David W. and {Lissauer}, Jack J. and {Torres}, Guillermo and {Brown}, Timothy M. and {Gilliland}, Ronald L. and {Buchhave}, Lars A. and {Caldwell}, Douglas A. and {Christensen-Dalsgaard}, J{\o}rgen and {Ciardi}, David and {Fressin}, Francois and {Haas}, Michael R. and {Howell}, Steve B. and {Kjeldsen}, Hans and {Seager}, Sara and {Rogers}, Leslie and {Sasselov}, Dimitar D. and {Steffen}, Jason H. and {Basri}, Gibor S. and {Charbonneau}, David and {Christiansen}, Jessie and {Clarke}, Bruce and {Dupree}, Andrea and {Fabrycky}, Daniel C. and {Fischer}, Debra A. and {Ford}, Eric B. and {Fortney}, Jonathan J. and {Tarter}, Jill and {Girouard}, Forrest R. and {Holman}, Matthew J. and {Johnson}, John Asher and {Klaus}, Todd C. and {Machalek}, Pavel and {Moorhead}, Althea V. and {Morehead}, Robert C. and {Ragozzine}, Darin and {Tenenbaum}, Peter and {Twicken}, Joseph D. and {Quinn}, Samuel N. and {Isaacson}, Howard and {Shporer}, Avi and {Lucas}, Philip W. and {Walkowicz}, Lucianne M. and {Welsh}, William F. and {Boss}, Alan and {Devore}, Edna and {Gould}, Alan and {Smith}, Jeffrey C. and {Morris}, Robert L. and {Prsa}, Andrej and {Morton}, Timothy D. and {Still}, Martin and {Thompson}, Susan E. and {Mullally}, Fergal and {Endl}, Michael and {MacQueen}, Phillip J.},
        title = "{Planet Occurrence within 0.25 AU of Solar-type Stars from Kepler}",
      journal = {\apjs},
         year = 2012,
        month = aug,
       volume = {201},
       number = {2},
          eid = {15},
        pages = {15},
          doi = {10.1088/0067-0049/201/2/15},
archivePrefix = {arXiv},
       eprint = {1103.2541},
 primaryClass = {astro-ph.EP},
       adsurl = {https://ui.adsabs.harvard.edu/abs/2012ApJS..201...15H}
}

@ARTICLE{Sayeed_2025,
       author = {{Sayeed}, Maryum and {Angus}, Ruth and {Berger}, Travis A. and {Lu}, Yuxi(Lucy) and {Christiansen}, Jessie L. and {Foreman-Mackey}, Daniel and {Ness}, Melissa K.},
        title = "{Exoplanet Occurrence Rate with Age for FGK Stars in Kepler}",
      journal = {\aj},
         year = 2025,
        month = feb,
       volume = {169},
       number = {2},
          eid = {112},
        pages = {112},
          doi = {10.3847/1538-3881/ada8a1},
archivePrefix = {arXiv},
       eprint = {2501.13809},
 primaryClass = {astro-ph.EP},
       adsurl = {https://ui.adsabs.harvard.edu/abs/2025AJ....169..112S}
}

@phdthesis{Windemuth_2020,
  author  = {{Windemuth}, Diana},
  title   = "{Characterizing Eclipsing Binaries and the Population of Planets Orbiting Around Them}",
  school  = "{University of Washington, College of Arts and Sciences}",
  year    = 2020
}

@ARTICLE{Georgakarakos_2024,
       author = {{Georgakarakos}, Nikolaos and {Eggl}, Siegfried and {Ali-Dib}, Mohamad and {Dobbs-Dixon}, Ian},
        title = "{Empirical Stability Criteria for 3D Hierarchical Triple Systems. I. Circumbinary Planets}",
      journal = {\aj},
         year = 2024,
        month = nov,
       volume = {168},
       number = {5},
          eid = {224},
        pages = {224},
          doi = {10.3847/1538-3881/ad7a78},
archivePrefix = {arXiv},
       eprint = {2404.13746},
 primaryClass = {astro-ph.EP},
       adsurl = {https://ui.adsabs.harvard.edu/abs/2024AJ....168..224G}
}

@ARTICLE{Fabrycky_2007,
       author = {{Fabrycky}, Daniel and {Tremaine}, Scott},
        title = "{Shrinking Binary and Planetary Orbits by Kozai Cycles with Tidal Friction}",
      journal = {\apj},
         year = 2007,
        month = nov,
       volume = {669},
       number = {2},
        pages = {1298-1315},
          doi = {10.1086/521702},
archivePrefix = {arXiv},
       eprint = {0705.4285},
 primaryClass = {astro-ph},
       adsurl = {https://ui.adsabs.harvard.edu/abs/2007ApJ...669.1298F}
}

@ARTICLE{Mogan_2025,
       author = {{Mogan}, Saahit and {Zanazzi}, J.~J.},
        title = "{Concealing Circumbinary Planets with Tidal Shrinkage}",
      journal = {\apj},
         year = 2025,
        month = aug,
       volume = {988},
       number = {2},
          eid = {150},
        pages = {150},
          doi = {10.3847/1538-4357/ade2e8},
archivePrefix = {arXiv},
       eprint = {2412.01616},
 primaryClass = {astro-ph.EP},
       adsurl = {https://ui.adsabs.harvard.edu/abs/2025ApJ...988..150M}
}

@ARTICLE{Farhat_2025,
       author = {{Farhat}, Mohammad and {Touma}, Jihad},
        title = "{Capture into Apsidal Resonance and the Decimation of Planets around Inspiraling Binaries}",
      journal = {\apjl},
         year = 2025,
        month = dec,
       volume = {995},
       number = {1},
          eid = {L23},
        pages = {L23},
          doi = {10.3847/2041-8213/ae21d8},
archivePrefix = {arXiv},
       eprint = {2512.03940},
 primaryClass = {astro-ph.EP},
       adsurl = {https://ui.adsabs.harvard.edu/abs/2025ApJ...995L..23F}
}

@ARTICLE{He_2020,
       author = {{He}, Matthias Y. and {Ford}, Eric B. and {Ragozzine}, Darin and {Carrera}, Daniel},
        title = "{Architectures of Exoplanetary Systems. III. Eccentricity and Mutual Inclination Distributions of AMD-stable Planetary Systems}",
      journal = {\aj},
         year = 2020,
        month = dec,
       volume = {160},
       number = {6},
          eid = {276},
        pages = {276},
          doi = {10.3847/1538-3881/abba18},
archivePrefix = {arXiv},
       eprint = {2007.14473},
 primaryClass = {astro-ph.EP},
       adsurl = {https://ui.adsabs.harvard.edu/abs/2020AJ....160..276H}
}

@ARTICLE{Masuda_2020,
       author = {{Masuda}, Kento and {Winn}, Joshua N. and {Kawahara}, Hajime},
        title = "{Mutual Orbital Inclinations between Cold Jupiters and Inner Super-Earths}",
      journal = {\aj},
         year = 2020,
        month = feb,
       volume = {159},
       number = {2},
          eid = {38},
        pages = {38},
          doi = {10.3847/1538-3881/ab5c1d},
archivePrefix = {arXiv},
       eprint = {1912.00387},
 primaryClass = {astro-ph.EP},
       adsurl = {https://ui.adsabs.harvard.edu/abs/2020AJ....159...38M}
}

@ARTICLE{Gehrels_1986,
       author = {{Gehrels}, N.},
        title = "{Confidence Limits for Small Numbers of Events in Astrophysical Data}",
      journal = {\apj},
         year = 1986,
        month = apr,
       volume = {303},
        pages = {336},
          doi = {10.1086/164079},
       adsurl = {https://ui.adsabs.harvard.edu/abs/1986ApJ...303..336G}
}

@ARTICLE{Czekala2019,
       author = {{Czekala}, Ian and {Chiang}, Eugene and {Andrews}, Sean M. and {Jensen}, Eric L.~N. and {Torres}, Guillermo and {Wilner}, David J. and {Stassun}, Keivan G. and {Macintosh}, Bruce},
        title = "{The Degree of Alignment between Circumbinary Disks and Their Binary Hosts}",
      journal = {\apj},
         year = 2019,
        month = sep,
       volume = {883},
       number = {1},
          eid = {22},
        pages = {22},
          doi = {10.3847/1538-4357/ab287b},
archivePrefix = {arXiv},
       eprint = {1906.03269},
 primaryClass = {astro-ph.EP},
       adsurl = {https://ui.adsabs.harvard.edu/abs/2019ApJ...883...22C}
}

@ARTICLE{Czekala2021,
       author = {{Czekala}, Ian and {Ribas}, {\'A}lvaro and {Cuello}, Nicol{\'a}s and {Chiang}, Eugene and {Mac{\'\i}as}, Enrique and {Duch{\^e}ne}, Gaspard and {Andrews}, Sean M. and {Espaillat}, Catherine C.},
        title = "{A Coplanar Circumbinary Protoplanetary Disk in the TWA 3 Triple M Dwarf System}",
      journal = {\apj},
         year = 2021,
        month = may,
       volume = {912},
       number = {1},
          eid = {6},
        pages = {6},
          doi = {10.3847/1538-4357/abebe3},
archivePrefix = {arXiv},
       eprint = {2102.11875},
 primaryClass = {astro-ph.EP},
       adsurl = {https://ui.adsabs.harvard.edu/abs/2021ApJ...912....6C}
}

@ARTICLE{Welsh_2012,
       author = {{Welsh}, William F. and {Orosz}, Jerome A. and {Carter}, Joshua A. and {Fabrycky}, Daniel C. and {Ford}, Eric B. and {Lissauer}, Jack J. and {Pr{\v{s}}a}, Andrej and {Quinn}, Samuel N. and {Ragozzine}, Darin and {Short}, Donald R. and {Torres}, Guillermo and {Winn}, Joshua N. and {Doyle}, Laurance R. and {Barclay}, Thomas and {Batalha}, Natalie and {Bloemen}, Steven and {Brugamyer}, Erik and {Buchhave}, Lars A. and {Caldwell}, Caroline and {Caldwell}, Douglas A. and {Christiansen}, Jessie L. and {Ciardi}, David R. and {Cochran}, William D. and {Endl}, Michael and {Fortney}, Jonathan J. and {Gautier}, III, Thomas N. and {Gilliland}, Ronald L. and {Haas}, Michael R. and {Hall}, Jennifer R. and {Holman}, Matthew J. and {Howard}, Andrew W. and {Howell}, Steve B. and {Isaacson}, Howard and {Jenkins}, Jon M. and {Klaus}, Todd C. and {Latham}, David W. and {Li}, Jie and {Marcy}, Geoffrey W. and {Mazeh}, Tsevi and {Quintana}, Elisa V. and {Robertson}, Paul and {Shporer}, Avi and {Steffen}, Jason H. and {Windmiller}, Gur and {Koch}, David G. and {Borucki}, William J.},
        title = "{Transiting circumbinary planets Kepler-34 b and Kepler-35 b}",
      journal = {\nat},
         year = 2012,
        month = jan,
       volume = {481},
       number = {7382},
        pages = {475-479},
          doi = {10.1038/nature10768},
archivePrefix = {arXiv},
       eprint = {1204.3955},
 primaryClass = {astro-ph.EP},
       adsurl = {https://ui.adsabs.harvard.edu/abs/2012Natur.481..475W}
}

@ARTICLE{Martin_2019_BEBOP,
       author = {{Martin}, David V. and {Triaud}, Amaury H.~M.~J. and {Udry}, St{\'e}phane and {Marmier}, Maxime and {Maxted}, Pierre F.~L. and {Collier Cameron}, Andrew and {Hellier}, Coel and {Pepe}, Francesco and {Pollacco}, Don and {S{\'e}gransan}, Damien and {West}, Richard},
        title = "{The BEBOP radial-velocity survey for circumbinary planets. I. Eight years of CORALIE observations of 47 single-line eclipsing binaries and abundance constraints on the masses of circumbinary planets}",
      journal = {\aap},
         year = 2019,
        month = apr,
       volume = {624},
          eid = {A68},
        pages = {A68},
          doi = {10.1051/0004-6361/201833669},
archivePrefix = {arXiv},
       eprint = {1901.01627},
 primaryClass = {astro-ph.EP},
       adsurl = {https://ui.adsabs.harvard.edu/abs/2019A&A...624A..68M}
}

@INPROCEEDINGS{Baycroft_2024,
       author = {{Baycroft}, Thomas A. and {Triaud}, Amaury H.~M.~J. and {Lai}, Dong},
        title = "{Progress report on the BEBOP search for circumbinary planets with radial velocities}",
    booktitle = {Complex Planetary Systems II: Latest Methods for an Interdisciplinary Approach},
         year = 2024,
       editor = {{Lemaitre}, Anne and {Libert}, Anne-Sophie},
       series = {IAU Symposium},
       volume = {382},
        month = jan,
        pages = {51-58},
          doi = {10.1017/S1743921323004180},
archivePrefix = {arXiv},
       eprint = {2409.04191},
 primaryClass = {astro-ph.EP},
       adsurl = {https://ui.adsabs.harvard.edu/abs/2024IAUS..382...51B}
}

@ARTICLE{Glusman_2026,
       author = {{Glusman}, Rowen I. and {Ca{\~n}as}, Caleb I. and {Kanodia}, Shubham and {Han}, Te and {Fernandes}, Rachel B. and {Stef{\'a}nsson}, Gu{\dj}mundur and {Ford}, Eric B. and {Maney}, Marissa and {Monson}, Andrew and {Hotnisky}, Andrew and {Mahadevan}, Suvrath and {Rodruck}, Michael and {Ment}, Kristo and {McWilliam}, Andrew and {Cochran}, William D. and {Col{\'o}n}, Knicole D. and {Giovinazzi}, Mark R. and {Alvarado-Montes}, Jaime A. and {Bender}, Chad F. and {Blake}, Cullen H. and {Boone}, Alexandra and {Diddams}, Scott A. and {Gupta}, Arvind F. and {Halverson}, Samuel and {Krolikowski}, Daniel and {Lin}, Andrea S.~J. and {Ninan}, Joe P. and {Robertson}, Paul and {Roy}, Arpita and {Schwab}, Christian and {Terrien}, Ryan and {Teske}, Johanna and {Wright}, Jason T.},
        title = "{Searching for GEMS: The Occurrence of Giant Planets Orbiting M Dwarfs within 100 pc}",
      journal = {\aj},
         year = 2026,
        month = mar,
       volume = {171},
       number = {3},
          eid = {146},
        pages = {146},
          doi = {10.3847/1538-3881/ae2fbb},
archivePrefix = {arXiv},
       eprint = {2509.08773},
 primaryClass = {astro-ph.EP},
       adsurl = {https://ui.adsabs.harvard.edu/abs/2026AJ....171..146G}
}

@ARTICLE{Sabotta_2021,
       author = {{Sabotta}, S. and {Schlecker}, M. and {Chaturvedi}, P. and {Guenther}, E.~W. and {Mu{\~n}oz Rodr{\'\i}guez}, I. and {Mu{\~n}oz S{\'a}nchez}, J.~C. and {Caballero}, J.~A. and {Shan}, Y. and {Reffert}, S. and {Ribas}, I. and {Reiners}, A. and {Hatzes}, A.~P. and {Amado}, P.~J. and {Klahr}, H. and {Morales}, J.~C. and {Quirrenbach}, A. and {Henning}, Th. and {Dreizler}, S. and {Pall{\'e}}, E. and {Perger}, M. and {Azzaro}, M. and {Jeffers}, S.~V. and {Kaminski}, A. and {K{\"u}rster}, M. and {Lafarga}, M. and {Montes}, D. and {Passegger}, V.~M. and {Zechmeister}, M.},
        title = "{The CARMENES search for exoplanets around M dwarfs. Planet occurrence rates from a subsample of 71 stars}",
      journal = {\aap},
         year = 2021,
        month = sep,
       volume = {653},
          eid = {A114},
        pages = {A114},
          doi = {10.1051/0004-6361/202140968},
archivePrefix = {arXiv},
       eprint = {2107.03802},
 primaryClass = {astro-ph.EP},
       adsurl = {https://ui.adsabs.harvard.edu/abs/2021A&A...653A.114S}
}

@ARTICLE{Endl_2006,
       author = {{Endl}, Michael and {Cochran}, William D. and {K{\"u}rster}, Martin and {Paulson}, Diane B. and {Wittenmyer}, Robert A. and {MacQueen}, Phillip J. and {Tull}, Robert G.},
        title = "{Exploring the Frequency of Close-in Jovian Planets around M Dwarfs}",
      journal = {\apj},
         year = 2006,
        month = sep,
       volume = {649},
       number = {1},
        pages = {436-443},
          doi = {10.1086/506465},
archivePrefix = {arXiv},
       eprint = {astro-ph/0606121},
 primaryClass = {astro-ph},
       adsurl = {https://ui.adsabs.harvard.edu/abs/2006ApJ...649..436E}
}

@ARTICLE{Mignon_2025,
       author = {{Mignon}, L. and {Delfosse}, X. and {Meunier}, N. and {Chaverot}, G. and {Burn}, R. and {Bonfils}, X. and {Bouchy}, F. and {Astudillo-Defru}, N. and {Lo Curto}, G. and {Gaisne}, G. and {Udry}, S. and {Forveille}, T. and {Segransan}, D. and {Lovis}, C. and {Santos}, N.~C. and {Mayor}, M.},
        title = "{Radial velocity homogeneous analysis of M dwarfs observed with HARPS: II. Detection limits and planetary occurrence statistics}",
      journal = {\aap},
         year = 2025,
        month = aug,
       volume = {700},
          eid = {A146},
        pages = {A146},
          doi = {10.1051/0004-6361/202451142},
archivePrefix = {arXiv},
       eprint = {2502.06553},
 primaryClass = {astro-ph.EP},
       adsurl = {https://ui.adsabs.harvard.edu/abs/2025A&A...700A.146M}
}

@ARTICLE{Ment2023,
       author = {{Ment}, Kristo and {Charbonneau}, David},
        title = "{The Occurrence Rate of Terrestrial Planets Orbiting Nearby Mid-to-late M Dwarfs from TESS Sectors 1-42}",
      journal = {\aj},
         year = 2023,
        month = jun,
       volume = {165},
       number = {6},
          eid = {265},
        pages = {265},
          doi = {10.3847/1538-3881/acd175},
archivePrefix = {arXiv},
       eprint = {2302.04242},
 primaryClass = {astro-ph.EP},
       adsurl = {https://ui.adsabs.harvard.edu/abs/2023AJ....165..265M}
}

@ARTICLE{Hsu2020,
       author = {{Hsu}, Danley C. and {Ford}, Eric B. and {Terrien}, Ryan},
        title = "{Occurrence rates of planets orbiting M Stars: applying ABC to Kepler DR25, Gaia DR2, and 2MASS data}",
      journal = {\mnras},
         year = 2020,
        month = oct,
       volume = {498},
       number = {2},
        pages = {2249-2262},
          doi = {10.1093/mnras/staa2391},
archivePrefix = {arXiv},
       eprint = {2002.02573},
 primaryClass = {astro-ph.EP},
       adsurl = {https://ui.adsabs.harvard.edu/abs/2020MNRAS.498.2249H}
}

@ARTICLE{Laughlin_2004,
       author = {{Laughlin}, Gregory and {Bodenheimer}, Peter and {Adams}, Fred C.},
        title = "{The Core Accretion Model Predicts Few Jovian-Mass Planets Orbiting Red Dwarfs}",
      journal = {\apjl},
         year = 2004,
        month = sep,
       volume = {612},
       number = {1},
        pages = {L73-L76},
          doi = {10.1086/424384},
archivePrefix = {arXiv},
       eprint = {astro-ph/0407309},
 primaryClass = {astro-ph},
       adsurl = {https://ui.adsabs.harvard.edu/abs/2004ApJ...612L..73L}
}

@ARTICLE{Coleman2023,
       author = {{Coleman}, Gavin A.~L. and {Nelson}, Richard P. and {Triaud}, Amaury H.~M.~J.},
        title = "{Global N-body simulations of circumbinary planet formation around Kepler-16 and -34 analogues I: Exploring the pebble accretion scenario}",
      journal = {\mnras},
         year = 2023,
        month = jul,
       volume = {522},
       number = {3},
        pages = {4352-4373},
          doi = {10.1093/mnras/stad833},
archivePrefix = {arXiv},
       eprint = {2303.09899},
 primaryClass = {astro-ph.EP},
       adsurl = {https://ui.adsabs.harvard.edu/abs/2023MNRAS.522.4352C}
}

@ARTICLE{Coleman2024,
       author = {{Coleman}, Gavin A.~L. and {Nelson}, Richard P. and {Triaud}, Amaury H.~M.~J. and {Standing}, Matthew R.},
        title = "{Constraining the formation history of the TOI-1338/BEBOP-1 circumbinary planetary system}",
      journal = {\mnras},
         year = 2024,
        month = jan,
       volume = {527},
       number = {1},
        pages = {414-427},
          doi = {10.1093/mnras/stad3216},
archivePrefix = {arXiv},
       eprint = {2310.11898},
 primaryClass = {astro-ph.EP},
       adsurl = {https://ui.adsabs.harvard.edu/abs/2024MNRAS.527..414C}
}

@ARTICLE{Cox2017,
       author = {{Cox}, Erin G. and {Harris}, Robert J. and {Looney}, Leslie W. and {Chiang}, Hsin-Fang and {Chandler}, Claire and {Kratter}, Kaitlin and {Li}, Zhi-Yun and {Perez}, Laura and {Tobin}, John J.},
        title = "{Protoplanetary Disks in {\ensuremath{\rho}} Ophiuchus as Seen from ALMA}",
      journal = {\apj},
         year = 2017,
        month = dec,
       volume = {851},
       number = {2},
          eid = {83},
        pages = {83},
          doi = {10.3847/1538-4357/aa97e2},
archivePrefix = {arXiv},
       eprint = {1711.03974},
 primaryClass = {astro-ph.SR},
       adsurl = {https://ui.adsabs.harvard.edu/abs/2017ApJ...851...83C}
}

@ARTICLE{Flaherty2025,
       author = {{Flaherty}, Kevin and {Knowlton}, Peter and {Smith-Gandy}, Tasan and {Hughes}, A. Meredith and {Kounkel}, Marina and {Jensen}, Eric and {Muzerolle}, James and {Covey}, Kevin},
        title = "{The Influence of Tight Binaries on Protoplanetary Disk Masses}",
      journal = {\aj},
         year = 2025,
        month = dec,
       volume = {170},
       number = {6},
          eid = {330},
        pages = {330},
          doi = {10.3847/1538-3881/ae112b},
archivePrefix = {arXiv},
       eprint = {2510.09823},
 primaryClass = {astro-ph.EP},
       adsurl = {https://ui.adsabs.harvard.edu/abs/2025AJ....170..330F}
}

@article{Kounkel_2019,
doi = {10.3847/1538-3881/ab13b1},
url = {https://dx.doi.org/10.3847/1538-3881/ab13b1},
year = {2019},
month = {apr},
publisher = {The American Astronomical Society},
volume = {157},
number = {5},
pages = {196},
author = {Kounkel, Marina and Covey, Kevin and Moe, Maxwell and Kratter, Kaitlin M. and Suárez, Genaro and Stassun, Keivan G. and Román-Zúñiga, Carlos and Hernandez, Jesus and Kim, Jinyoung Serena and Ramírez, Karla Peña and Roman-Lopes, Alexandre and Stringfellow, Guy S and Jaehnig, Karl O and Borissova, Jura and Tofflemire, Benjamin and Krolikowski, Daniel and Rizzuto, Aaron and Kraus, Adam and Badenes, Carles and Longa-Peña, Penélope and Maqueo Chew, Yilen Gómez and Barba, Rodolfo and Nidever, David L. and Brown, Cody and De Lee, Nathan and Pan, Kaike and Bizyaev, Dmitry and Oravetz, Daniel and Oravetz, Audrey},
title = {Close Companions around Young Stars},
journal = {The Astronomical Journal}
}

@ARTICLE{Gan2025,
       author = {{Gan}, Tianjun and {Theissen}, Christopher A. and {Wang}, Sharon X. and {Burgasser}, Adam J. and {Mao}, Shude},
        title = "{Metallicity Dependence of Giant Planets around M Dwarfs}",
      journal = {\apjs},
         year = 2025,
        month = feb,
       volume = {276},
       number = {2},
          eid = {47},
        pages = {47},
          doi = {10.3847/1538-4365/ad9c65},
archivePrefix = {arXiv},
       eprint = {2412.06137},
 primaryClass = {astro-ph.EP},
       adsurl = {https://ui.adsabs.harvard.edu/abs/2025ApJS..276...47G}
}

@INPROCEEDINGS{Offner_2023,
       author = {{Offner}, S.~S.~R. and {Moe}, M. and {Kratter}, K.~M. and {Sadavoy}, S.~I. and {Jensen}, E.~L.~N. and {Tobin}, J.~J.},
        title = "{The Origin and Evolution of Multiple Star Systems}",
    booktitle = {Protostars and Planets VII},
         year = 2023,
       editor = {{Inutsuka}, S. and {Aikawa}, Y. and {Muto}, T. and {Tomida}, K. and {Tamura}, M.},
       series = {Astronomical Society of the Pacific Conference Series},
       volume = {534},
        month = jul,
        pages = {275},
          doi = {10.48550/arXiv.2203.10066},
archivePrefix = {arXiv},
       eprint = {2203.10066},
 primaryClass = {astro-ph.SR},
       adsurl = {https://ui.adsabs.harvard.edu/abs/2023ASPC..534..275O}
}

@ARTICLE{Offner_2010,
       author = {{Offner}, Stella S.~R. and {Kratter}, Kaitlin M. and {Matzner}, Christopher D. and {Krumholz}, Mark R. and {Klein}, Richard I.},
        title = "{The Formation of Low-mass Binary Star Systems Via Turbulent Fragmentation}",
      journal = {\apj},
         year = 2010,
        month = dec,
       volume = {725},
       number = {2},
        pages = {1485-1494},
          doi = {10.1088/0004-637X/725/2/1485},
archivePrefix = {arXiv},
       eprint = {1010.3702},
 primaryClass = {astro-ph.SR},
       adsurl = {https://ui.adsabs.harvard.edu/abs/2010ApJ...725.1485O}
}

@ARTICLE{Tokovinin_2006,
       author = {{Tokovinin}, A. and {Thomas}, S. and {Sterzik}, M. and {Udry}, S.},
        title = "{Tertiary companions to close spectroscopic binaries}",
      journal = {\aap},
         year = 2006,
        month = may,
       volume = {450},
       number = {2},
        pages = {681-693},
          doi = {10.1051/0004-6361:20054427},
archivePrefix = {arXiv},
       eprint = {astro-ph/0601518},
 primaryClass = {astro-ph},
       adsurl = {https://ui.adsabs.harvard.edu/abs/2006A&A...450..681T}
}

@ARTICLE{Rappaport2013,
       author = {{Rappaport}, S. and {Deck}, K. and {Levine}, A. and {Borkovits}, T. and {Carter}, J. and {El Mellah}, I. and {Sanchis-Ojeda}, R. and {Kalomeni}, B.},
        title = "{Triple-star Candidates among the Kepler Binaries}",
      journal = {\apj},
         year = 2013,
        month = may,
       volume = {768},
       number = {1},
          eid = {33},
        pages = {33},
          doi = {10.1088/0004-637X/768/1/33},
archivePrefix = {arXiv},
       eprint = {1302.0563},
 primaryClass = {astro-ph.SR},
       adsurl = {https://ui.adsabs.harvard.edu/abs/2013ApJ...768...33R}
}

@ARTICLE{Borkovits_2025,
       author = {{Borkovits}, T. and {Rappaport}, S.~A. and {Mitnyan}, T. and {B{\'\i}r{\'o}}, I.~B. and {Cs{\'a}nyi}, I. and {Forg{\'a}cs-Dajka}, E. and {Forr{\'o}}, A. and {Hajdu}, T. and {Seli}, B. and {Sztakovics}, J. and {G{\"o}bly{\"o}s}, A. and {P{\'a}l}, A.},
        title = "{Then and now: A new look at the eclipse timing variations of hierarchical triple star candidates in the primordial Kepler field, revisited by TESS}",
      journal = {\aap},
         year = 2025,
        month = mar,
       volume = {695},
          eid = {A209},
        pages = {A209},
          doi = {10.1051/0004-6361/202453616},
archivePrefix = {arXiv},
       eprint = {2502.09480},
 primaryClass = {astro-ph.SR},
       adsurl = {https://ui.adsabs.harvard.edu/abs/2025A&A...695A.209B}
}

@ARTICLE{Munoz_2015,
       author = {{Mu{\~n}oz}, Diego J. and {Lai}, Dong},
        title = "{Survival of planets around shrinking stellar binaries}",
      journal = {Proceedings of the National Academy of Science},
         year = 2015,
        month = jul,
       volume = {112},
       number = {30},
        pages = {9264-9269},
          doi = {10.1073/pnas.1505671112},
archivePrefix = {arXiv},
       eprint = {1505.05514},
 primaryClass = {astro-ph.EP},
       adsurl = {https://ui.adsabs.harvard.edu/abs/2015PNAS..112.9264M}
}

@ARTICLE{Hamers_2016,
       author = {{Hamers}, Adrian S. and {Perets}, Hagai B. and {Portegies Zwart}, Simon F.},
        title = "{A triple origin for the lack of tight coplanar circumbinary planets around short-period binaries}",
      journal = {\mnras},
         year = 2016,
        month = jan,
       volume = {455},
       number = {3},
        pages = {3180-3200},
          doi = {10.1093/mnras/stv2447},
archivePrefix = {arXiv},
       eprint = {1506.02039},
 primaryClass = {astro-ph.EP},
       adsurl = {https://ui.adsabs.harvard.edu/abs/2016MNRAS.455.3180H}
}

@ARTICLE{Bate_2000,
       author = {{Bate}, Matthew R.},
        title = "{Predicting the properties of binary stellar systems: the evolution of accreting protobinary systems}",
      journal = {MNRAS},
         year = 2000,
        month = may,
       volume = {314},
       number = {1},
        pages = {33-53},
          doi = {10.1046/j.1365-8711.2000.03333.x},
archivePrefix = {arXiv},
       eprint = {astro-ph/0002143},
 primaryClass = {astro-ph},
       adsurl = {https://ui.adsabs.harvard.edu/abs/2000MNRAS.314...33B}
}

@ARTICLE{Bate_2002,
       author = {{Bate}, Matthew R. and {Bonnell}, Ian A. and {Bromm}, Volker},
        title = "{The formation of close binary systems by dynamical interactions and orbital decay}",
      journal = {\mnras},
         year = 2002,
        month = nov,
       volume = {336},
       number = {3},
        pages = {705-713},
          doi = {10.1046/j.1365-8711.2002.05775.x},
archivePrefix = {arXiv},
       eprint = {astro-ph/0212403},
 primaryClass = {astro-ph},
       adsurl = {https://ui.adsabs.harvard.edu/abs/2002MNRAS.336..705B}
}

@ARTICLE{Zhao2013,
       author = {{Zhao}, Bo and {Li}, Zhi-Yun},
        title = "{Orbital and Mass Ratio Evolution of Protobinaries Driven by Magnetic Braking}",
      journal = {\apj},
         year = 2013,
        month = jan,
       volume = {763},
       number = {1},
          eid = {7},
        pages = {7},
          doi = {10.1088/0004-637X/763/1/7},
archivePrefix = {arXiv},
       eprint = {1210.2308},
 primaryClass = {astro-ph.SR},
       adsurl = {https://ui.adsabs.harvard.edu/abs/2013ApJ...763....7Z}
}

@ARTICLE{Fleming2018,
       author = {{Fleming}, David P. and {Barnes}, Rory and {Graham}, David E. and {Luger}, Rodrigo and {Quinn}, Thomas R.},
        title = "{On the Lack of Circumbinary Planets Orbiting Isolated Binary Stars}",
      journal = {\apj},
         year = 2018,
        month = may,
       volume = {858},
       number = {2},
          eid = {86},
        pages = {86},
          doi = {10.3847/1538-4357/aabd38},
archivePrefix = {arXiv},
       eprint = {1804.03676},
 primaryClass = {astro-ph.SR},
       adsurl = {https://ui.adsabs.harvard.edu/abs/2018ApJ...858...86F}
}

@ARTICLE{Liu_2026,
       author = {{Liu}, Bin and {Lai}, Dong},
        title = "{Planetary Desert around Compact Binaries: Dynamical Instability Triggered by Resonance-Induced Eccentricity Excitation}",
      journal = {arXiv e-prints},
         year = 2026,
        month = feb,
          eid = {arXiv:2603.00547},
        pages = {arXiv:2603.00547},
          doi = {10.48550/arXiv.2603.00547},
archivePrefix = {arXiv},
       eprint = {2603.00547},
 primaryClass = {astro-ph.EP},
       adsurl = {https://ui.adsabs.harvard.edu/abs/2026arXiv260300547L}
}

@ARTICLE{Lubow2024,
       author = {{Lubow}, Stephen H. and {Childs}, Anna C. and {Martin}, Rebecca G.},
        title = "{Inclination instability of circumbinary planets}",
      journal = {\mnras},
         year = 2024,
        month = may,
       volume = {530},
       number = {3},
        pages = {2852-2865},
          doi = {10.1093/mnras/stae1040},
archivePrefix = {arXiv},
       eprint = {2404.10080},
 primaryClass = {astro-ph.EP},
       adsurl = {https://ui.adsabs.harvard.edu/abs/2024MNRAS.530.2852L}
}

@ARTICLE{RMartin_2017,
       author = {{Martin}, Rebecca G. and {Lubow}, Stephen H.},
        title = "{Polar Alignment of a Protoplanetary Disk around an Eccentric Binary}",
      journal = {\apjl},
         year = 2017,
        month = feb,
       volume = {835},
       number = {2},
          eid = {L28},
        pages = {L28},
          doi = {10.3847/2041-8213/835/2/L28},
archivePrefix = {arXiv},
       eprint = {1702.00545},
 primaryClass = {astro-ph.EP},
       adsurl = {https://ui.adsabs.harvard.edu/abs/2017ApJ...835L..28M}
}

@ARTICLE{RMartin_2018,
       author = {{Martin}, Rebecca G. and {Lubow}, Stephen H.},
        title = "{Polar alignment of a protoplanetary disc around an eccentric binary - II. Effect of binary and disc parameters}",
      journal = {\mnras},
         year = 2018,
        month = sep,
       volume = {479},
       number = {1},
        pages = {1297-1308},
          doi = {10.1093/mnras/sty1648},
archivePrefix = {arXiv},
       eprint = {1806.08388},
 primaryClass = {astro-ph.EP},
       adsurl = {https://ui.adsabs.harvard.edu/abs/2018MNRAS.479.1297M}
}

@article{Prsa_2005,
doi = {10.1086/430591},
url = {https://doi.org/10.1086/430591},
year = {2005},
month = {jul},
publisher = {},
volume = {628},
number = {1},
pages = {426},
author = {Pr\v{s}a, A. and Zwitter, T.},
title = {A Computational Guide to Physics of Eclipsing Binaries. I. Demonstrations and Perspectives},
journal = {The Astrophysical Journal}
}

@article{Prsa_2016,
doi = {10.3847/1538-4365/227/2/29},
url = {https://doi.org/10.3847/1538-4365/227/2/29},
year = {2016},
month = {dec},
publisher = {The American Astronomical Society},
volume = {227},
number = {2},
pages = {29},
author = {Pr\v{s}a, A. and Conroy, K. E. and Horvat, M. and Pablo, H. and Kochoska, A. and Bloemen, S. and Giammarco, J. and Hambleton, K. M. and Degroote, P.},
title = {PHYSICS OF ECLIPSING BINARIES. II. TOWARD THE INCREASED MODEL FIDELITY},
journal = {The Astrophysical Journal Supplement Series}
}

@ARTICLE{Orosz_2000,
       author = {{Orosz}, J.~A. and {Hauschildt}, P.~H.},
        title = "{The use of the NextGen model atmospheres for cool giants in a light curve synthesis code}",
      journal = {\aap},
         year = 2000,
        month = dec,
       volume = {364},
        pages = {265-281},
          doi = {10.48550/arXiv.astro-ph/0010114},
archivePrefix = {arXiv},
       eprint = {astro-ph/0010114},
 primaryClass = {astro-ph},
       adsurl = {https://ui.adsabs.harvard.edu/abs/2000A&A...364..265O}
}

@ARTICLE{Powell_2021,
       author = {{Powell}, Brian P. and {Kostov}, Veselin B. and {Rappaport}, Saul A. and {Borkovits}, Tam{\'a}s and {Zasche}, Petr and {Tokovinin}, Andrei and {Kruse}, Ethan and {Latham}, David W. and {Montet}, Benjamin T. and {Jensen}, Eric L.~N. and {Jayaraman}, Rahul and {Collins}, Karen A. and {Ma{\v{s}}ek}, Martin and {Hellier}, Coel and {Evans}, Phil and {Tan}, Thiam-Guan and {Schlieder}, Joshua E. and {Torres}, Guillermo and {Smale}, Alan P. and {Friedman}, Adam H. and {Barclay}, Thomas and {Gagliano}, Robert and {Quintana}, Elisa V. and {Jacobs}, Thomas L. and {Gilbert}, Emily A. and {Kristiansen}, Martti H. and {Col{\'o}n}, Knicole D. and {LaCourse}, Daryll M. and {Olmschenk}, Greg and {Omohundro}, Mark and {Schnittman}, Jeremy D. and {Schwengeler}, Hans M. and {Barry}, Richard K. and {Terentev}, Ivan A. and {Boyd}, Patricia and {Schmitt}, Allan R. and {Quinn}, Samuel N. and {Vanderburg}, Andrew and {Palle}, Enric and {Armstrong}, James and {Ricker}, George R. and {Vanderspek}, Roland and {Seager}, S. and {Winn}, Joshua N. and {Jenkins}, Jon M. and {Caldwell}, Douglas A. and {Wohler}, Bill and {Shiao}, Bernie and {Burke}, Christopher J. and {Daylan}, Tansu and {Villase{\~n}or}, Joel},
        title = "{TIC 168789840: A Sextuply Eclipsing Sextuple Star System}",
      journal = {\aj},
         year = 2021,
        month = apr,
       volume = {161},
       number = {4},
          eid = {162},
        pages = {162},
          doi = {10.3847/1538-3881/abddb5},
archivePrefix = {arXiv},
       eprint = {2101.03433},
 primaryClass = {astro-ph.SR},
       adsurl = {https://ui.adsabs.harvard.edu/abs/2021AJ....161..162P}
}

@ARTICLE{Gaia_2016,
       author = {{Gaia Collaboration} and {Prusti}, T. and {de Bruijne}, J.~H.~J. and {Brown}, A.~G.~A. and {Vallenari}, A. and {Babusiaux}, C. and {Bailer-Jones}, C.~A.~L. and {Bastian}, U. and {Biermann}, M. and {Evans}, D.~W. and {Eyer}, L. and {Jansen}, F. and {Jordi}, C. and {Klioner}, S.~A. and {Lammers}, U. and {Lindegren}, L. and {Luri}, X. and {Mignard}, F. and {Milligan}, D.~J. and {Panem}, C. and {Poinsignon}, V. and {Pourbaix}, D. and {Randich}, S. and {Sarri}, G. and {Sartoretti}, P. and {Siddiqui}, H.~I. and {Soubiran}, C. and {Valette}, V. and {van Leeuwen}, F. and {Walton}, N.~A. and {Aerts}, C. and {Arenou}, F. and {Cropper}, M. and {Drimmel}, R. and {H{\o}g}, E. and {Katz}, D. and {Lattanzi}, M.~G. and {O'Mullane}, W. and {Grebel}, E.~K. and {Holland}, A.~D. and {Huc}, C. and {Passot}, X. and {Bramante}, L. and {Cacciari}, C. and {Casta{\~n}eda}, J. and {Chaoul}, L. and {Cheek}, N. and {De Angeli}, F. and {Fabricius}, C. and {Guerra}, R. and {Hern{\'a}ndez}, J. and {Jean-Antoine-Piccolo}, A. and {Masana}, E. and {Messineo}, R. and {Mowlavi}, N. and {Nienartowicz}, K. and {Ord{\'o}{\~n}ez-Blanco}, D. and {Panuzzo}, P. and {Portell}, J. and {Richards}, P.~J. and {Riello}, M. and {Seabroke}, G.~M. and {Tanga}, P. and {Th{\'e}venin}, F. and {Torra}, J. and {Els}, S.~G. and {Gracia-Abril}, G. and {Comoretto}, G. and {Garcia-Reinaldos}, M. and {Lock}, T. and {Mercier}, E. and {Altmann}, M. and {Andrae}, R. and {Astraatmadja}, T.~L. and {Bellas-Velidis}, I. and {Benson}, K. and {Berthier}, J. and {Blomme}, R. and {Busso}, G. and {Carry}, B. and {Cellino}, A. and {Clementini}, G. and {Cowell}, S. and {Creevey}, O. and {Cuypers}, J. and {Davidson}, M. and {De Ridder}, J. and {de Torres}, A. and {Delchambre}, L. and {Dell'Oro}, A. and {Ducourant}, C. and {Fr{\'e}mat}, Y. and {Garc{\'\i}a-Torres}, M. and {Gosset}, E. and {Halbwachs}, J. -L. and {Hambly}, N.~C. and {Harrison}, D.~L. and {Hauser}, M. and {Hestroffer}, D. and {Hodgkin}, S.~T. and {Huckle}, H.~E. and {Hutton}, A. and {Jasniewicz}, G. and {Jordan}, S. and {Kontizas}, M. and {Korn}, A.~J. and {Lanzafame}, A.~C. and {Manteiga}, M. and {Moitinho}, A. and {Muinonen}, K. and {Osinde}, J. and {Pancino}, E. and {Pauwels}, T. and {Petit}, J. -M. and {Recio-Blanco}, A. and {Robin}, A.~C. and {Sarro}, L.~M. and {Siopis}, C. and {Smith}, M. and {Smith}, K.~W. and {Sozzetti}, A. and {Thuillot}, W. and {van Reeven}, W. and {Viala}, Y. and {Abbas}, U. and {Abreu Aramburu}, A. and {Accart}, S. and {Aguado}, J.~J. and {Allan}, P.~M. and {Allasia}, W. and {Altavilla}, G. and {{\'A}lvarez}, M.~A. and {Alves}, J. and {Anderson}, R.~I. and {Andrei}, A.~H. and {Anglada Varela}, E. and {Antiche}, E. and {Antoja}, T. and {Ant{\'o}n}, S. and {Arcay}, B. and {Atzei}, A. and {Ayache}, L. and {Bach}, N. and {Baker}, S.~G. and {Balaguer-N{\'u}{\~n}ez}, L. and {Barache}, C. and {Barata}, C. and {Barbier}, A. and {Barblan}, F. and {Baroni}, M. and {Barrado y Navascu{\'e}s}, D. and {Barros}, M. and {Barstow}, M.~A. and {Becciani}, U. and {Bellazzini}, M. and {Bellei}, G. and {Bello Garc{\'\i}a}, A. and {Belokurov}, V. and {Bendjoya}, P. and {Berihuete}, A. and {Bianchi}, L. and {Bienaym{\'e}}, O. and {Billebaud}, F. and {Blagorodnova}, N. and {Blanco-Cuaresma}, S. and {Boch}, T. and {Bombrun}, A. and {Borrachero}, R. and {Bouquillon}, S. and {Bourda}, G. and {Bouy}, H. and {Bragaglia}, A. and {Breddels}, M.~A. and {Brouillet}, N. and {Br{\"u}semeister}, T. and {Bucciarelli}, B. and {Budnik}, F. and {Burgess}, P. and {Burgon}, R. and {Burlacu}, A. and {Busonero}, D. and {Buzzi}, R. and {Caffau}, E. and {Cambras}, J. and {Campbell}, H. and {Cancelliere}, R. and {Cantat-Gaudin}, T. and {Carlucci}, T. and {Carrasco}, J.~M. and {Castellani}, M. and {Charlot}, P. and {Charnas}, J. and {Charvet}, P. and {Chassat}, F. and {Chiavassa}, A. and {Clotet}, M. and {Cocozza}, G. and {Collins}, R.~S. and {Collins}, P. and {Costigan}, G.},
        title = "{The Gaia mission}",
      journal = {\aap},
         year = 2016,
        month = nov,
       volume = {595},
          eid = {A1},
        pages = {A1},
          doi = {10.1051/0004-6361/201629272},
archivePrefix = {arXiv},
       eprint = {1609.04153},
 primaryClass = {astro-ph.IM},
       adsurl = {https://ui.adsabs.harvard.edu/abs/2016A&A...595A...1G}
}

@ARTICLE{Christiansen_2015,
       author = {{Christiansen}, Jessie L. and {Clarke}, Bruce D. and {Burke}, Christopher J. and {Seader}, Shawn and {Jenkins}, Jon M. and {Twicken}, Joseph D. and {Catanzarite}, Joseph D. and {Smith}, Jeffrey C. and {Batalha}, Natalie M. and {Haas}, Michael R. and {Thompson}, Susan E. and {Campbell}, Jennifer R. and {Sabale}, Anima and {Kamal Uddin}, AKM},
        title = "{Measuring Transit Signal Recovery in the Kepler Pipeline II: Detection Efficiency as Calculated in One Year of Data}",
      journal = {\apj},
         year = 2015,
        month = sep,
       volume = {810},
       number = {2},
          eid = {95},
        pages = {95},
          doi = {10.1088/0004-637X/810/2/95},
archivePrefix = {arXiv},
       eprint = {1507.05097},
 primaryClass = {astro-ph.EP},
       adsurl = {https://ui.adsabs.harvard.edu/abs/2015ApJ...810...95C}
}

@book{brent1973algorithms,
  title={Algorithms for minimization without derivatives},
  author={Brent, Richard P},
  year={1973},
  publisher={Courier Corporation}
}

@ARTICLE{Kostov_2020_toi1338,
       author = {{Kostov}, Veselin B. and {Orosz}, Jerome A. and {Feinstein}, Adina D. and {Welsh}, William F. and {Cukier}, Wolf and {Haghighipour}, Nader and {Quarles}, Billy and {Martin}, David V. and {Montet}, Benjamin T. and {Torres}, Guillermo and {Triaud}, Amaury H.~M.~J. and {Barclay}, Thomas and {Boyd}, Patricia and {Briceno}, Cesar and {Cameron}, Andrew Collier and {Correia}, Alexandre C.~M. and {Gilbert}, Emily A. and {Gill}, Samuel and {Gillon}, Micha{\"e}l and {Haqq-Misra}, Jacob and {Hellier}, Coel and {Dressing}, Courtney and {Fabrycky}, Daniel C. and {Furesz}, Gabor and {Jenkins}, Jon M. and {Kane}, Stephen R. and {Kopparapu}, Ravi and {Hod{\v{z}}i{\'c}}, Vedad Kunovac and {Latham}, David W. and {Law}, Nicholas and {Levine}, Alan M. and {Li}, Gongjie and {Lintott}, Chris and {Lissauer}, Jack J. and {Mann}, Andrew W. and {Mazeh}, Tsevi and {Mardling}, Rosemary and {Maxted}, Pierre F.~L. and {Eisner}, Nora and {Pepe}, Francesco and {Pepper}, Joshua and {Pollacco}, Don and {Quinn}, Samuel N. and {Quintana}, Elisa V. and {Rowe}, Jason F. and {Ricker}, George and {Rose}, Mark E. and {Seager}, S. and {Santerne}, Alexandre and {S{\'e}gransan}, Damien and {Short}, Donald R. and {Smith}, Jeffrey C. and {Standing}, Matthew R. and {Tokovinin}, Andrei and {Trifonov}, Trifon and {Turner}, Oliver and {Twicken}, Joseph D. and {Udry}, St{\'e}phane and {Vanderspek}, Roland and {Winn}, Joshua N. and {Wolf}, Eric T. and {Ziegler}, Carl and {Ansorge}, Peter and {Barnet}, Frank and {Bergeron}, Joel and {Huten}, Marc and {Pappa}, Giuseppe and {van der Straeten}, Timo},
        title = "{TOI-1338: TESS' First Transiting Circumbinary Planet}",
      journal = {\aj},
         year = 2020,
        month = jun,
       volume = {159},
       number = {6},
          eid = {253},
        pages = {253},
          doi = {10.3847/1538-3881/ab8a48},
archivePrefix = {arXiv},
       eprint = {2004.07783},
 primaryClass = {astro-ph.EP},
       adsurl = {https://ui.adsabs.harvard.edu/abs/2020AJ....159..253K}
}

@ARTICLE{Kostov_2021,
       author = {{Kostov}, Veselin B. and {Powell}, Brian P. and {Orosz}, Jerome A. and {Welsh}, William F. and {Cochran}, William and {Collins}, Karen A. and {Endl}, Michael and {Hellier}, Coel and {Latham}, David W. and {MacQueen}, Phillip and {Pepper}, Joshua and {Quarles}, Billy and {Sairam}, Lalitha and {Torres}, Guillermo and {Wilson}, Robert F. and {Bergeron}, Serge and {Boyce}, Pat and {Bieryla}, Allyson and {Buchheim}, Robert and {Ben Christiansen}, Caleb and {Ciardi}, David R. and {Collins}, Kevin I. and {Conti}, Dennis M. and {Dixon}, Scott and {Guerra}, Pere and {Haghighipour}, Nader and {Herman}, Jeffrey and {Hintz}, Eric G. and {Howard}, Ward S. and {Jensen}, Eric L.~N. and {Kielkopf}, John F. and {Kruse}, Ethan and {Law}, Nicholas M. and {Martin}, David and {Maxted}, Pierre F.~L. and {Montet}, Benjamin T. and {Murgas}, Felipe and {Nelson}, Matt and {Olmschenk}, Greg and {Otero}, Sebastian and {Quimby}, Robert and {Richmond}, Michael and {Schwarz}, Richard P. and {Shporer}, Avi and {Stassun}, Keivan G. and {Stephens}, Denise C. and {Triaud}, Amaury H.~M.~J. and {Ulowetz}, Joe and {Walter}, Bradley S. and {Wiley}, Edward and {Wood}, David and {Yenawine}, Mitchell and {Agol}, Eric and {Barclay}, Thomas and {Beatty}, Thomas G. and {Boisse}, Isabelle and {Caldwell}, Douglas A. and {Christiansen}, Jessie and {Col{\'o}n}, Knicole D. and {Deleuil}, Magali and {Doyle}, Laurance and {Fausnaugh}, Michael and {F{\H{u}}r{\'e}sz}, G{\'a}bor and {Gilbert}, Emily A. and {H{\'e}brard}, Guillaume and {James}, David J. and {Jenkins}, Jon and {Kane}, Stephen R. and {Kidwell}, Jr., Richard C. and {Kopparapu}, Ravi and {Li}, Gongjie and {Lissauer}, Jack J. and {Lund}, Michael B. and {Majewski}, Steve R. and {Mazeh}, Tsevi and {Quinn}, Samuel N. and {Quintana}, Elisa and {Ricker}, George and {Rodriguez}, Joseph E. and {Rowe}, Jason and {Santerne}, Alexander and {Schlieder}, Joshua and {Seager}, Sara and {Standing}, Matthew R. and {Stevens}, Daniel J. and {Ting}, Eric B. and {Vanderspek}, Roland and {Winn}, Joshua N.},
        title = "{TIC 172900988: A Transiting Circumbinary Planet Detected in One Sector of TESS Data}",
      journal = {\aj},
         year = 2021,
        month = dec,
       volume = {162},
       number = {6},
          eid = {234},
        pages = {234},
          doi = {10.3847/1538-3881/ac223a},
archivePrefix = {arXiv},
       eprint = {2105.08614},
 primaryClass = {astro-ph.EP},
       adsurl = {https://ui.adsabs.harvard.edu/abs/2021AJ....162..234K}
}

@ARTICLE{Kovacs_2002,
       author = {{Kov{\'a}cs}, G. and {Zucker}, S. and {Mazeh}, T.},
        title = "{A box-fitting algorithm in the search for periodic transits}",
      journal = {\aap},
         year = 2002,
        month = aug,
       volume = {391},
        pages = {369-377},
          doi = {10.1051/0004-6361:20020802},
archivePrefix = {arXiv},
       eprint = {astro-ph/0206099},
 primaryClass = {astro-ph},
       adsurl = {https://ui.adsabs.harvard.edu/abs/2002A&A...391..369K}
}

@ARTICLE{Tamburo_2023AJ,
       author = {{Tamburo}, Patrick and {Muirhead}, Philip S. and {Dressing}, Courtney D.},
        title = "{Predicting the Yield of Small Transiting Exoplanets around Mid-M and Ultracool Dwarfs in the Nancy Grace Roman Space Telescope Galactic Bulge Time Domain Survey}",
      journal = {\aj},
         year = 2023,
        month = jun,
       volume = {165},
       number = {6},
          eid = {251},
        pages = {251},
          doi = {10.3847/1538-3881/acd1de},
archivePrefix = {arXiv},
       eprint = {2303.09959},
 primaryClass = {astro-ph.EP},
       adsurl = {https://ui.adsabs.harvard.edu/abs/2023AJ....165..251T}
}

@ARTICLE{Coleman2024_freefloating,
       author = {{Coleman}, Gavin A.~L.},
        title = "{On the properties of free-floating planets originating in circumbinary planetary systems}",
      journal = {\mnras},
         year = 2024,
        month = may,
       volume = {530},
       number = {1},
        pages = {630-644},
          doi = {10.1093/mnras/stae903},
archivePrefix = {arXiv},
       eprint = {2403.18481},
 primaryClass = {astro-ph.EP},
       adsurl = {https://ui.adsabs.harvard.edu/abs/2024MNRAS.530..630C}
}
\bibliographystyle{aasjournalv7}


\end{document}